\documentclass[
reprint,
superscriptaddress,
pra,
aps,
showpacs,
10pt,
letterpaper
]{revtex4-1}

\usepackage[T1]{fontenc}
\usepackage[utf8]{inputenc}

\usepackage{amsmath}
\usepackage{amssymb}
\usepackage{amsthm}

\usepackage{cases}
\usepackage{mathrsfs}
\usepackage{braket}
\usepackage{bm}

\usepackage{appendix}

\usepackage{graphicx}
\usepackage{dcolumn}
\usepackage{xcolor}
\usepackage{comment}

\usepackage{subfigure}
\usepackage{capt-of}
\usepackage{etoolbox}

\makeatletter
\patchcmd{\normalsize}{11.5}{12}
{}
{\PackageWarning{layout}{Could not patch normalsize}}
\makeatother

\definecolor{linkteal}{RGB}{0,128,128}

\usepackage[
colorlinks=true,
linkcolor=linkteal,
anchorcolor=black,
citecolor=linkteal,
urlcolor=linkteal
]{hyperref}

\newtheorem{definition}{Definition}
\newtheorem{theorem}{Theorem}

\newtheorem{proposition}{Proposition}

\begin{document}
	
	\title{Operational characterization of dynamic imaginarity}
	
	\author{Chuanfa Wu}
	\affiliation{
		Department of Mathematics,
		Nanchang University, Nanchang 330031, China
	}
	\affiliation{
		Institute of Quantum Computing and Software,
		School of Computer Science and Engineering,
		Sun Yat-sen University, Guangzhou 510006, China
	}
	
	\author{Zhaoqi Wu}
	\email{wuzhaoqi@ncu.edu.cn}
	\affiliation{
		Department of Mathematics,
		Nanchang University, Nanchang 330031, China
	}
	
	\author{Mao-Sheng Li}
	\affiliation{
		School of Mathematics,
		South China University of Technology,
		Guangzhou 510641, China
	}
	
	\author{Fei Shi}
	\email{shif26@mail.sysu.edu.cn}
	\affiliation{
		Institute of Quantum Computing and Software,
		School of Computer Science and Engineering,
		Sun Yat-sen University, Guangzhou 510006, China
	}
	
	\author{Shao-Ming Fei}
	\email{feishm@cnu.edu.cn}
	\affiliation{
		School of Mathematics,
		Capital Normal University, Beijing 100048, China
	}
	\affiliation{
		Max-Planck-Institute for Mathematics in the Sciences,
		04103 Leipzig, Germany
	}

\begin{abstract}
The central goal of quantum resource theories is to uncover the fundamental origin of the quantum advantages in quantum information science. Although the resource theory of state imaginarity has been extensively explored, the quantification and operational meaning of imaginarity for quantum channels remain comparatively underexplored. In this work, we investigate the channel imaginarity and its operational significance via state-based and Choi-based quantifiers. We derive a lower bound on the success probability of channel discrimination and show that the imaginarity endows a discrimination advantage. We further obtain an upper bound on the exact conversion rate of quantum channels. Focusing on one-shot exact distillation and dilution of dynamic imaginarity, we derive an analytical expression for the distillation rate, cast the dilution cost as a semidefinite program, and establish their interrelations. Our findings elucidate the roles played  by the dynamic imaginarity as a resource in quantum information processing tasks.
\end{abstract}
\maketitle

\section{Introduction}

The quantum advantages originate from the information processing capabilities provided by quantum states or quantum channels under specific physical restrictions. Once the allowed preparations of quantum states, physical transformations, and measurements are determined, those physical objects that cannot be freely obtained under the corresponding restrictions may become valuable quantum resources \cite{CEGG,BFGG}. Quantum resource theory, characterized by the three fundamental concepts of free states, free channels and resource measures, provides a unified framework for studying such resources. It has been widely applied to nonclassical properties such as quantum entanglement \cite{HRHP,ZKHP,VV,NL,SF2025,WDSF2026}, quantum coherence \cite{BTCM,WAYD,ASGA,CNTR}, and quantum magic \cite{VVMS,LLOS,ZLAW}, not only revealing the common mathematical structures shared by different quantum resources but also providing theoretical tools for analyzing resource costs and operational advantages in quantum computing \cite{FACJ}, quantum precision metrology \cite{LGMB,GVLS}, and other information processing tasks.

The role played by complex numbers in quantum mechanics is a fundamental issue~\cite{LAAF,RK,GN,FML}. Although the probabilities directly observed in experiments are real numbers, quantum states and quantum channels are generally described by complex matrices. While the global phase has no observable effect, the relative phases between probability amplitudes can alter interference patterns, quantum correlations, and the final measurement statistics \cite{CEGG,ZMAS,JJ}. Consequently, the complex structure not only participates in the representation of quantum states but also pervades the processes of quantum state preparation, dynamic evolution and measurement. Recent theoretical and experimental studies on quantum networks have shown that, under the same assumption of local composition, certain correlations allowed by complex quantum theory cannot be realized by the corresponding real Hilbert space models \cite{RMTD,WDJY,ZLYM}. This indicates that, in scenarios subject to physical constraints such as locality and independent preparation, complex phases are not merely a matter of mathematical convenience, but may constitute a physical element capable of influencing realizable quantum correlations and information processing capabilities. Hence, how to quantitatively characterize the resource provided by complex phases and in which tasks such a resource can offer operational advantages has become an important question in quantum information theory.

Based on this understanding, Hickey and Gour \cite{HAGG} established the quantum resource theory of imaginarity. In a given reference basis, quantum states whose density matrices are real are regarded as free states, while quantum states with nonzero imaginary parts possess imaginarity resources. Regarding the quantification of quantum state imaginarity, various measures have been proposed, including the relative entropy of imaginarity \cite{XSGJ}, robustness of imaginarity \cite{HAGG,WKKT} and geometric-like imaginarity \cite{GMLB}. Furthermore, imaginarity witnesses \cite{FCWR,ZLLN}, characterization methods based on Bargmann invariants~\cite{LMTY,ZLXB,XJ26,XJ26+}, set imaginarity~\cite{GY2026}, the behavior of imaginarity under coherent superpositions of orthogonal pure states~\cite{XMLD2026}, and theories of imaginarity resource conversion and manipulation~\cite{WKKT,GMLB,KTDC,SDZB,SWML} have been developed. These works systematically reveal the resource implications of complex phase structures in quantum states.

However, the carriers of practical quantum information processing are not only quantum states but also quantum channels that implement state evolution and information transmission~\cite{LGSI,ASHV}. Unlike a quantum state, which describes the properties of a system at a given instant, a quantum channel characterizes the complete dynamic action on all possible inputs~\cite{GGMM,CHBP,PHXQ}. The imaginarity of states alone provides only a partial description of the resource content of a quantum process, and thus cannot fully characterize its dynamic imaginarity. Therefore, it is of considerable physical significance to study the dynamic imaginarity as a resource theory at the quantum channel level~\cite{RKKW,BSCY,KJEC}.

At present, compared with the quantum state imaginarity~\cite{HAGG,XSGJ,WKKT,GMLB,XMLD2026,KTDC,SDZB,SWML}, the research on dynamic imaginarity is very scarce. The main difficulty lies in the fact that a quantum channel is not a single static object but a dynamic map describing the complete input and output relationship, whose structure is much more complicated than that of a quantum state~\cite{HRHP,ASGA,ASHV,MK}. Moreover, the role of dynamic quantum imaginarity resources in information processing tasks remains unclear. 

To fill this gap, we first construct quantifiers of quantum channel imaginarity from two perspectives. One is state-based, and the other is Choi-based. We investigate the operational meanings of these two types of dynamic imaginarity measures, respectively. The measure based on quantum states can provide quantitative guarantees for quantum channel discrimination performance and reveal the advantages of probe states with imaginarity over real probe states, while Choi-based measures characterize exact channel conversion and the one-shot dynamic imaginarity distillation and dilution.

The remainder of this paper is organized as follows. In Section~\ref{secII}, we introduce the basic notation and review the resource theories of imaginarity for quantum states and quantum channels. In Section~\ref{secIII}, we study channel imaginarity measures based on quantum states and on the Choi state. In Section~\ref{secIV}, we analyze the role of imaginarity in quantum channel discriminations. In Section~\ref{secV}, we construct an exact conversion framework for dynamic imaginarity and investigate the corresponding conversion rates. In Section~\ref{secVI}, we study the one-shot dynamic imaginarity distillation and dilution, and present their relation. We conclude and discuss in Section~\ref{secVII}.

\section{PRELIMINARIES}\label{secII}

Throughout this paper, all Hilbert spaces are finite-dimensional. Capital letters $A, B,...$ label quantum systems associated Hilbert spaces $\mathcal{H}_A, \mathcal{H}_B,...$ of dimensions $d_A, d_B,...$, respectively. For a composite system $AB$, we write $\mathcal{H}_{AB} = \mathcal{H}_A \otimes \mathcal{H}_B$ and $d_{AB} = d_A d_B$. Primed labels such as $A'$ and $B'$ denote the isomorphic copies of the corresponding systems. A reference orthonormal basis is fixed for each system, with corresponding bases chosen for isomorphic copies.

We denote by $\mathcal{L}(\mathcal{H}_A)$, $\mathcal{L}(\mathcal{H}_B),...$ and $\mathcal{D}(\mathcal{H}_A)$, $\mathcal{D}(\mathcal{H}_B),...$ the spaces of linear operators and the sets of density operators on $\mathcal{H}_A$, $\mathcal{H}_B,...$, respectively. System subscripts are omitted when the underlying spaces are clear from the context. The set of completely positive and trace-preserving (CPTP) linear maps from $\mathcal{L}(\mathcal{H}_A)$ to $\mathcal{L}(\mathcal{H}_B)$ is denoted by $\mathcal{C}(A,B)$. Every quantum channel $\mathcal{E} \in \mathcal{C}(A,B)$ admits a Kraus representation, i.e., $\mathcal{E}(\cdot) = \sum_n K_n (\cdot) K_n^\dagger$, in which $\{K_n\}$ is a set of Kraus operators satisfying $\sum_{n} K_{n}^{\dagger}K_{n}= \mathbf{I}_A$. 
We use $\mathbf{I}_A$ and $\mathbb{I}_A$ to denote the identity operator and the identity channel on $A$, respectively. We adopt similar notations for other physical systems. 

We denote by $\mathcal{SC}(AB \to CD)$ the set of superchannels transforming channels in $\mathcal{C}(A,B)$ into channels in $\mathcal{C}(C,D)$. For an operator $X$, $\overline{X}$, $X^T$ and $X^\dagger$ denote its entrywise complex conjugate, transpose and adjoint, respectively, with complex conjugation and transposition taken in the fixed reference bases. In addition, let $\mathbb R$ be the set of real numbers, and
$\mathbb N=\{1,2,\ldots\}$ and $\mathbb N_0=\{0,1,2,\ldots\}$ be the sets of positive and nonnegative integers, respectively,
and all logarithms are taken to base $2$.
\subsection{Static quantum resource of imaginarity}
We recall the quantum resource theory of imaginarity. Let $\{|i\rangle \}_{i=0}^{d-1}$ be an orthonormal basis in a $d$-dimensional Hilbert space $\mathcal{H}$. The resource theory of imaginarity in $\mathcal{H}$ is defined as a tuple $(\mathcal{F}, \mathcal{FC},M)$. Here $\mathcal{F}\subseteq\mathcal{D}(\mathcal{H})$, $\mathcal{FC}$ and $M$ denote free states, free channels and the resource measures of states, respectively. $ \rho\in\mathcal{D}(\mathcal{H})$ is a free state (real state) if 
\begin{align} \label{eq1}
	\rho=\sum\limits_{ij} \rho_{ij} \ |i \rangle  \langle j|,
\end{align}
where ${\rho_{ij}} \in  \mathbb{R} $ for arbitrary $i$,$j \in \{ 0,1,\dots,d-1\}$~\cite{HAGG}. We use $ \mathcal {F}_A$, $ \mathcal {F}_B,...$ to denote the sets of free states on Hilbert space $\mathcal{H}_A$ and $\mathcal{H}_B,...$, respectively.

Then free quantum channels are defined as the quantum channels that map real states to real states. The most trivial free quantum channels are those with $\mathrm{RKO}$ (real Kraus operators) which satisfy $\langle j|K_{n}| i \rangle \in \mathbb {R} \notag $ for arbitrary $i,j$ and $n$, where $K_n$ are the Kraus operators of the quantum channel~\cite{XSGJ,WKKT,WKKT2}. And the largest set of free channels is $\mathrm{RNG}$ (resource non-generating) channels~\cite{HAGG}. We use $\mathcal{FC}(A,B)$ to represent the set of free channels from $\mathcal{D}(\mathcal{H}_{A})$ to $\mathcal{D}(\mathcal{H}_{B})$. If the input and output spaces are clear from the context, we denote it by $\mathcal{FC}$ for simplicity. In the following, we shall refer to a quantum channel as free (or real) if it belongs to the class of $\mathrm{RKO}$ channels.

An imaginarity measure of quantum states is a functional $M$: $\mathcal{D}(\mathcal{H})\rightarrow [0,+\infty)$ satisfying the following properties~\cite{HAGG}:\\
\indent (M1) Faithfulness. $M(\rho)\ge0$, and $M(\rho)= 0$ if and only if $\rho \in \mathcal{F}$.\\
\indent (M2) Monotonicity. $M(\Lambda (\rho)) \le M(\rho)$ whenever $\Lambda$ is a free channel.

(M3) Strong monotonicity. $M(\rho) \ge  \sum_{j}p_{j}M(\rho_{j})$, where $p_{j}=\mathrm{tr}(K_{j} \rho K_{j}^{\dagger})$, $ \rho_{j}=K_{j} \rho K_{j}^{\dagger}/p_{j}  $ with free Kraus operators $K_{j}$.

(M4) Convexity. $\sum_{j}p_{j}M(\rho_{j}) \ge M(\sum_{j}p_{j} \rho_{j}) $ for any set of states $\{\rho_{j}\}$ and any  $p_{j} \ge 0$ with $\sum_{j}p_{j}=1 $.

\indent Similar to the coherence resource theory~\cite{YXZD}, the following property (M5) has also been proposed.

(M5) Additivity. $M(p\rho_{1} \oplus (1-p)\rho_{2})=pM(\rho_{1})+(1-p)M(\rho_2)$, where $p \in(0,1).$\\
\indent In Ref.~\cite{XSGJ}, it has been demonstrated that (M2) and (M5) are equivalent to (M3) and (M4).

\begin{figure}[h]
  	\centering
  	\includegraphics[width=6cm]{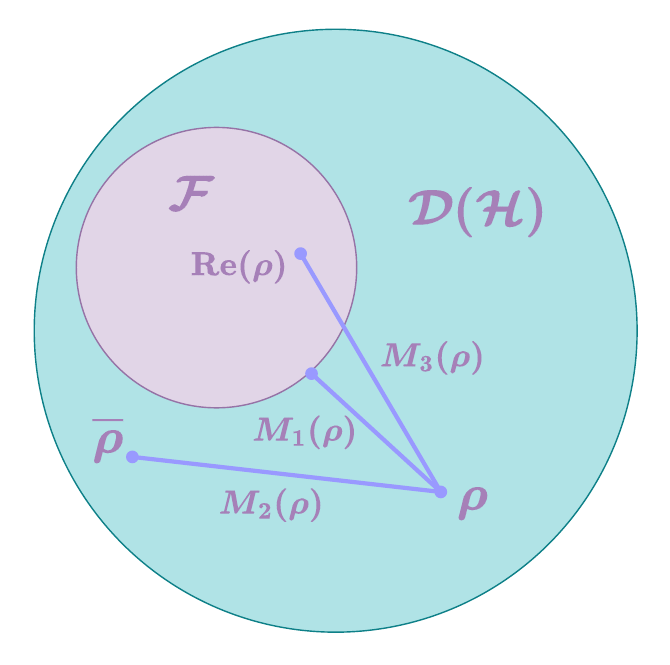}\\
  	\caption{Three approaches in constructing imaginarity measures. \label{fig:Fig1}}
  \end{figure}

There are three commonly used methods to construct imaginarity measures. In the following, $D$ is said to be a distance measure if it satisfies $D(\rho,\sigma)\ge0$ and $D(\rho,\sigma)=0$ iff $\rho=\sigma$ for any $\rho$, $\sigma\in\mathcal{D}(\mathcal{H})$. The most conventional way is to quantify the imaginarity of a state $\rho$ based on the distance between $\rho$ and $\mathcal{F}$,
\begin{align}\label{eq2}
	M_1(\rho)=\min\limits_{\sigma\in \mathcal{F}} D(\rho,\sigma),
\end{align}
where the minimum is taken over all free states~\cite{HAGG,XSGJ,WKKT}. For example, the relative entropy of imaginarity is defined by~\cite{XSGJ}
\begin{align}\label{eq3}
	M_r(\rho)=\min_{\sigma\in\mathcal F} S(\rho\|\sigma)=S\left(\mathrm{Re}(\rho)\right)-S(\rho),
\end{align} the geometric imaginarity is defined by~\cite{WKKT,KTDC}
\begin{align}\label{eq4}
	M_{G}(\rho) = 1-\max_{\sigma\in\mathcal F} F(\rho,\sigma)=\frac{1-\sqrt{F(\rho,\rho^T)}}{2}.
\end{align}
and the trace norm of imaginarity is defined by~\cite{HAGG}
\begin{align}
	M_{\mathrm{tr}}(\rho)=\min_{\sigma\in\mathcal{F}}\lVert\rho-\sigma\rVert_1=\frac{1}{2}\lVert\rho-\rho^T\rVert_1,
\end{align} 
where $S(\rho||\sigma)=\mathrm{tr}(\rho\log \rho)-\mathrm{tr}(\rho\log \sigma)$ is the relative entropy, $\mathrm{Re}(\rho)=\frac{1}{2}(\rho+\rho^T)$ and $S(\rho)=-\mathrm{tr}(\rho\log \rho)$ is the von Neumann entropy, $\sqrt{F(\rho,\sigma)}= \mathrm{tr} \sqrt{\sqrt{\rho} \sigma \sqrt{\rho}} $ is the Uhlmann fidelity, and $\lVert X \rVert_1$ denotes the trace norm of $X$.
Moreover, the robustness of imaginarity is defined by~\cite{HAGG,WKKT}
\begin{align}\label{eq6}
	M_R(\rho)=\min_{\tau\in\mathcal{D}(\mathcal{H})}
	\left\{ s\ge 0 \;:\; \frac{\rho + s\tau}{1+s} \in \mathcal{F} \right\}
	=\frac12\bigl\|\rho-\rho^T\bigr\|_1.
\end{align} 
In fact, the analytical expressions for the trace norm of imaginarity and the robustness of imaginarity are the same.
Alternatively, with respect to (\ref{eq2}), one has $M_2(\rho)= D(\rho,\overline{\rho})$~\cite{XJ24,WCWZ2} and
$M_3(\rho)= D(\rho,\mathrm{Re}(\rho))$~\cite{LJSY} for appropriate distance measure $D$. The relations among $M_1(\rho)$, $M_2(\rho)$ and $M_3(\rho)$ are illustrated schematically in Fig. \ref{fig:Fig1}.

\subsection{Dynamic quantum resource of imaginarity}

Similar to the imaginarity resource of quantum states, the dynamic one is also defined as a tuple $(\mathcal{FC},\mathcal{FSC},C)$. Here, $\mathcal{FSC}$ is the set of free superchannels, and $C$ is the imaginarity measure of quantum channels. An arbitrary superchannel $\Theta$ that transforms a quantum channel $\mathcal{N}_{A\rightarrow B}$ to a quantum channel $\mathcal{M}_{C\rightarrow D}$ can be written as~\cite{GCGM,GCGM+}
\begin{equation}\label{eq8}
	\Theta(\mathcal{N}_{A\rightarrow B})= \mathcal{Q}_{EB\rightarrow D}\circ(\mathbb{I}_E\otimes \mathcal{N}_{A\rightarrow B})\circ \mathcal{P}_{C\rightarrow EA} =\mathcal{M}_{C\rightarrow D},
\end{equation}
where $\mathbb{I}_E$ is the identity channel on the ancillary system $E$, and $\mathcal{Q}_{EB\rightarrow D}$ and $\mathcal{P}_{C\rightarrow EA}$ are quantum channels. In the following, the subscripts will be dropped whenever the meaning is clear from the context, that is, we rewrite Eq.~\eqref{eq8} as $	\Theta(\mathcal{N})= \mathcal{Q}\circ(\mathbb{I}_E\otimes \mathcal{N})\circ \mathcal{P}$. 

The set of free superchannels is a subset of superchannels that map free channels to free channels.
\begin{definition}\label{def1}
 $\Theta$ is a free superchannel if $\mathcal{Q}$ and $\mathcal{P}$ in Eq. \eqref{eq8} are RKO channels. 
 \end{definition}
 
Here we choose $\mathcal{Q}$ and $\mathcal{P}$ as RKO channels rather than RNG channels, since all the imaginarity measures of quantum states mentioned in Eqs.~(\ref{eq3})-(\ref{eq6}) are formulated within the RKO setting.
 
Recently, Ref.~\cite{TPSY} introduced a class of free superchannels,  with the definition given as follows.

\begin{definition}\label{def2}
	A superchannel $\Theta $ is said to be free if it commutes with the channel conjugation, i.e., for any quantum channel $\mathcal{N}$, it holds that 
\begin{align}\label{eq9}
		\overline{\Theta(\mathcal{N})}=\Theta(\overline{\mathcal{N}}) ,
\end{align}
where the conjugate channel $\overline{\mathcal{N}}$ of $\mathcal{N}$ is defined with respect to Kraus representation $\mathcal{N}(\cdot) = \sum_\alpha K_\alpha (\cdot) K_\alpha^\dagger$ as
	\begin{align}
			\overline{\mathcal{N}}(\cdot) = \sum_\alpha \overline{K}_\alpha (\cdot) \overline{K}_\alpha^\dagger.
	\end{align}
\end{definition}

In consistent with the full space supermap convention stated above, the pre- and post-processing maps in Definition~\ref{def1} remain RKO channels, while the corresponding realization formula applies to any completely positive input map, without requiring the input map to be trace-preserving. Accordingly, in Definition \ref{def2} we require that Eq.~\eqref{eq9} holds for any CP maps, not only for quantum channels.
We clarify the relationship between the two classes of free superchannels introduced above.

\begin{proposition}\label{prop1}
	Definition \ref{def1} and Definition \ref{def2} are equivalent.
\end{proposition}
The proof is provided in Appendix~\hyperref[app:A1]{A1}.

Then an imaginarity measure of quantum channel $C$ is a real-valued functional satisfying the following properties:\\
\indent (C1) Nonnegativity. $C(\mathcal{N})\ge0$ for any quantum channel $\mathcal{N}$, and $C(\mathcal{N})= 0$ if $\mathcal{N} $ is a free channel.\\
\indent (C2) Monotonicity. $C[\Theta (\mathcal{N})]\le C(\mathcal{N})$ for any quantum channel $\mathcal{N}$ and $\Theta \in\mathcal{FSC}$. 

\subsection{Choi-Jamio\l kowski isomorphism}
The Choi-Jamio\l kowski isomorphism associates each channel with a quantum state while preserving the complete information about its input-output action, thereby allowing techniques developed for state resource theories to be adapted to dynamic settings. The Choi state of any $\mathcal{N}\in \mathcal{C}(A,B)$ is given by \cite{AJ,MDC},
\begin{equation}\label{eq31}
J(\mathcal{N})=(\mathbb{I}_{A}\otimes \mathcal{N}_{A'\rightarrow B} )(\phi_{A A'}),
\end{equation}
where $\phi_{AA' }=\frac{1}{d_{A'}}\sum_{i,j=0}^{d_A'-1}|ii\rangle \langle jj|_{AA'}$ is a maximally entangled state.
Here $J(\mathcal{N})$ is defined for quantum channels $\mathcal{N}$, but we allow it to be extended to all linear maps. For a quantum channel $\mathcal{N}$, it means that $J(\mathcal{N})$ is positive semidefinite, i.e., $J(\mathcal{N})\succeq 0$ and $\mathrm{tr}_{B}[J(\mathcal{N})]=\frac{1}{d_A}\mathbf{I}_{A}$.

Given a Choi state, the corresponding quantum channel is given by~\cite{GCGM} 
\begin{equation}\label{eq35-}
	J^{-1}(J(\mathcal{N}))(\rho)=\mathcal{N}(\rho)=d_A \mathrm{tr}_{A}\left[ (\rho^T\otimes \mathbf{I}_{B})J(\mathcal{N})\right].
\end{equation}
For any quantum superchannel $\Theta \in \mathcal{SC}(AB\rightarrow CD)$ and quantum channel $\mathcal{N}$, there exists a map 
\begin{equation}\label{eq33}
\mathcal{E}_{\Theta}=J\circ \Theta\circ J^{-1} : J(\mathcal{N})\rightarrow J(\Theta(\mathcal{N}))
\end{equation}
such that
\begin{equation}\label{eq34}
	J(\Theta(\mathcal{N}))=\mathcal{E}_{\Theta}(J(\mathcal{N})).
\end{equation}
Here $\mathcal{E}_{\Theta}$  is defined on the entire operator space and maps $\mathcal{L}(\mathcal{H}_A \otimes \mathcal{H}_B)$ to $\mathcal{L}(\mathcal{H}_C \otimes \mathcal{H}_D)$~\cite{GCGM+,MHHY}.

It is worth noting that $\mathcal{E}_{\Theta}$ is CP, but not necessarily CPTP~\cite{GCGM+}. And $\mathcal{E}_{\Theta}$ is TP if and only if its Choi operator $J(\mathcal{E}_{\Theta})$ satisfies
\begin{equation}
	\mathrm{tr}_{CD}[J(\mathcal{E}_{\Theta})]=\frac{\mathbf{I}_{A}}{d_{A}} \otimes \frac{\mathbf{I}_{B}}{d_{B}}.
\end{equation}
We present below another characterization of $\mathcal{E}_{\Theta}$ defined by Eq.~\eqref{eq33}.

\begin{proposition}\label{prop2-}
	For any quantum superchannel $\Theta$ realized by Eq.~\eqref{eq8}, the map $\mathcal{E}_{\Theta}$ induced by $\Theta$ is CPTP if and only if
	\begin{align}\label{eq39--}
		\operatorname{tr}_E \mathcal{P}\left(\frac{\mathbf{I}_C}{d_C}\right) = \frac{\mathbf{I}_A}{d_A}.
	\end{align} 
\end{proposition}
The proof is detailed in Appendix~\hyperref[app:A2]{A2}.

Based on the above characterization, we define a class of free superchannels $\mathcal{FSC}_{\mathrm{CPTP}}$,
\begin{align}\label{eq39-}
	 \mathcal{FSC}_{\mathrm{CPTP}}=\{ \Theta\in\mathcal{FSC}: \mathcal{E}_{\Theta}=J\circ \Theta\circ J^{-1} \notag\\ \text{ is trace-preserving} \}.
\end{align}
As an example of $\mathcal{FSC}_{\mathrm{CPTP}}$, consider the following subset of free superchannels $\mathcal{LFSC}$,
\begin{align}
	\mathcal{LFSC}=\{\Theta\in\mathcal{FSC}: \Theta({\mathcal{N})=\mathcal{Q}\circ(\mathbb{I}_E\otimes\mathcal{N})} \text{ for any } \mathcal{N}\}.
\end{align}
It is evident that $\mathcal{LFSC}$ is obtained by setting 
$\mathcal{P}=\mathbb{I} $ in Definition~\ref{def1}. Here, it should be emphasized that if a channel $\mathcal{N}$ can be converted to $\mathcal{M}$ via $\mathcal{LFSC}$, it necessarily implies that the input dimension of $\mathcal{M}$ is not smaller than that of $\mathcal{N}$.

For any $\Theta\in \mathcal{LFSC}$, we have 
\begin{align}
	J(\Theta(\mathcal{N}))&=J(\mathcal{Q}\circ(\mathbb{I}_E\otimes\mathcal{N})) \notag\\
	&=(\mathbb{I}_{EA} \otimes \mathcal{Q})J(\mathbb{I}_E\otimes\mathcal{N})\notag \\
	&=(\mathbb{I}_{EA} \otimes \mathcal{Q})[J(\mathbb{I}_E)\otimes J(\mathcal{N})].
\end{align}
Taking $\mathcal{E}_{\Theta}(\cdot)=(\mathbb{I}_{EA} \otimes \mathcal{Q})[J(\mathbb{I}_E)\otimes (\cdot)]$, it is easy to see that $\mathcal{E}_{\Theta}$ is a quantum channel. It follows that $\mathcal{LFSC}\subseteq 	\mathcal{FSC}_{\mathrm{CPTP}}$. In general, the class of free superchannels is too broad for us to study operational dynamic quantum resources. However, if we focus only on this special case, the situation becomes considerably simpler. In Section~\ref{secVI}, we show that, for quantum channel dilution under $\mathcal{FSC}$, it suffices to consider operations in $\mathcal{FSC}_{\mathrm{CPTP}}$. This substantially simplifies the calculation of the dilution cost.

Note that the following superchannel
\begin{align}
	\mathcal{SC}_{\mathrm{CPTP}}=\{ \Theta\in\mathcal{SC}: \mathcal{E}_{\Theta}=J\circ \Theta\circ J^{-1}  \notag\\
	\text{ is trace-preserving} \}
\end{align}
has been studied in Refs.~\cite{MHHY,GG2018,SP2026}. Clearly, $\mathcal{FSC}_{\mathrm{CPTP}}\subseteq\mathcal{SC}_{\mathrm{CPTP}}$.
The class of quantum superchannels defined by Eq.~\eqref{eq39-} connects information loss caused by physical processing with recoverability and provides a theoretical basis for approximate quantum error correction~\cite{MJ2018,JC2019}. The quantum channel \(\mathcal{E}_{\Theta}\) induced by superchannels $\Theta$ defined in Eq.~\eqref{eq33} plays an important role in quantum channel recovery~\cite{SP2026}. The following property of $\mathcal{E}_{\Theta}$ induced by superchannel $\Theta$ will be used later.

\begin{proposition}\label{prop2}
Let $\Theta\in\mathcal{FSC}_{\mathrm{CPTP}}$ be a free superchannel. Then $\mathcal{E}_{\Theta}$ defined by Eq.~\eqref{eq33} is a free channel.
\end{proposition}
The proof is provided in Appendix~\hyperref[app:A3]{A3}.

\section{Quantifying imaginarity of quantum channels}\label{secIII}
In this section, we introduce two classes of imaginarity measures of quantum channels, the state-based and the Choi-based imaginarity measures, and discuss some elegant properties and relations between them.
\subsection{State-based imaginarity measure of quantum channels}
The state-based imaginarity measure of quantum channels is induced from an imaginarity measure of quantum states.
Following Refs. \cite{TMRS,WZZL,LLBK,LYYX,LZWA,ZJLY}, we define imaginarity measures of a quantum channel $\mathcal{N}$ as 
\begin{equation}\label{eq58}
	C^g_M(\mathcal{N})=\max_{\rho\in\mathcal{F}} M[\mathcal{N}(\rho)],
\end{equation}
 where $M(\cdot)$ is an imaginarity measure of states defined by Eq.~\eqref{eq2}. For example, taking $M$ as the robustness of imaginarity, we have
 \begin{align}
 C_{M_R}^g(\mathcal{N})=\frac{1}{2} \max_{\rho\in\mathcal{F}}\lVert \mathcal{N}(\rho)-[\mathcal{N}(\rho)]^T\rVert_1.
 \end{align}

Alternatively, as in Ref.~\cite{ZJLY}, the imaginarity measures of quantum channels can also be defined as
\begin{equation}\label{eq59}
	C^b_M(\mathcal{N})=\max_{\rho\in \mathcal{D(H})} \{M[\mathcal{N}(\rho)]-M(\rho)\},
\end{equation}
 where $M(\cdot)$ is an imaginarity measure of states defined by Eq.~\eqref{eq2}.
 

Note that $C^g_M$ defined by Eq.~\eqref{eq58} and $C^b_M$ defined by Eq.~\eqref{eq59} are monotonic under memoryless free superchannels $(\mathcal{MLFSC})$, i.e., free superchannel $\Theta$ admitting a realization of the form,
\begin{align}
\Theta(\mathcal{N}) = \mathcal{Q} \circ\mathcal{N}\circ \mathcal{P},
\end{align}
where $\mathcal{Q}$ and $\mathcal{P}$ are RKO channels.

\begin{proposition}\label{prop3}
$C^g_M$ defined by Eq.~\eqref{eq58} and $C^b_M$ defined by Eq.~\eqref{eq59} are dynamic imaginarity measures under $\mathcal{MLFSC}$.
\end{proposition}

The proof is provided in Appendix~\hyperref[app:B1]{B1}.
Eqs.~\eqref{eq58} and~\eqref{eq59} characterize the maximum extent to which the imaginarity resource can generate after a quantum state undergoes a quantum channel.
It follows from Ref.~\cite[Lemma~2]{LLBK} that if $D$ in Eq.~\eqref{eq59} satisfies the triangle inequality and the data-processing inequality, one obtains that $	C^g_M(\mathcal{N})=C^b_M(\mathcal{N})$ for any channel $\mathcal{N}$.  We next present the relation between the relative entropy of imaginarity of quantum channels $C^g_{M_r}(\mathcal{N})$ and geometric imaginarity of quantum channels $C^g_{M_G}(\mathcal{N})$.

\begin{proposition}\label{prop4}
For any quantum channel $\mathcal{N}$, the imaginarity measures of quantum channels $C^g_{M_r}(\mathcal{N})$ and $C^g_{M_G}(\mathcal{N})$ satisfy 
\begin{align}\label{eq66}
	2C^g_{M_G}(\mathcal{N}) \le C^g_{M_r}(\mathcal{N}).
\end{align}
\end{proposition}

The proof is provided in Appendix~\hyperref[app:B2]{B2}.
Proposition~\ref{prop4} establishes a rigorous quantitative order between the two dynamic imaginarity measures, one based on relative entropy and the other on geometric distance. Specifically, it shows that the relative entropy of imaginarity is uniformly bounded from below by twice the geometric imaginarity, thereby providing a precise hierarchy between the two resource quantifiers. This result not only deepens our understanding of the hierarchical structure of imaginarity resource in quantum channels, but also provides a fundamental comparative framework for investigating the equivalence and discrepancy among different measures in more general channel settings. 

\begin{figure}[h]
	\centering
	\includegraphics[width=\linewidth]{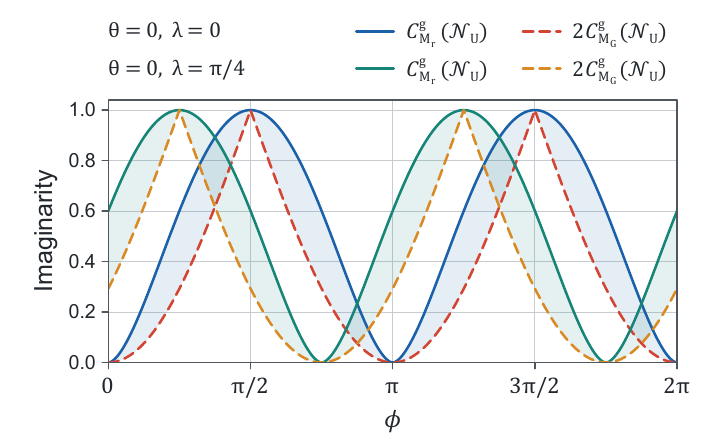}\\
\caption{
	Comparison of two dynamic imaginarity quantifiers for the channel $\mathcal{N}_U$, with $\theta=0$ and $\phi\in[0,2\pi]$. The solid and dashed curves represent
	$C_{M_r}^{g}(\mathcal{N}_U)$ and $2C_{M_G}^{g}(\mathcal{N}_U)$, respectively. The blue solid and red dashed curves correspond to $\lambda=0$,
	whereas the teal solid and orange dashed curves correspond to $\lambda=\pi/4$.
	The shaded regions indicate the gap between the two quantifiers. For $\lambda=0$, equality holds at
	$\phi=0,\pi/2,\pi,3\pi/2,2\pi$;
	for $\lambda=\pi/4$, equality holds at
	$\phi=\pi/4,3\pi/4,5\pi/4,7\pi/4$.
	At all other points in the respective cases,
	$2C_{M_G}^{g}(\mathcal{N}_U)<C_{M_r}^{g}(\mathcal{N}_U)$.
}\label{fig:Fig2}
\end{figure}

For a more intuitive illustration of Proposition~\ref{prop4}, we consider a specific example. It is known that a general qubit unitary channel $\mathcal{N}_{U}$ can be induced by a unitary operator of the following form~\cite{NMCI},
\begin{align}\label{eq81}
	U(\theta, \phi, \lambda) =
	\begin{pmatrix}
		e^{-\mathrm{i}(\phi + \lambda)/2} \cos \frac{\theta}{2} & -e^{\mathrm{i}(\phi - \lambda)/2} \sin \frac{\theta}{2} \\
		e^{-\mathrm{i}(\phi - \lambda)/2} \sin \frac{\theta}{2} & e^{\mathrm{i}(\phi + \lambda)/2} \cos \frac{\theta}{2}
	\end{pmatrix},
\end{align} where $\theta, \phi, \lambda\in[0,2\pi]$. Setting $\theta=0$, we consider the channel
$\mathcal N_U(\rho)=U(0,\phi,\lambda)\rho U^\dagger(0,\phi,\lambda)$.
By direct calculations, we obtain that
\begin{align}
	C^g_{M_r}(\mathcal{N}_U)=h\left(\frac{1+|\cos(\phi+\lambda)|}{2}\right)
\end{align} and
\begin{align}
	C^g_{M_G}(\mathcal{N}_U)=\frac{1-|\cos(\phi+\lambda)|}{2},
\end{align} 
where $h(x)$ is the binary Shannon entropy defined by
$h(x) = -x \log x - (1-x) \log(1-x)$ with $x\in[0,1]$. 
The plots of $C^g_{M_r}(\mathcal{N}_U)$ and $2C^g_{M_G}(\mathcal{N}_U)$ are shown in Fig.~\ref{fig:Fig2}.

\subsection{Choi-based imaginarity measure of quantum channels}
Using the Choi-Jamio\l kowski isomorphism to characterize the resource of quantum channels has been extensively and deeply studied in coherence resource theory and even in general resource theories~\cite{EZCM}. The main idea is to define the resource of a quantum channel as the resource of its Choi state. In the vast majority of existing studies such as~\cite{XJ19}, resource measures of quantum channels based on the Choi state are constructed by quantifying the distance from the Choi state of the channel to the set of free states. 

In the resource theory of imaginarity, Refs.~\cite{CXLQ,FSDT,TPSY} characterize the imaginarity of a quantum channel by quantifying the distance between the Choi states of the channel and that of its conjugate channel, while Ref.~\cite{WCWZ} addresses this quantification from the perspectives of detecting and creating imaginarity. Furthermore, the imaginarity of Gaussian channels in infinite dimensional settings has also been investigated in Ref.~\cite{ZTHJ}. Recently, in Ref.~\cite{TPSY,MHHY} the resource measure of quantum channels is constructed by quantifying the distance from the Choi state of the channel to the set of Choi states of free channels, which is closely related to the conversion of quantum resources. Inspired by Ref.~\cite{MHHY}, we define the following Choi-based quantifiers for quantum channel level,
\begin{align}\label{eq84}
C^J_D(\mathcal{N})=\min_{\mathcal{M}\in \mathcal{FC}}D(J(\mathcal{N}),J(\mathcal{M})),
\end{align}
where $D$ is a distance measure. 

\begin{proposition}\label{prop5}
	$C^J_D(\mathcal{N})$ defined by Eq.~\eqref{eq84} is an imaginarity measure of quantum channels under $\mathcal{FSC}_{\mathrm{CPTP}}$ if the distance measure $D$ satisfies the data-processing inequality.
\end{proposition}

The proof is provided in Appendix~\hyperref[app:B3]{B3}.
In the resource theory of coherence, Ref.~\cite{TTDE} points out that, although the Choi-Jamio\l kowski isomorphism faithfully represents arbitrary quantum channels, it does not allow a direct reduction of the resource theory of coherence for channels to that for states. In other words, the coherence of the Choi state cannot be directly equated with the coherence of the corresponding quantum channels. The origin of this limitation lies in the fact that the Choi images of incoherent channels do not coincide with the set of incoherent states. Consider, for example, the identity channel $\mathbb{I}$, it is clearly an incoherent channel, as it maps every incoherent state to itself. However, its corresponding Choi state is a maximally entangled state and thus possesses coherence resource. Therefore, directly defining the coherence of a quantum channel via the coherence of its Choi state would inevitably violate the physical reasonableness required by the resource theory.
Nevertheless, within the framework of the resource theory of imaginarity, a quantum channel is a RKO channel if and only if its corresponding Choi state is a free (real) state~\cite{FSDT,EZCM}. Therefore, characterizing the imaginarity resource of a quantum channel via its Choi state, as in Eq.~\eqref{eq84}, is mathematically sound and physically reasonable. 

Taking $D$ as the relative entropy and the geometric distance in Eq. \eqref{eq84}, respectively, we have
\begin{equation}\label{eq85}
	C^J_S(\mathcal{N})=\min_{\mathcal{M}\in \mathcal{FC}}S(J(\mathcal{N})\Vert J(\mathcal{M}))
\end{equation}
and 
\begin{equation}\label{eq86}
		C^J_G(\mathcal{N})=1-\max_{\mathcal{M}\in \mathcal{FC}}F(J(\mathcal{N}),J(\mathcal{M})).
\end{equation}

We now provide the analytical expression for $C^J_S(\mathcal{N})$ along with some of its elegant properties.

\begin{proposition}\label{prop6}
For any quantum channel $\mathcal{N}\in\mathcal{C}(A,B)$, we have
\begin{align}\label{eq90}
C^J_S(\mathcal{N})=S[\mathrm{Re}(J(\mathcal{N}))]-S[J(\mathcal{N})].
\end{align}
\end{proposition}

The proof of Proposition~\ref{prop6} is given in Appendix~\hyperref[app:B4]{B4}, from which it follows that for any channel $\mathcal{N}$, $\frac{\mathcal{N}+\overline{\mathcal{N}}}{2}$ is a free channel, and its Choi state is $\mathrm{Re}[J(\mathcal{N})]$. Thus, it is observed that
\begin{align}
	C_{\mathrm{tr}}^J(\mathcal{N})&=\min_{\mathcal{M}\in\mathcal{FC}}\lVert J(\mathcal{N})-J(\mathcal{M})\rVert_1 \notag\\
	&=\frac{1}{2}\lVert J(\mathcal{N})-J(\mathcal{N})^T\rVert_1.
\end{align}

We now clarify the relations among the above-mentioned imaginarity measures of quantum channels.
\begin{proposition}\label{prop10}
	For any quantum channel $\mathcal{N}\in\mathcal{C}(A,B)$, we have
	\begin{align}\label{eq120}
		&\frac{1-\sqrt{F[J(\mathcal{N}),J(\mathcal{N})^T]}}{2}\le C^J_G(\mathcal{N})\notag \\
		&\qquad\le 1-F[J(\mathcal{N}),\mathrm{Re}(J(\mathcal{N}))].
	\end{align}
\end{proposition}

The proof is provided in Appendix~\hyperref[app:B8]{B5}.
Proposition~\ref{prop10} provides upper and lower bounds for $C^J_G(\mathcal{N})$. Although obtaining a closed-form expression of $C^J_G(\mathcal{N})$ is generally challenging, it can be computed via the following semidefinite programming (SDP).

\begin{proposition}\label{prop11-}
For any $\mathcal{N}\in\mathcal{C}(A,B)$, we have
	\begin{equation}
		C_G^J(\mathcal N)=1-u^2,
	\end{equation}
	where $u$ is the optimal value of the following
	SDP:
	\begin{align}\label{eq130}
			\text{maximize}\quad&
			\operatorname{tr}[\operatorname{Re}(X)]\notag\\
			\text{subject\ to}\quad&
			\begin{pmatrix}
				J(\mathcal N) & X \notag\\
				X^\dagger & \sigma
			\end{pmatrix}\succeq0,\notag\\
			&\sigma=\sigma^T,\notag\\
			&\operatorname{tr}_B\sigma=\frac{\mathbf{I}_A}{d_A},\notag\\
			&X\in\mathcal{L}
			(\mathcal{H}_A\otimes\mathcal{H}_B),\notag\\
			&\sigma\in\mathcal{D}
			(\mathcal H_A\otimes\mathcal H_B).
	\end{align}
\end{proposition}
The proof is provided in Appendix~\hyperref[app:B9]{B6}.

\section{ Quantum channel discrimination tasks }\label{secIV}
The quantum channel discrimination is one of the most fundamental tasks in quantum information theory.
In this section, we present the applications of the state-based imaginarity measures of quantum channels in quantum channel discrimination tasks. The central objective is to identify an unknown quantum dynamic process, drawn from a known set of candidate channels, with the maximal possible success probability by appropriately designing the initial probe state and the subsequent measurement scheme. In other words, this task amounts to utilizing the information encoded in the evolution of quantum states to infer the identification of the unknown process.

Given a finite set of candidate channels $\{\mathcal{N}_i\}_{i=1}^n$, each occurring with a known prior probability $p_i$. The task is to choose an initial probe state $\rho$, send it through the unknown channel, and perform an $n$-outcome positive operator-valued measure (POVM) $\{\Pi_i\}$ on the output state, so as to maximize the average success probability 
$\sum_{i=1}^n p_i \, \mathrm{tr}[\mathcal{N}_i(\rho) \Pi_i]$ of inferring that the channel is $\mathcal{N}_i$ when the measurement outcome is $i$~\cite{WJ}.

When $n = 2$ and the prior probabilities are equal, the problem reduces to the most basic scenario of binary channel discrimination. In this case, the success probability is given by 
\begin{align}\label{eq134}
	&p_{\mathrm{succ}}(\mathcal{N},\mathcal{M},\rho)\notag \\ &=\max_{\{\Pi,\mathbf{I}-\Pi\}} \left\{  \frac{1}{2}\mathrm{tr}\left[\mathcal{N}(\rho)\Pi\right ]+\frac{1}{2}\mathrm{tr}\left[\mathcal{M}(\rho)(\mathbf{I}-\Pi)\right] \right\},
\end{align} where the maximum is taken over all POVMs given by $\{\Pi,\mathbf{I}-\Pi\}$~\cite{WJ}. With the aid of Holevo-Helstrom theorem~\cite{HCW}, the optimization problem in Eq.~\eqref{eq134} can be solved exactly as
\begin{align}
	p_{\mathrm{succ}}(\mathcal{N},\mathcal{M},\rho)=\frac{1}{2}+\frac{1}{4}\lVert\mathcal{N}(\rho)-\mathcal{M}(\rho)\rVert_1.
\end{align}
Therefore, the maximal success probability over all probe states is given by
\begin{align}\label{eq137}
p_{\mathrm{succ}}(\mathcal{N}, \mathcal{M}, \mathcal{D}(\mathcal{H})) = \max_{\rho \in \mathcal{D}(\mathcal{H})} p_{\mathrm{succ}}(\mathcal{N}, \mathcal{M}, \rho).
\end{align} 
When the probe states are restricted to the set of free states $\mathcal{F}$, the success probability of distinguishing two given channels $\mathcal{N}$ and $\mathcal{M}$ is given \cite{WCWZ}
\begin{align}
p_{\mathrm{succ}}(\mathcal{N}, \mathcal{M}, \mathcal{F}) = \max_{\rho \in \mathcal{F}} p_{\mathrm{succ}}(\mathcal{N}, \mathcal{M}, \rho).
\end{align}
To determine further whether an unknown channel belongs to a given set of free channels $\mathcal{FC}$, following Refs.~\cite{,LLBK,WCWZ} we define
\begin{align}\label{eq152}
p_{\mathrm{succ}}(\mathcal{N}, \mathcal{FC}, \mathcal{F}) = \max_{\rho \in \mathcal{F}} \min_{\mathcal{M} \in \mathcal{FC}} p_{\text{succ}}(\mathcal{N}, \mathcal{M}, \rho).
\end{align}

Utilizing the imaginarity measure $C^g_{M_R}(\mathcal{N})$ of quantum channels, we establish a lower bound of $p_{\mathrm{succ}}(\mathcal{N},\mathcal{M},\mathcal{F})$.

\begin{theorem}\label{thm1}
For any quantum channels $\mathcal{N}$ and $\mathcal{M}$, we have
\begin{align}\label{eq37}
	&\frac{1}{2}+\frac{1}{4}|C^g_{M_R}(\mathcal{N})-C^g_{M_R}(\mathcal{M})|\le p_{\mathrm{succ}}(\mathcal{N},\mathcal{M},\mathcal{F}).
\end{align}
\end{theorem}

The proof is provided in Appendix~\hyperref[app:C1]{C1}.
This theorem provides a computable estimation of the discrimination probability using quantum channel  imaginarity resource measure. In general, evaluating $p_{\mathrm{succ}}(\mathcal{N},\mathcal{M},\mathcal{F})$ requires a simultaneous optimization over real probe states and output measurements. Theorem~\ref{thm1}, however, demonstrates that once the imaginarity resource of the two channels are known, one obtains a clear lower bound without directly solving the full channel discrimination optimization problem. Hence, $C_{M_R}^g(\mathcal{N})$ serves not only as a measure of the imaginarity resources contained in a single channel, but also as a tool for comparing the behavioral differences of distinct channels in actual information processing tasks.

\begin{figure}[h]
	\centering
	\includegraphics[width=\linewidth]{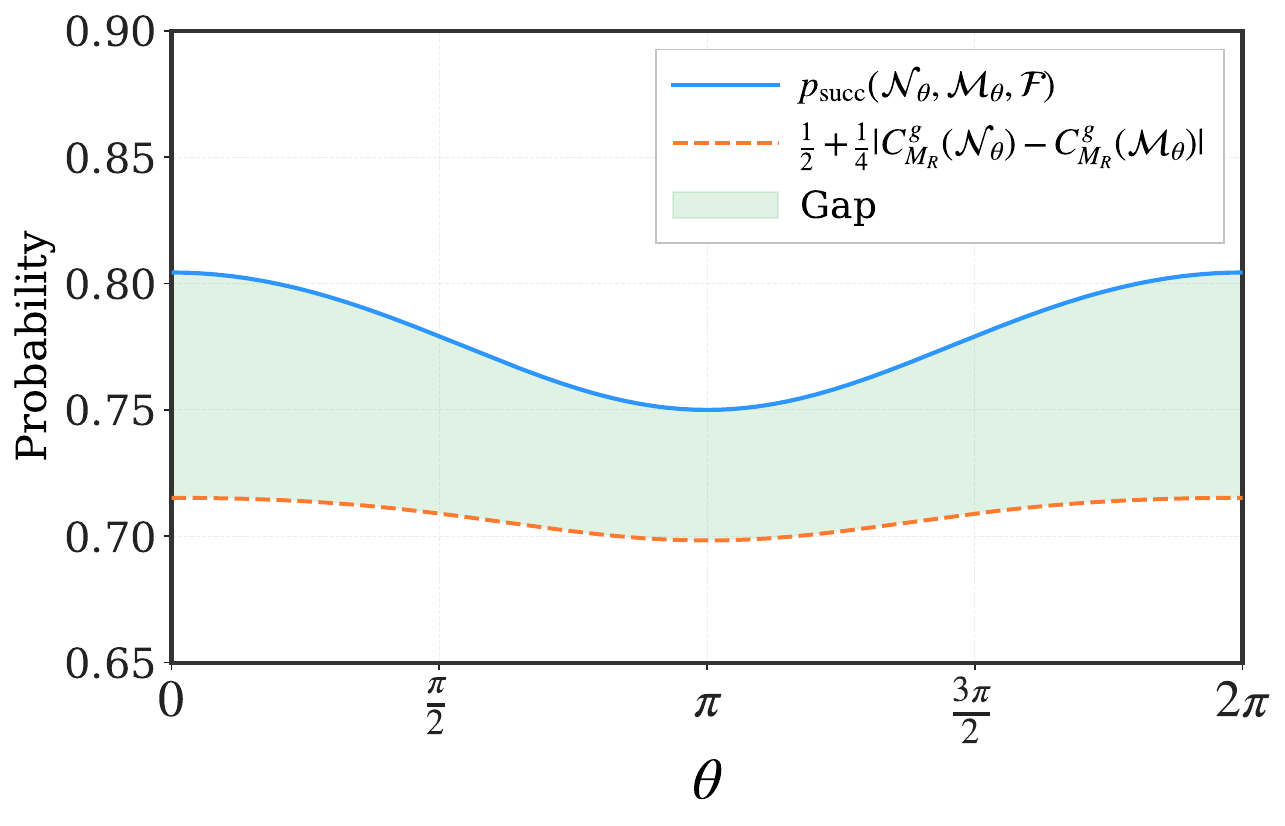}\\
	\caption{The success probability $p_{\mathrm{succ}}(\mathcal{N}_{\theta},\mathcal{M}_{\theta},\mathcal{F})$ (solid blue line) and its lower bound (dashed orange line) are plotted as functions of the parameter $\theta$. The light green shaded region between them indicates the corresponding gap. \label{fig:Fig3}}
\end{figure}

To illustrate the result of Theorem~\ref{thm1}, consider the qubit unitary channel. Taking $\mathcal{N}_{\theta}(\cdot)=U(\theta,\frac{25\pi}{24},0)(\cdot)U^{\dagger}(\theta,\frac{25\pi}{24},0)$ and $\mathcal{M}_{\theta}(\cdot)=U(\theta,\frac{17\pi}{12},\frac{\pi}{24})(\cdot)U^{\dagger}(\theta,\frac{17\pi}{12},\frac{\pi}{24})$ in Eq.~\eqref{eq81}, we plot the success probability $p_{\mathrm{succ}}(\mathcal{N}_{\theta},\mathcal{M}_{\theta},\mathcal{F})$ for discriminating between channels $\mathcal{N}_{\theta}$ and $\mathcal{M}_{\theta}$ using the set of free states as probe states, together with its lower bound, in Fig.~\ref{fig:Fig3}.

\begin{theorem}\label{thm2}
	For any quantum channel $\mathcal{N} \in RNG \setminus RKO$, there exists some RKO channel $\mathcal{M}$ such that 
	\begin{align}\label{eq141}
		p_{\mathrm{succ}}(\mathcal{N},\mathcal{M},\mathcal{F})<p_{\mathrm{succ}}(\mathcal{N},\mathcal{M},\mathcal{D}(\mathcal{H})).
	\end{align}
\end{theorem}

The proof is provided in Appendix~\hyperref[app:C2]{C2}.
Theorem~\ref{thm2} reveals the activating role of imaginarity resources in quantum channel discrimination. It rigorously demonstrates that there exists channel pairs $(\mathcal{N},\mathcal{M})$ such that the success probability of distinguishing between them using free states $\mathcal{F}$ as probes is strictly smaller than that using $\mathcal{D(H)}$ as probes. This indicates that the imaginarity resource can enhance the success probability in channel discrimination. Furthermore, let $\mathcal{N}\in\mathcal{C}$. Then the success probability of distinguishing $\mathcal{N}$ from $\mathrm{RKO}$ by free states is given by
\begin{align}
p_{\mathrm{succ}}(\mathcal{N},\mathrm{RKO},\mathcal{F})=\frac{1}{2}+\frac{1}{4}C^g_{M_R}(\mathcal{N})=\frac{1}{2}+\frac{1}{4}C^b_{M_R}(\mathcal{N}).
\end{align} 
This result reveals that the success probability $p_{\mathrm{succ}}(\mathcal{N},\mathrm{RKO},\mathcal{F})$ only depends on the dynamic imaginarity resource of $\mathcal{N}$. 

We next compare our results with related works. Ref.~\cite{WCWZ} introduced quantum channel imaginarity measures based on imaginarity detection and creation and applied them to channel discrimination, obtaining an upper bound on the discrimination success probability. Since this bound involves an operator norm, obtaining an analytical expression is challenging. In contrast, we derive a lower bound on the discrimination success probability that is comparatively straightforward to evaluate. Additionally, using channel divergences, Ref.~\cite{TPSY} established exact relations between trace norm measures of a quantum channel divergences from the set of real channels or its conjugate channel and the optimal success probabilities in the corresponding discrimination tasks, allowing ancillary assistance or fixing the probe to a maximally entangled state. Here, we investigate the role of probe state imaginarity in channel discrimination without ancillary assistance. Moreover, for discrimination between a given channel and the set of real channels, we adopt the optimization order $\max_{\rho\in\mathcal{F}}\min_{\mathcal{M}\in\mathrm{RKO}}$ in Eq.~\eqref{eq152}, optimizing the real probe to maximize the discrimination success probability against the worst case of real channels. Ref.~\cite{TPSY} instead uses $\min_{\mathcal{M}\in\mathrm{RKO}}\max_{\rho}$, optimizing the joint probe for each alternative real channel before minimizing over real channels. Thus, the two studies differ in both the admissible probe states and the order of optimization, corresponding to operational settings with distinct physical significance.

\section{Dynamic imaginarity resource conversion}\label{secV}
Another important problem in quantum resource theory is the interconversion of quantum resources under a prescribed class of free transformations. While conventional resource theories mainly concern the manipulation of static resources in quantum states~\cite{KKHY20}, many practically relevant resources are intrinsically dynamic and are carried by quantum channels. In this setting, the fundamental question is whether multiple uses of a given resource channel can be converted into uses of another target channel by means of free superchannels, with the conversion error vanishing in the asymptotic limit. The corresponding optimal conversion rate quantifies the amount of the target dynamic resource that can be extracted per use of the initial channel. It therefore provides a unified description of dynamic resource distillation, dilution and channel simulation~\cite{LYYX,MHHY,KFXW}.

In the resource theory of imaginarity, dynamic resource conversion concerns the manipulation of quantum channels whose action cannot be represented entirely by real matrices in a fixed reference basis. Such channels provide access to complex phases that cannot be generated by free channels alone. To investigate their exact conversion, we employ dynamic imaginarity measures defined through the Choi states of quantum channels. We define the exact conversion rate of quantum channels under $\mathcal{FSC}$ as
\begin{align}\label{eq156}
r(\mathcal N\to\mathcal M)=\sup R(\mathcal N,\mathcal M),
\end{align}
where
\begin{align}\label{eq157}
	&R(\mathcal N,\mathcal M)=\notag\\
   &\Bigg\{\frac{m}{n}:n\in\mathbb N, m\in\mathbb N_0,\exists\,\Theta\in\mathcal{FSC}\text{ s.t. }\Theta(\mathcal N^{\otimes n})=\mathcal M^{\otimes m}\Bigg\}.
\end{align}
We use the convention
$\mathcal M^{\otimes0}=\mathbb{I}_1$, namely, the identity
channel on a one-dimensional system. Thus, $m=0$ means that no copy of the target channel is produced. 

\begin{figure}[h]
	\centering
	\includegraphics[width=\linewidth]{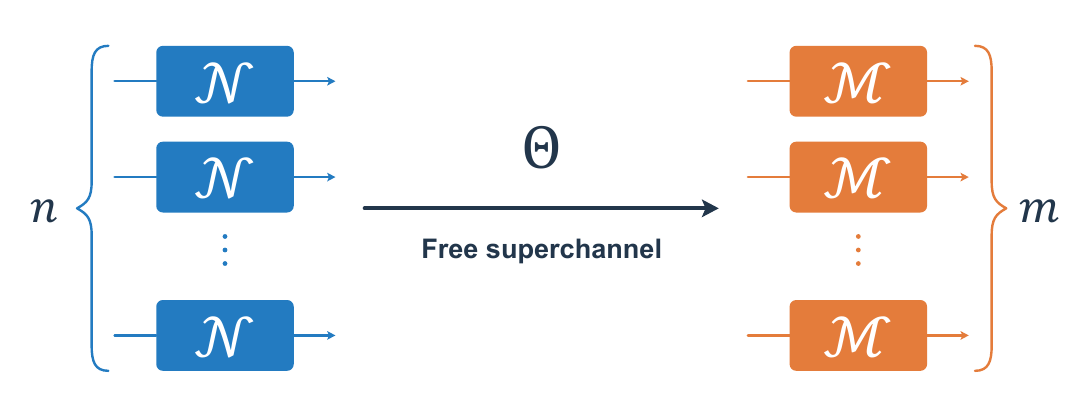}\\
	\caption{The schematic diagram of quantum channels conversion. \label{fig:Fig4}}
\end{figure}

In contrast to conversion with asymptotically vanishing error,
Eq.~\eqref{eq156} requires an exact channel equality. As illustrated in Fig.~\ref{fig:Fig4}, $n$ copies of
$\mathcal N$ are converted into $m$ copies of $\mathcal M$ by a free superchannel. 

Replacing $\Theta_n\in\mathcal{FSC}$ in Eq.~\eqref{eq157} by
$\Theta_n\in\mathcal{FSC}_{\mathrm{CPTP}}$ defines
$r^*(\mathcal N\to\mathcal M)$,
\begin{equation}
	r^*(\mathcal N\to\mathcal M)
	\le r(\mathcal N\to\mathcal M).
\end{equation}
The supremum defining $r$ and $r^*$ are taken in $[0,+\infty]$.
In particular, the conversion rate is defined to be zero
when no positive rate is achievable.
We next examine how a deterministic free conversion between two
channels orders their conversion rates relative to a third channel.

\begin{proposition}\label{prop11}
	Let $\mathcal N_1,\mathcal N_2$ and $\mathcal M$ be quantum channels. If there exists $\Phi\in\mathcal{FSC}$ such that
	$\Phi(\mathcal N_1)=\mathcal N_2$, then
	\begin{align}\label{eq159}
		r(\mathcal M\to\mathcal N_1)
		&\le r(\mathcal M\to\mathcal N_2),\\
		r(\mathcal N_2\to\mathcal M)
		&\le r(\mathcal N_1\to\mathcal M).\label{eq160}
	\end{align}
	In particular, if $\Phi\in\mathcal{FSC}_{\mathrm{CPTP}}$, then
	\begin{align}\label{eq161}
		r^*(\mathcal M\to\mathcal N_1)
		&\le r^*(\mathcal M\to\mathcal N_2),\\
		r^*(\mathcal N_2\to\mathcal M)
		&\le r^*(\mathcal N_1\to\mathcal M).\label{eq162}
	\end{align}
\end{proposition}

The proof is provided in Appendix~\hyperref[app:D1]{D1}.

Proposition~\ref{prop11} therefore expresses two complementary manifestations. We know that the dynamic imaginarity resource content of channel $\mathcal{N}_1$ is no less than that of channel $\mathcal{N}_2$, since 	$\Phi(\mathcal N_1)=\mathcal N_2$ and imaginarity measures of quantum are monotonic under free superchannels. Proposition~\ref{prop11} thus shows that there exists an intrinsic connection between the exact conversion rate of quantum channels and their dynamic imaginarity resource. These properties constitute a principle in exact dynamic imaginarity conversion.

Now consider another interesting question: what is the relation between the exact conversion rates of obtaining $\mathcal{N}_3$ via the two paths $\mathcal{N}_1 \to \mathcal{N}_2 \to \mathcal{N}_3$ and $\mathcal{N}_1 \to \mathcal{N}_3$? We adopt below the convention $0\cdot+\infty=+\infty\cdot0=0$.

\begin{proposition}\label{prop12}
	Let $\mathcal N_1$, $\mathcal N_2$ and $\mathcal N_3$ be quantum
	channels. We have
	\begin{align}\label{eq168}
		r(\mathcal N_1\to\mathcal N_3)
		&\ge r(\mathcal N_1\to\mathcal N_2)
		r(\mathcal N_2\to\mathcal N_3),\\
		r^*(\mathcal N_1\to\mathcal N_3)
		&\ge r^*(\mathcal N_1\to\mathcal N_2)
		r^*(\mathcal N_2\to\mathcal N_3)\label{eq169}.
	\end{align}
\end{proposition}

The proof is provided in Appendix~\hyperref[app:D2]{D2}.
Proposition~\ref{prop12} establishes a concatenation rule for exact dynamic imaginarity conversion, which is consistent with our intuition. Since the direct conversion rate $r(\mathcal{N}_1 \to \mathcal{N}_3)$ is optimized over all admissible protocols, it cannot be smaller than the rate achieved by this particular two step strategy.

Finally, we derive a Choi-based upper bound appropriate to exact conversion. For this purpose, denote
\begin{align}
	\mathcal L^J(\mathcal N)=-\log F\left[J(\mathcal{N}),J(\mathcal{N})^T\right],
\end{align}
where $-\log 0=+\infty$.

\begin{theorem}\label{thm3}
	Let $\mathcal N_1$ and $\mathcal N_2$ be quantum channels. If $0<\mathcal L^J(\mathcal N_2)<+\infty$, then
	\begin{equation}\label{eq179}
		r^*(\mathcal N_1\to\mathcal N_2)
		\le
		\frac{\mathcal L^J(\mathcal N_1)}
		{\mathcal L^J(\mathcal N_2)}.
	\end{equation}
\end{theorem}

The proof is provided in Appendix~\hyperref[app:D3]{D3}.
Theorem~\ref{thm3} provides a concise and effective way to estimate the exact conversion rate $r^*(\mathcal{N}_1 \to \mathcal{N}_2) $ in experiments. We then present a family of channels which attain the upper bound in Theorem~\ref{thm3}. To begin, we define the following family of replacement channels: for any $X\in\mathcal{L}(\mathcal{H}_C)$,
\begin{align}\label{eq188}
	\mathcal{P}_s(X) = \operatorname{tr}(X) |\psi_s\rangle\langle\psi_s|,
\end{align}
where $|\psi_s\rangle = \sqrt{\frac{1+s}{2}}\,|0\rangle + \mathrm{i}\sqrt{\frac{1-s}{2}}\,|1\rangle$ and $0 < s < 1$.

\begin{proposition}\label{prop14}
For any replacement channels defined by Eq.~\eqref{eq188}, with $s_1,s_2\in(0,1)$, we have
    \begin{align}\label{eq189}
	r(\mathcal{P}_{s_1} \to \mathcal{P}_{s_2}) &= r^*(\mathcal{P}_{s_1} \to \mathcal{P}_{s_2})= \frac{\mathcal{L}^J(\mathcal{P}_{s_1})}{\mathcal{L}^J(\mathcal{P}_{s_2})} = \frac{\log s_1}{\log s_2}.
	\end{align}
\end{proposition} 

The proof is provided in Appendix~\hyperref[app:D4]{D4}.

Moreover, for this family the geometric Choi measure satisfies
\begin{equation}\label{eq214}
	C_G^J(\mathcal P_s)=\frac{1-s}{2}.
\end{equation}
Indeed, we find that for any free Choi state $\omega$, 
\begin{align}
	F(J(\mathcal{P}_s), \omega) &\le F(|\psi_s\rangle\langle \psi_s|, \mathrm{tr}_{\mathrm{in}}\omega) \notag\\
	&= \langle \psi_s | \mathrm{tr}_{\mathrm{in}}\omega | \psi_s \rangle,
\end{align}
where $\mathrm{tr}_{\mathrm{in}}$ represents the partial trace taken over the input space of the channel associated with $\omega$, and the inequality holds because the fidelity is non-decreasing under quantum channels. Denote $\mathrm{tr}_{\mathrm{in}}\omega=\widetilde{\omega}$ and $\widetilde{\omega}_{ij}=\langle i|\widetilde{\omega}|j\rangle$. Then we have 
\begin{align}
	&\langle \psi_s | \widetilde{\omega} | \psi_s \rangle \notag\\
	&= \frac{1 + s}{2} \widetilde{\omega}_{00} + \frac{1-s}{2} \widetilde{\omega}_{11} + \frac{\mathrm{i}\sqrt{1-s^2}}{2} \widetilde{\omega}_{01} - \frac{\mathrm{i}\sqrt{1 - s^2}}{2} \widetilde{\omega}_{10}\notag \\
    &=\frac{1 + s}{2} \widetilde{\omega}_{00} + \frac{1-s}{2} \widetilde{\omega}_{11} \notag \\
    &\le \frac{1 + s}{2},
\end{align}
  where the second equality follows from the fact that $\widetilde\omega_{10}=\widetilde\omega_{01}$,
  since $\widetilde\omega$ is real and Hermitian, and the inequality comes from $\widetilde{\omega}_{00},\widetilde{\omega}_{11}\ge0$ and $\widetilde{\omega}_{00}+\widetilde{\omega}_{11}=1$.
  
  On the other hand, let $\mathcal{Z}(X)=\mathrm{tr}(X)|0\rangle\langle0|$ for any $X\in \mathcal{L}(\mathcal{H}_C)$. Through direct calculations, we obtain that 
  \begin{align}
  	F\left( J(\mathcal{P}_s), J(\mathcal{Z}) \right) =F\left( J(\mathcal{P}_s), \frac{\mathbf{I}_C}{d_C} \otimes |0\rangle\langle 0| \right) = \frac{1 + s}{2}.
  \end{align}
  By the definition of $C_G^J(\mathcal P_s)$, we have that $C_G^J(\mathcal P_s)=1-\frac{1 + s}{2}=\frac{1-s}{2}$. We thus derive Eq.~\eqref{eq214}.
 Consequently,
\begin{equation}
	r(\mathcal P_{s_1}\to\mathcal P_{s_2})
	=r^*(\mathcal{P}_{s_1} \to \mathcal{P}_{s_2})=
	\frac{\log[1-2C_G^J(\mathcal P_{s_1})]}
	{\log[1-2C_G^J(\mathcal P_{s_2})]}.
\end{equation}

\section{One-shot exact dynamic imaginarity distillation and dilution}\label{secVI}
In quantum resource theories, distillation and dilution are
commonly formulated as two particular tasks of channel conversion~\cite{ZQ2018,RB2018,CS2019}.
Distillation transforms a given channel into a chosen optimal resource
channel through free superchannels, whereas dilution uses the
reference resource to simulate a given channel~\cite{LYYX,LZWA}. When multiple channels
are available, these tasks are typically quantified by the number of
reference resource channels extracted or consumed~\cite{LYYX,ZQ2018,RB2018,CS2019,ZQ2019,RB2021,YX2024}. 

In the resource theory of imaginarity, following the view of Ref.~\cite{LYYX} channels generating more powerful resource states possess more resources, implying that for the optimal imaginary channel, there exists at least one real input such that its output is the maximally imaginary state. Furthermore, its dynamic imaginary resource content should be maximal, i.e., the imaginarity measure of the quantum channel should attain its maximum when the input is the optimal imaginary channel, and the optimal resource channel should be physically consistent with the corresponding imaginarity resource framework.

The phase channel capturing the capability that a channel provides beyond real channels is defined by
\begin{equation}
	\mathcal{S}(\rho)=S\rho S^\dagger,
	\quad
	S=
	\begin{pmatrix}
		1 & 0\\
		0 & \mathrm{i}
	\end{pmatrix}.
\end{equation}
It introduces a relative phase of $\pi/2$ between the two basis states and maps the free states
\begin{equation}
	|+\rangle=\frac{|0\rangle+|1\rangle}{\sqrt{2}} \text{ and } |-\rangle=\frac{|0\rangle-|1\rangle}{\sqrt{2}}
\end{equation}
to the maximally imaginary states
\begin{equation}
	S|+\rangle=|+\mathrm{i}\rangle
	=\frac{|0\rangle+\mathrm{i}|1\rangle}{\sqrt{2}}\text{ and }	S|-\rangle=|-\mathrm{i}\rangle
	=\frac{|0\rangle-\mathrm{i}|1\rangle}{\sqrt{2}},
\end{equation}
respectively. This property makes the phase channel a natural reference for dynamic
imaginarity: it converts the real off-diagonal elements of a free input
into imaginary ones, thereby generating the maximally imaginary states that cannot be
produced from a free input by real channels. 

When the input is the phase channel $\mathcal{S}$, we have $C_S^J(\mathcal{S})=1$, 
which is the maximum value attainable when both the input and output dimensions of the channel are 2.
Choosing the optimal phase channel, we consider resource conversion
with a fixed number of channel uses. Taking only the optimal channel as a
fixed distillation target addresses whether the optimal phase channel
can be obtained exactly. However, imperfect phase channels can also
generate imaginarity from real inputs, and their resource value merits
a finer characterization in the single use setting. We therefore
introduce a reference family with a continuous parameter to
characterize both the quality of the standard phase channel obtainable
from a given channel and the quality required for its exact simulation.
This formulation complements copy number conversion tasks: with the
number of used channels fixed, we optimize the quality of the reference
resource.

Specifically, we choose the reference family,
\begin{equation}\label{eq222}
	\mathcal{G}_t
	=
	\frac{1+t}{2}\mathcal{S}
	+
	\frac{1-t}{2}\overline{\mathcal{S}},
	\quad 0\leq t\leq 1,
\end{equation}
where
\begin{equation}
	\overline{\mathcal{S}}(\rho)
	=
	\overline{S}\rho\overline{S}^{\dagger},
	\quad
	\overline{S}
	=
	\begin{pmatrix}
		1 & 0\\
		0 & -\mathrm{i}
	\end{pmatrix}.
\end{equation}
The phase gate and its complex conjugate introduce opposite relative
phases, $+\pi/2$ and $-\pi/2$, respectively. Both are resourceful channels
in imaginarity theory. Accordingly, $t$ is not the probability of applying
a resourceful channel. It is the difference between the statistical
weights of the two opposite phase channels. Since $\overline{S}=ZS$,
where $Z=\operatorname{diag}(1,-1)$, the channel can equivalently be
written as
\begin{equation}
	\mathcal{G}_t=\mathcal{D}_t\circ\mathcal{S},
	\quad
	\mathcal{D}_t(\rho)
	=
	\frac{1+t}{2}\rho
	+
	\frac{1-t}{2}Z\rho Z.
	\label{eq:phase_noise_decomposition}
\end{equation}
Thus, $\mathcal{G}_t$ describes the optimal phase gate followed by a phase-flip
error with probability $(1-t)/2$. The endpoint $t=1$ gives the optimal phase
channel, whereas $t=0$ gives the free completely dephasing channel.
Intermediate values of $t$ describe standard phase channels with different
noise levels. 

The physical meaning of $t$ can also be identified directly from the
output generated by a free input:
\begin{align}
	\mathcal{G}_t(|+\rangle\langle+|)
	&=
	\tau_t
	=
	\frac{\mathbf{I}_2+t\sigma_y}{2}
	\nonumber\\
	&=
	\frac{1+t}{2}|+\mathrm{i}\rangle\langle+\mathrm{i}|
	+
	\frac{1-t}{2}|-\mathrm{i}\rangle\langle-\mathrm{i}|,
\end{align}
Consequently, one has
\begin{equation}
	\operatorname{Tr}(\sigma_y\tau_t)=t,
	\text{ and }
	\langle+\mathrm{i}|\tau_t|+\mathrm{i}\rangle=\frac{1+t}{2}.
\end{equation}
The parameter $t$ therefore characterizes the imaginary component of the
standard output and its overlap with the ideal maximally imaginary state,
providing a concrete resource quality coordinate for this family.

Using this family as the reference, we define the one-shot exact dynamic imaginarity
distillation rate and dilution cost as
\begin{equation}\label{eq227}
	d(\mathcal{N})
	=
	\sup\left\{
	t\in[0,1]:
	\exists\,\Theta\in\mathcal{FSC}_{\mathrm{CPTP}}
	\ \text{s.t.}\
	\Theta(\mathcal{N})=\mathcal{G}_t
	\right\},
\end{equation}
and
\begin{equation}\label{eq228}
	c(\mathcal{N})
	=
	\inf\left\{
	t\in[0,1]:
	\exists\,\Theta\in\mathcal{FSC}_{\mathrm{CPTP}}
	\ \text{s.t.}\
	\Theta(\mathcal{G}_t)=\mathcal{N}
	\right\},
\end{equation}
respectively. The distillation rate quantifies the highest quality
standard phase resource obtainable from one use of the given channel,
while the dilution cost quantifies the lowest quality standard phase
resource sufficient for its exact simulation. 

\begin{theorem}\label{thm4}
	For any channel $\mathcal{N}\in \mathcal{C}(A,B)$, the one-shot exact dynamic imaginarity distillation rate is given by
	\begin{align}\label{eq235-}
	d(\mathcal{N}) = C_{\mathrm{tr}}^J(\mathcal{N}) = \frac{1}{2} \| J(\mathcal{N}) - J(\mathcal{N})^T \|_1.
	\end{align}
\end{theorem}
	
	The proof is provided in Appendix~\hyperref[app:E1]{E1}.
	Theorem~\ref{thm4} establishes an exact operational interpretation of the Choi-based trace norm imaginarity in one-shot exact dynamic imaginarity distillation. Under $\mathcal{FSC}_{\mathrm{CPTP}}$, $C_{\mathrm{tr}}^J(\mathcal{N})$ equals to the standard resource channel $\mathcal{G}_t$ with the largest parameter $t$ that can be obtained exactly from a single use of $\mathcal{N}$. The equality includes both a converse bound and an achievable protocol: no admissible conversion can produce a target with $t > C_{\mathrm{tr}}^J(\mathcal{N})$, while the boundary value is attained by the constructed free superchannel. In particular, $C_{\mathrm{tr}}^J(\mathcal{N}) = 0$ excludes every imaginarity target in the family $\{\mathcal{G}_t\}_{t \in [0,1]}$, whereas $C_{\mathrm{tr}}^J(\mathcal{N}) = 1$ permits exact distillation of the ideal phase gate $\mathcal{S}$. Intermediate values of $C_{\mathrm{tr}}^J(\mathcal{N})$ characterize the maximal achievable quality within this standard resource family.
	
Previously, we have presented the exact expression for the one-shot exact dynamic imaginarity distillation rate $d(\mathcal{N})$. In the following, we present a SDP for solving the one-shot exact dynamic imaginarity dilution rate $c(\mathcal{N})$. Denote for any  $\mathcal{N}\in\mathcal{C}(A,B)$,
	\begin{equation}
		W= \frac{J(\mathcal{N}) + J(\mathcal{N})^T}{2} \text{ and }  V= \frac{J(\mathcal{N}) - J(\mathcal{N})^T}{2}.
	\end{equation}
	
	\begin{theorem}\label{thm5}
		For any $\mathcal{N}\in\mathcal{C}(A,B)$ and
		 $t\in(0,1]$, it holds that
		\begin{equation}\label{eq268}
			\mathcal{G}_t \xrightarrow{\mathcal{FSC}_{\mathrm{CPTP}}} \mathcal{N} \iff W + \frac{V}{t} \succeq \textbf{0}.
		\end{equation}
	 Moreover, we have
		\begin{equation}\label{eq269}
			c(\mathcal{N}) = \min\{t \in [0, 1] : tW + V \succeq \textbf{0}\}.
		\end{equation}
		Correspondingly, the SDP for computing $c(\mathcal{N})$ is given by
		\begin{align}\label{eq270}
			\text{minimize} \quad & t \notag\\
			\text{subject to} \quad & tW + V \succeq \textbf{0}, \notag\\
			& t \ge 0, \notag\\
			& 1 - t \ge 0,\notag\\
			& t\in\mathbb{R}.
		\end{align}
	\end{theorem}

The proof is provided in Appendix~\hyperref[app:E2]{E2}.
We would particularly point out that although Theorem~\ref{thm5} is stated for $\mathcal{FSC}_{\mathrm{CPTP}}$, it also holds under the larger class $\mathcal{FSC}$ as can be seen from the proof. It follows that the one-shot exact dynamic imaginarity dilution cost is the same under $\mathcal{FSC}_{\mathrm{CPTP}}$ and $\mathcal{FSC}$. In other words, when considering the one-shot exact dynamic imaginarity dilution cost, it suffices to focus on the conversion under $\mathcal{FSC}_{\mathrm{CPTP}}$. 
Theorem~\ref{thm5} provides an exact characterization of one-shot exact channel conversion from the standard resource family $\{\mathcal{G}_t\}$. Consequently, $c(\mathcal{N})$ identifies the minimum admissible source strength and can be computed by a semidefinite program involving only the target Choi operator and one real variable.
	
Next, we characterize the one-shot exact dynamic imaginarity distillation rate $d(\mathcal{N})$ and one-shot exact dynamic imaginarity dilution cost $c(\mathcal{N})$ in terms of the previously defined $C_S^J(\mathcal{N})$ and $C_G^J(\mathcal{N})$. For this purpose, let
	\begin{equation}\label{eq312}
		f(t) = 1 - h\left(\frac{1+t}{2}\right)\text{ and } g(t) = \frac{1 - \sqrt{1-t^2}}{2},
	\end{equation}
	where $t \in [0,1]$.
	
	\begin{theorem}\label{thm6}
	  For any channel $\mathcal{N}\in\mathcal{C}(A,B)$, it holds that
		\begin{equation}\label{eq313}
			f(d(\mathcal{N})) \le C_S^J(\mathcal{N}) \le f(c(\mathcal{N})),
		\end{equation}
and
		\begin{equation}\label{eq314}
			g(d(\mathcal{N})) \le C_G^J(\mathcal{N}) \le g(c(\mathcal{N})),
		\end{equation}
		where $f(t)$ and $g(t)$ are defined by Eq.~\eqref{eq312}. Equivalently, we have
		\begin{equation}\label{eq315}
			d(\mathcal{N}) \le f^{-1}(C_S^J(\mathcal{N})) \le c(\mathcal{N}),
		\end{equation}
and
		\begin{equation}\label{eq316}
			d(\mathcal{N}) \le 2\sqrt{C_G^J(\mathcal{N})(1 - C_G^J(\mathcal{N}))} \le c(\mathcal{N}).
		\end{equation}
	\end{theorem}
The proof is provided in Appendix~\hyperref[app:E3]{E3}.

As a concrete example, consider the following quantum channel,
 \begin{align}\label{eq347}
 \mathcal{W}_\theta(\rho) = 0.9 R_x(\theta) \mathcal{A}_{0.5}(\rho) R_x(\theta)^\dagger + 0.1 \frac{\mathbf{I}_2}{2},
 \end{align} 
 where $R_x(\theta)=\begin{pmatrix}
 	\cos \frac{\theta}{2} & -\mathrm{i}\sin \frac{\theta}{2} \\
 	-\mathrm{i}\sin \frac{\theta}{2} & \cos \frac{\theta}{2}
 \end{pmatrix}$ with $\theta\in[0,\pi]$ is the $X$-rotation gate, and $\mathcal{A}_\gamma(\rho) = K_0\rho K_0^\dagger + K_1\rho K_1^\dagger$ with $K_0 = \begin{pmatrix}
 1 & 0 \\
 0 & \sqrt{1-\gamma}
 \end{pmatrix}\quad\text{and}\quad
 K_1 = \begin{pmatrix}
 0 & \sqrt{\gamma} \\
 0 & 0
 \end{pmatrix}$ is the amplitude damping channel.
 
\begin{figure*}[t]
	\centering
	\includegraphics[width=\linewidth]{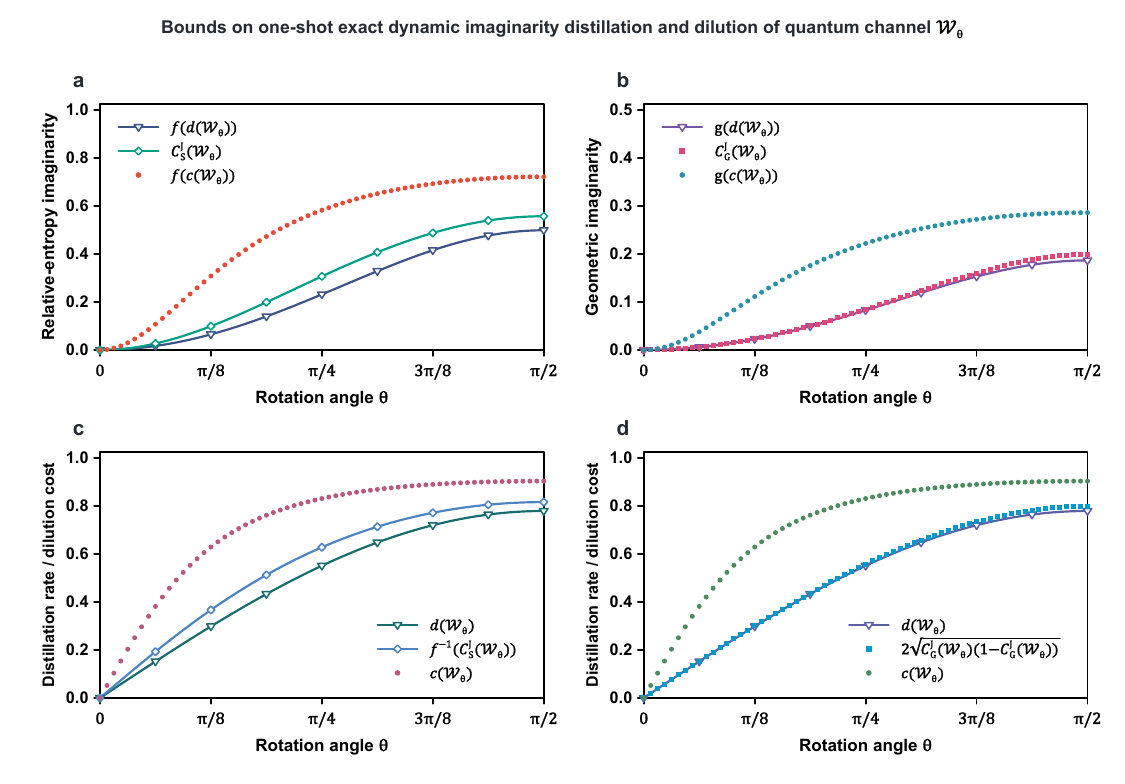}
		\caption{
		Relations between Choi-based imaginarity measures and one-shot exact
		distillation rates and dilution costs for a noisy qubit rotation channel.
		We consider the channel
		$\mathcal{W}_{\theta}(\rho)$ defined by Eq.~\eqref{eq347},
		with $\theta\in[0,\pi/2]$.
		Panels (a)--(d) compare, respectively,
		(a) $C_S^J(\mathcal{W}_{\theta})$ with
		$f[d(\mathcal{W}_{\theta})]$ and
		$f[c(\mathcal{W}_{\theta})]$;
		(b) $C_G^J(\mathcal{W}_{\theta})$ with
		$g[d(\mathcal{W}_{\theta})]$ and
		$g[c(\mathcal{W}_{\theta})]$;
		(c) $f^{-1}[C_S^J(\mathcal{W}_{\theta})]$ with
		$d(\mathcal{W}_{\theta})$ and
		$c(\mathcal{W}_{\theta})$; and
		(d) $2\sqrt{C_G^J(\mathcal{W}_{\theta})
			[1-C_G^J(\mathcal{W}_{\theta})]}$ with
		$d(\mathcal{W}_{\theta})$ and
		$c(\mathcal{W}_{\theta})$.
		Solid lines with open downward-pointing triangles represent quantities
		derived from $d(\mathcal{W}_{\theta})$, whereas solid lines with open
		diamonds represent quantities derived from
		$C_S^J(\mathcal{W}_{\theta})$.
		Filled circles represent $c(\mathcal{W}_{\theta})$ and its corresponding
		functions, obtained from the semidefinite program in
		Theorem~\ref{thm5}; filled squares represent
		$C_G^J(\mathcal{W}_{\theta})$ and its corresponding function,
		obtained using the semidefinite program in
		Proposition~\ref{prop11-}.
	}
	\label{fig:Fig6}
\end{figure*}

 Equivalently, the dynamical evolution described by Eq.~\eqref{eq347} corresponds to a quantum state undergoing, in sequence, an amplitude damping channel with parameter $\gamma=0.5$, an $X$-rotation gate, and the depolarizing channel $\mathcal{D}(\rho)=0.9\rho+0.1\frac{\mathbf{I}_2}{2}$. 
 By making use of the analytical expression of $d(\mathcal{W}_{\theta})$ given by Eq.~\eqref{eq235-}, the semidefinite program formulated in Eq.~\eqref{eq130} for solving $C_{G}^J(\mathcal{N})$, as well as the semidefinite program in Eq.~\eqref{eq270} for solving $c(\mathcal{N})$, we are capable of reproducing the results presented in Theorem~\ref{thm6}, see Fig.~\ref{fig:Fig6} for $\mathcal{W}_{\theta}$ with respect to Eqs.~(\ref{eq313})-(\ref{eq316}).
 
\section{Conclusion and discussion}\label{secVII}
In this work, we focus on the roles of dynamic quantum imaginarity played in channel discrimination, exact channel conversion, and one-shot exact dynamic imaginarity distillation and dilution. We have connected quantum channel imaginarity with discrimination and resource conversion through state-based and Choi-based quantifiers. We derived the lower bound on discrimination success probabilities and the upper bound on exact channel conversion rates, demonstrating the discrimination advantages enabled by imaginarity. For one-shot exact dynamic imaginarity distillation and dilution, we obtained an analytical expression for the distillation rate, a semidefinite program for the dilution cost, and quantitative relations between them. 

Several issues remained resolved. We have only obtained an upper bound on the quantum channel exact conversion rate for a restricted class of free superchannels $\mathcal{FSC}_{\mathrm{CPTP}}$, whether analogous results can be established in more general free superchannel frameworks remains unclear. Moreover, the exact distillation and dilution also constitute an important class of quantum information processing tasks in practical experiments, and thus require further investigation.

\begin{acknowledgments}
The core ideas and overall approach of this work were developed by the authors. The authors used generative AI tools during the derivation of Proposition~\ref{prop14} and the proofs of sufficiency for Theorem~\ref{thm4} and Theorem~\ref{thm5}. The authors independently composed the entire initial manuscript and take full responsibility for the correctness of the proofs and conclusions presented in this paper.
	
The authors would like to thank Prof. Jianwei Xu, and Prof. ZhenPeng Xu for helpful discussions. Zhaoqi Wu was supported by National Natural Science Foundation of China (Grant Nos. 12561084) and Natural Science Foundation of Jiangxi Province (Grant No. 20232ACB211003). Mao-Sheng Li was supported by National Natural Science Foundation of China (Grant No. 12371458). Fei Shi was supported by National Natural Science Foundation of China (Grant No. 62601907 and No. 12571493), and the 2026 Basic Start-up Fund of Sun Yat-sen University (Grant No. 67000-12266020). 
\end{acknowledgments}

\begin{widetext}
	\begin{appendices}
		\renewcommand{\theequation}{\Alph{section}\arabic{equation}}
		\setcounter{equation}{0}
		\begin{center}
			\textbf{Appendix A: Proof of Section~\ref{secII} \label{app:A}}
		\end{center}
		
		\renewcommand{\theequation}{A\arabic{equation}}
		\setcounter{equation}{0}
\begin{center}
	\textbf{A1. Proof of Proposition \ref{prop1}}\label{app:A1}
\end{center}
\begin{proof}
	Definition \ref{def2} is understood to require Eq.~\eqref{eq9} for any completely positive (CP) map, not necessarily trace-preserving. The pre- and post-processing maps in Definition \ref{def1} remain RKO channels. We retain the notation $\mathcal{N}$ for an arbitrary CP input map. We begin by showing that
	\begin{align}\label{eq11}
		\overline{\mathcal{N}_2 \circ \mathcal{N}_1} = \overline{\mathcal{N}_2} \circ \overline{\mathcal{N}_1} \quad \text{and} \quad \overline{\mathcal{N}_1 \otimes \mathcal{N}_2} = \overline{\mathcal{N}_1} \otimes \overline{\mathcal{N}_2}.
	\end{align}
	
	In fact, let $\mathcal{N}_1$ and $\mathcal{N}_2$ be arbitrary CP maps, with compatible input and output spaces when composed, and with Kraus operators $\{K_{\alpha}^1\}$ and $\{K_{\beta}^2\}$, respectively. Their conjugate maps $\overline{\mathcal{N}_1}$ and $\overline{\mathcal{N}_2}$ have Kraus operators $\{\overline{K_{\alpha}^1}\}$ and $\{\overline{K_{\beta}^2}\}$, respectively. The composition $\mathcal{N}_2\circ\mathcal{N}_1$ has the Kraus representation
	\begin{align}
		\mathcal{N}_2\circ\mathcal{N}_1(\cdot)
		&=\sum_{\alpha,\beta}K_{\beta}^{2}K_{\alpha}^{1}(\cdot)
		\left(K_{\alpha}^{1}\right)^\dagger\left(K_{\beta}^{2}\right)^\dagger\notag\\
		&=\sum_{\alpha,\beta}K_{\beta}^{2}K_{\alpha}^{1}(\cdot)
		\left(K_{\beta}^{2}K_{\alpha}^{1}\right)^\dagger,
	\end{align}
	which yields that
	\begin{align}
		\overline{\mathcal{N}_2\circ\mathcal{N}_1}(\cdot)
		&=\sum_{\alpha,\beta}\overline{K_{\beta}^{2}K_{\alpha}^{1}}(\cdot)
		\overline{K_{\beta}^{2}K_{\alpha}^{1}}^{\dagger}\notag\\
		&=\sum_{\alpha,\beta}\overline{K_{\beta}^{2}}\,\overline{K_{\alpha}^{1}}(\cdot)
		\overline{K_{\alpha}^{1}}^\dagger\overline{K_{\beta}^{2}}^\dagger\notag\\
		&=\overline{\mathcal{N}_2}\circ\overline{\mathcal{N}_1}(\cdot).
	\end{align}
	It follows that $\overline{\mathcal{N}_2\circ\mathcal{N}_1}=\overline{\mathcal{N}_2}\circ\overline{\mathcal{N}_1}$. The tensor-product identity follows analogously.
	
	Next we prove that Definition \ref{def1} implies Definition \ref{def2}. Suppose that $\Theta$ admits a fixed realization with RKO pre- and post-processing channels. Its action on any CP map $\mathcal{N}$ is given by
	\begin{align}
		\Theta(\mathcal{N})=\mathcal{Q}\circ(\mathbb{I}_{E}\otimes \mathcal{N})\circ\mathcal{P},
	\end{align}
	where $\mathcal{Q}$ and $\mathcal{P}$ are RKO channels. In addition, it follows from Ref.~\cite[Lemma~1]{TPSY} that a quantum channel $\mathcal{X}$ is a real channel if and only if $\mathcal{X} = \overline{\mathcal{X}} $. We thus know that $\mathcal{Q}=\overline{\mathcal{Q}}$ and $\mathcal{P}=\overline{\mathcal{P}}$. Therefore, one gets
	\begin{align}
		\overline{\Theta(\mathcal{N})}
		&=\overline{\mathcal{Q}\circ(\mathbb{I}_{E}\otimes\mathcal{N})\circ\mathcal{P}}\notag\\
		&=\overline{\mathcal{Q}}\circ\overline{\mathbb{I}_{E}\otimes\mathcal{N}}\circ\overline{\mathcal{P}}\notag\\
		&=\overline{\mathcal{Q}}\circ(\overline{\mathbb{I}_{E}}\otimes\overline{\mathcal{N}})\circ\overline{\mathcal{P}}\notag\\
		&=\overline{\mathcal{Q}}\circ(\mathbb{I}_{E}\otimes\overline{\mathcal{N}})\circ\overline{\mathcal{P}}\notag\\
		&=\mathcal{Q}\circ(\mathbb{I}_{E}\otimes\overline{\mathcal{N}})\circ\mathcal{P}\notag\\
		&=\Theta(\overline{\mathcal{N}}).
	\end{align}
	Here the second and third equalities follow from Eq.~\eqref{eq11}, and the fourth uses the fact that $\mathbb{I}_{E}$ is a RKO channel. This proves the required covariance for all CP input maps.
	
	Conversely, suppose that $\overline{\Theta(\mathcal{N})}=\Theta(\overline{\mathcal{N}})$ for any CP map $\mathcal{N}$. Fix an arbitrary pre- and post-processing realization of the deterministic supermap $\Theta$, with CPTP maps $\mathcal{W}:\mathcal{L}(\mathcal{H}_C)\to\mathcal{L}(\mathcal{H}_E\otimes\mathcal{H}_A)$ and $\mathcal{V}:\mathcal{L}(\mathcal{H}_E\otimes\mathcal{H}_B)\to\mathcal{L}(\mathcal{H}_D)$. Thus, for any CP input map,
	\begin{align}\label{eq16}
		\Theta(\mathcal{N})=\mathcal{V}\circ(\mathbb{I}_{E}\otimes \mathcal{N})\circ\mathcal{W}.
	\end{align}
	
	Based on Eq.~\eqref{eq16}, we construct a superchannel $\widetilde{\Theta}$ with RKO pre- and post-processing channels, and then show that $\widetilde{\Theta}=\Theta$ on the full space of linear maps. To this end, introduce a two-dimensional flag system $F$, with the maximally imaginary states
	\begin{equation}
		I_+=|+\rangle \langle +|\quad \text{and}\quad I_-=|-\rangle \langle -|,
	\end{equation}
	where $|+\rangle=(|0\rangle+\mathrm{i}|1\rangle)/\sqrt{2}$ and $|-\rangle=(|0\rangle-\mathrm{i}|1\rangle)/\sqrt{2}$ belong to $\mathcal{H}_F$. Define the pre-processing channel $\Gamma_{\mathrm{pre}}:\mathcal{L}(\mathcal{H}_C)\to\mathcal{L}(\mathcal{H}_F\otimes\mathcal{H}_E\otimes\mathcal{H}_A)$ by
	\begin{align}
		\Gamma_{\text{pre}}(\rho) = \frac{1}{2} I_+ \otimes \mathcal{W}(\rho) + \frac{1}{2} I_- \otimes \overline{\mathcal{W}}(\rho).
	\end{align}
	This map is CPTP because $\mathcal{W}$ and $\overline{\mathcal{W}}$ are channels and $I_\pm$ are normalized states. Since $\overline{I_+}=I_-$, we have $\overline{\Gamma_{\mathrm{pre}}}=\Gamma_{\mathrm{pre}}$, so $\Gamma_{\mathrm{pre}}$ is a RKO channel.
	
	Let $\{V_k\}$ be Kraus operators of $\mathcal{V}$, where $V_k:\mathcal{H}_E\otimes\mathcal{H}_B\to\mathcal{H}_D$ and $\sum_k V_k^\dagger V_k=\mathbf{I}_{EB}$. Define
	\begin{align}
		M_{k,+} = V_{k}(\langle + | \otimes \mathbf{I}_{EB}) \quad \text{and} \quad M_{k,-} =\overline{ V_{k}}(\langle - | \otimes \mathbf{I}_{EB}).
	\end{align}
	It is obvious that
	\begin{align}\label{eq20}
		M_{k,+}=\overline{M_{k,-}}.
	\end{align}
	Although these operators need not individually be real, we can define
	\begin{align}
		L_{k,1} = \frac{M_{k,+} + M_{k,-}}{\sqrt{2}} \quad \text{and} \quad L_{k,2} = \frac{M_{k,+} - M_{k,-}}{\mathrm{i}\sqrt{2}}.
	\end{align}
	Define the post-processing map $\Gamma_{\mathrm{post}}:\mathcal{L}(\mathcal{H}_F\otimes\mathcal{H}_E\otimes\mathcal{H}_B)\to\mathcal{L}(\mathcal{H}_D)$ by
	\begin{align}
		\Gamma_{\text{post}}(\rho)=\sum_{k} 	L_{k,1} \rho L_{k,1}^{\dagger} +\sum_{k} 	L_{k,2} \rho L_{k,2}^{\dagger}.
	\end{align}
	It follows from Eq.~\eqref{eq20} that $L_{k,1}$ and $L_{k,2}$ are real for any $k$.
	We now verify that the set of Kraus operators defined above satisfies the completeness relation. Noting that
	\begin{align}
		L_{k,1}^{\dagger} L_{k,1} 
		&= \frac{1}{2} (M_{k,+}^{\dagger} + M_{k,-}^{\dagger})(M_{k,+} + M_{k,-})\notag \\
		&= \frac{1}{2} \left( M_{k,+}^{\dagger} M_{k,+} + M_{k,+}^{\dagger} M_{k,-} + M_{k,-}^{\dagger} M_{k,+} + M_{k,-}^{\dagger} M_{k,-} \right),
	\end{align}
	and
	\begin{align}
		L_{k,2}^{\dagger} L_{k,2}
		&= \frac{1}{(-\mathrm{i}\sqrt{2})(\mathrm{i}\sqrt{2})}
		\left( M_{k,+}^{\dagger} - M_{k,-}^{\dagger} \right)
		\left( M_{k,+} - M_{k,-} \right) \notag\\
		&= \frac{1}{2} \Big(
		M_{k,+}^{\dagger} M_{k,+}
		- M_{k,+}^{\dagger} M_{k,-}
		- M_{k,-}^{\dagger} M_{k,+}
		+ M_{k,-}^{\dagger} M_{k,-}
		\Big),
	\end{align}
	we get
	\begin{align}\label{eq25}
		L_{k,1}^{\dagger}L_{k,1}+L_{k,2}^{\dagger}L_{k,2}
		&=M_{k,+}^{\dagger}M_{k,+}+M_{k,-}^{\dagger}M_{k,-}\notag\\
		&=(|+\rangle\otimes\mathbf{I}_{EB})V_k^\dagger V_k
		(\langle+|\otimes\mathbf{I}_{EB})+(|-\rangle\otimes\mathbf{I}_{EB})\overline{V_k}^\dagger\overline{V_k}
		(\langle-|\otimes\mathbf{I}_{EB}),
	\end{align}
	which yields that
	\begin{align}
		\sum_k\left(L_{k,1}^{\dagger}L_{k,1}+L_{k,2}^{\dagger}L_{k,2}\right)
		&=\sum_k\Big[(|+\rangle\otimes\mathbf{I}_{EB})V_k^\dagger V_k
		(\langle+|\otimes\mathbf{I}_{EB})
		+(|-\rangle\otimes\mathbf{I}_{EB})\overline{V_k}^\dagger\overline{V_k}
		(\langle-|\otimes\mathbf{I}_{EB})\Big]\notag\\
		&=|+\rangle\langle+|\otimes\mathbf{I}_{EB}
		+|-\rangle\langle-|\otimes\mathbf{I}_{EB}\notag\\
		&=\mathbf{I}_{FEB}.
	\end{align}
	Thus the completeness relation holds, and $\Gamma_{\mathrm{post}}$ is a RKO channel. The enlarged memory is $F\otimes E$, so the identity channel accompanying the input map acts on both $F$ and $E$. For any CP input map $\mathcal{N}$, we have
	\begin{align}
		(\mathbb{I}_{FE}\otimes\mathcal{N})\circ\Gamma_{\mathrm{pre}}(\rho)
		=\frac12 I_+\otimes\big[(\mathbb{I}_{E}\otimes\mathcal{N})\circ\mathcal{W}\big](\rho)
		+\frac12 I_-\otimes\big[(\mathbb{I}_{E}\otimes\mathcal{N})\circ\overline{\mathcal{W}}\big](\rho).
	\end{align}
	
	For any input state $\rho$, write $	\widetilde{\rho}=\big[(\mathbb{I}_{FE}\otimes\mathcal{N})\circ\Gamma_{\mathrm{pre}}\big](\rho)$.
	This operator need not be normalized, since $\mathcal{N}$ need not be trace-preserving. By cancellation of the cross terms, it follows from the definitions of $L_{k,1}$ and $L_{k,2}$ that
	\begin{align}
		\sum_{j=1}^{2}L_{k,j}X L_{k,j}^{\dagger}
		=M_{k,+}X M_{k,+}^{\dagger}+M_{k,-}X M_{k,-}^{\dagger}
	\end{align}
	for any operator $X$. Using this identity and the orthogonality of the flag states, we obtain
	\begin{align}
		\Gamma_{\mathrm{post}}\circ(\mathbb{I}_{FE}\otimes\mathcal{N})\circ\Gamma_{\mathrm{pre}}(\rho)
		&=\sum_k\left(M_{k,+}\widetilde{\rho}M_{k,+}^{\dagger}
		+M_{k,-}\widetilde{\rho}M_{k,-}^{\dagger}\right)\notag\\
		&=\frac12\mathcal{V}\circ(\mathbb{I}_{E}\otimes\mathcal{N})\circ\mathcal{W}(\rho)
		+\frac12\overline{\mathcal{V}}\circ(\mathbb{I}_{E}\otimes\mathcal{N})\circ\overline{\mathcal{W}}(\rho)\notag\\
		&=\frac12\Theta(\mathcal{N})(\rho)
		+\frac12\overline{\Theta(\overline{\mathcal{N}})}(\rho)\notag\\
		&=\frac12\Theta(\mathcal{N})(\rho)+\frac12\Theta(\mathcal{N})(\rho)\notag\\
		&=\Theta(\mathcal{N})(\rho),
	\end{align}
	where the fourth equality follows from Eq.~\eqref{eq9} applied to the CP map $\overline{\mathcal{N}}$. Hence, for any CP map $\mathcal{N}$,
	\begin{align}
		\Gamma_{\text{post}}\circ(\mathbb{I}_{FE} \otimes \mathcal{N})\circ\Gamma_{\text{pre}}=\mathcal{V}\circ(\mathbb{I}_E\otimes \mathcal{N})\circ\mathcal{W}.
	\end{align}
	Define the supermap
	\begin{align}
		\widetilde{\Theta}(\mathcal{M})
		=\Gamma_{\mathrm{post}}\circ(\mathbb{I}_{FE}\otimes\mathcal{M})\circ\Gamma_{\mathrm{pre}}
	\end{align}
	for any linear map $\mathcal{M}$. The preceding calculation shows that $\widetilde{\Theta}(\mathcal{N})=\Theta(\mathcal{N})$ for any CP map $\mathcal{N}$. Under the Choi isomorphism, CP maps correspond to positive semidefinite operators, whose complex linear span is the entire operator space. Thus CP maps span the full space of linear maps over the complex numbers. Since both $\widetilde{\Theta}$ and $\Theta$ are complex linear, they coincide on this full space.
	
	Consequently, any superchannel satisfying Eq.~\eqref{eq9} for all CP maps admits a realization of the form in Eq.~\eqref{eq8} with RKO pre- and post-processing channels and memory $F\otimes E$. Therefore, Definition \ref{def2} implies Definition \ref{def1}.
\end{proof}	

\begin{center}
	\textbf{A2. Proof of Proposition \ref{prop2-}}\label{app:A2}
\end{center}

\begin{proof}
	Since $\mathcal{E}_{\Theta}$ is CP,
	it suffices to show that it is TP if and only if
	Eq.~\eqref{eq39--} holds. For simplicity let
	\begin{equation}
		\tau_A= \operatorname{tr}_E
		\mathcal{P}\left(\frac{\mathbf{I}_C}{d_C}\right).
	\end{equation}
	For an arbitrary operator $X \in \mathcal{L}(\mathcal{H}_A \otimes \mathcal{H}_B)$, define
	$\Phi_X = J^{-1}(X)$.
	Here, $\Phi_X$ is a linear map and need not be a quantum channel.
	
	By the definition of the normalized Choi map, one has
	\begin{equation}\label{eq40-}
		\operatorname{tr} J(\Lambda)
		=\operatorname{tr}\frac{1}{d_C} \sum_{i,j} |i\rangle\langle j| \otimes \Lambda(|i\rangle\langle j|) =
		\operatorname{tr}\Lambda\left(\frac{\mathbf{I}_C}{d_C}\right)
	\end{equation}
	for any linear map $\Lambda$, where the second equality uses
	$\operatorname{tr}(|i\rangle\langle j|)=\delta_{ij}$. Therefore,
	\begin{align}
		\operatorname{tr}\mathcal{E}_{\Theta}(X)
		&=\operatorname{tr}[(J\circ \Theta\circ J^{-1})X] \notag \\
		&=
		\operatorname{tr}\left[
		\Theta(\Phi_X)\left(\frac{\mathbf{I}_C}{d_C}\right)
		\right] \notag\\
		&=
		\operatorname{tr}\mathcal{Q}\left[
		(\mathbb{I}_E \otimes \Phi_X)
		\left(
		\mathcal{P}\left(\frac{\mathbf{I}_C}{d_C}\right)
		\right)
		\right] \notag\\
		&=
		\operatorname{tr}\left[
		(\mathbb{I}_E \otimes \Phi_X)
		\left(
		\mathcal{P}\left(\frac{\mathbf{I}_C}{d_C}\right)
		\right)
		\right] \notag\\
		&=
		\operatorname{tr}\Phi_X(\tau_A),
	\end{align}
	where the second equality follows from Eq.~\eqref{eq40-}, the third
	one uses the realization of $\Theta$, the fourth one
	follows from the trace-preserving property of $\mathcal{Q}$,
	and the fifth one follows from
	$\operatorname{tr}_E[(\mathcal{I}_E\otimes\Phi_X)(Y_{EA})]
	=\Phi_X(\operatorname{tr}_E Y_{EA})$
	for any operator $Y_{EA}$.
	
	On the other hand, the inverse of the Choi map, i.e., Eq.~\eqref{eq35-}, gives
	\begin{equation}
		\operatorname{tr}\mathcal{E}_{\Theta}(X)
		=\operatorname{tr}\Phi_X(\tau_A)=
		d_A \operatorname{tr}
		\left[
		(\tau_A^T \otimes \mathbf{I}_B)X
		\right].
	\end{equation}
	It follows that $\mathcal{E}_{\Theta}$ is TP
	if and only if
	\begin{equation}\label{eq41-}
		d_A \operatorname{tr}
		\left[
		(\tau_A^T \otimes \mathbf{I}_B)X
		\right]
		=
		\operatorname{tr}X
	\end{equation}
	for all  $X \in \mathcal{L}(\mathcal{H}_{A} \otimes \mathcal{H}_{B})$. Eq.~\eqref{eq41-} is equivalent to
     $\mathrm{tr}[MX]=0$
	for any $X$, where $M=d_A \tau_A^T \otimes \mathbf{I}_B-\mathbf{I}_A \otimes \mathbf{I}_B$. Taking $X=M^{\dagger}$ yields
	\begin{align}
		\operatorname{tr}(M M^\dagger) = \sum_{i,j} |\langle i| M |j \rangle|^2=0,
	\end{align}
	which means that $M=\mathbf{0}$. Hence $d_A \tau_A^T \otimes \mathbf{I}_B=\mathbf{I}_A \otimes \mathbf{I}_B$, that is, $\tau_A = \frac{\mathbf{I}_A}{d_A}$.	
\end{proof}

\begin{center}
	\textbf{A3. Proof of Proposition \ref{prop2}}\label{app:A3}
\end{center}
	\begin{proof}
	Let
	\begin{align}\label{eq39}
		\Theta(\mathcal{N}_{A\rightarrow B})= \mathcal{Q}_{EB\rightarrow D}\circ(\mathbb{I}_E\otimes \mathcal{N}_{A\rightarrow B})\circ \mathcal{P}_{C\rightarrow EA}=\mathcal{M}_{C\to D}, 
	\end{align}
	where $\mathcal Q_{EB\to D}$ and $\mathcal P_{C\to EA}$
	are free channels. Since both channels admit real Kraus
	representations, we may write
	\begin{align}\label{40}
		\mathcal P(\cdot) = \sum_{\alpha} P_{\alpha} (\cdot) P_{\alpha}^{\dagger}\quad \text{and}\quad \mathcal Q(\cdot) = \sum_{\beta} Q_{\beta} (\cdot) Q_{\beta}^{\dagger},
	\end{align}
	where $P_{\alpha}:C\to E\otimes A$ and $Q_{\beta}:E\otimes B\to D$ are all real matrices. For clarity, fix reference bases for the respective systems: $\{|a\rangle_A\},\quad \{|b\rangle_B\},\quad \{|c\rangle_C\},\quad \{|d\rangle_D\},\quad \{|e\rangle_E\}$.
	Here, $a,b,c,d,e$ are the basis indices for systems $A,B,C,D,E$, respectively, and $f$ denotes another independent basis index for system $E$. Let
	\begin{align}\label{eq41}
		p_{ea;c}^{\alpha} = (\langle e| \otimes \langle a|) P_{\alpha} |c\rangle \quad \text{and} \quad q_{d;eb}^{\beta} = \langle d| Q_{\beta} (|e\rangle \otimes |b\rangle)
	\end{align}
	denote the matrix elements of $P_{\alpha}$ and $Q_{\beta}$, respectively. Then we have
	\begin{align}\label{eq42}
		P_{\alpha}|c\rangle = \sum_{e,a}|e,a\rangle\langle e,a|P_{\alpha}|c\rangle
		= \sum_{e,a} p_{ea;c}^{\alpha} |e\rangle \otimes |a\rangle.
	\end{align}
	Similarly, one gets $Q_{\beta}(|e\rangle \otimes |b\rangle) = \sum_{d} q_{d;eb}^{\beta} |d\rangle$. Eq.~\eqref{eq42} gives
	\begin{align}
		\mathcal{P}(|c\rangle\langle c'|)= \sum_{\alpha} P_{\alpha}|c\rangle\langle c'|P_{\alpha}^{\dagger} =\sum_{\alpha,e,f,a,a'} p_{ea;c}^{\alpha} \overline{p_{fa';c'}^{\alpha}} \, |e\rangle\langle f| \otimes |a\rangle\langle a'|,
	\end{align}
	which means that
	\begin{align}\label{eq44}
		(\mathbb{I}_E \otimes \mathcal{N})\bigl(\mathcal{P}(|c\rangle\langle c'|)\bigr) = \sum_{\alpha,e,f,a,a'} p_{ea;c}^{\alpha} \overline{p_{fa';c'}^{\alpha}} \, |e\rangle\langle f| \otimes \mathcal{N}(|a\rangle\langle a'|).
	\end{align}
	
	Next let
	\begin{align}\label{eq45}
		\mathcal{N}(|a\rangle\langle a'|) = \sum_{b,b'} n_{bb';aa'} |b\rangle\langle b'|,
	\end{align}
	where $n_{bb';aa'} = \langle b|\mathcal{N}(|a\rangle\langle a'|)|b'\rangle$ denotes the corresponding matrix element of the original channel output. Substituting Eq.~\eqref{eq45} into Eq.~\eqref{eq44}, we obtain that
	\begin{align}
		(\mathbb{I}_E \otimes \mathcal{N})\bigl(\mathcal{P}(|c\rangle\langle c'|)\bigr)
		=\sum_{\alpha,e,f,a,a',b,b'} p_{ea;c}^{\alpha} \overline{p_{fa';c'}^{\alpha}} \, n_{bb';aa'} |e,b\rangle\langle f,b'|.
	\end{align}
	
	Finally, we apply the post-processing channel $\mathcal{Q}$. Using Eqs.~\eqref{40} and Eq~\eqref{eq41},
	\begin{align}
		\mathcal{Q}(|e,b\rangle\langle f,b'|) = \sum_{\beta} Q_{\beta} |e,b\rangle\langle f,b'| Q_{\beta}^{\dagger} 
		= \sum_{\beta,d,d'} q_{d;eb}^{\beta} \overline{q_{d';fb'}^{\beta}} \, |d\rangle\langle d'|.
	\end{align}
	Therefore, the matrix elements of $\mathcal{M} = \Theta(\mathcal{N})$ satisfy
	\begin{align}\label{eq48}
		&\langle d|\mathcal{M}(|c\rangle\langle c'|)|d'\rangle 
		= \sum_{\alpha,\beta,e,f,a,a',b,b'} p_{ea;c}^{\alpha} \overline{p_{fa';c'}^{\alpha}} \, q_{d;eb}^{\beta} \overline{q_{d';fb'}^{\beta}} \, n_{bb';aa'}.
	\end{align}
	
	For convenience, denote $\rho=J(\mathcal{N})$ and $\tau=J(\mathcal{M})$. By the definition of Choi-Jamio\l kowski isomorphism, i.e., Eq.~\eqref{eq31}, one has
	\begin{align}\label{eq49}
		\rho = \frac{1}{d_A} \sum_{a,a'} |a\rangle\langle a'| \otimes \mathcal{N}(|a\rangle\langle a'|)\quad \text{and} \quad \tau = \frac{1}{d_C} \sum_{c,c'} |c\rangle\langle c'| \otimes \mathcal{M}(|c\rangle\langle c'|).
	\end{align}
	Therefore, from Eqs.~\eqref{eq45} and~\eqref{eq49}, we know that
	\begin{align}\label{eq50}
		\rho_{ab;a'b'} = \frac{1}{d_A} n_{bb';aa'}\quad \text{and}\quad \tau_{cd;c'd'} = \frac{1}{d_C} \langle d|\mathcal{M}(|c\rangle\langle c'|)|d'\rangle.
	\end{align}
	
	Combining Eqs.~\eqref{eq48} and~\eqref{eq50}, one has
	\begin{align}\label{eq51}
		\tau_{cd;c'd'} = \frac{d_A}{d_C} \sum_{\substack{\alpha,\beta,e,f,a,a',b,b'}} p_{ea;c}^{\alpha} \overline{p_{fa';c'}^{\alpha}} \times q_{d;eb}^{\beta} \overline{q_{d';fb'}^{\beta}} \rho_{ab;a'b'}.
	\end{align}
	
	We now construct a completely positive map
	$\mathcal{G}: \mathcal{L}(\mathcal{H}_{A}\otimes \mathcal{H}_{B})\to \mathcal{L}(\mathcal{H}_{C}\otimes \mathcal{H}_{D})$
	of the form
	\begin{align}
		\mathcal G(\cdot)=
		\sum_{\alpha,\beta}
		V_{\alpha\beta}(\cdot) V_{\alpha\beta}^{\dagger},
	\end{align}
	where the operators $V_{\alpha\beta}:\mathcal{H}_{A}\otimes \mathcal{H}_{B}\to \mathcal{H}_{C}\otimes \mathcal{H}_{D}$. For each pair $(\alpha,\beta)$, we define its matrix elements as
	\begin{align}\label{eq53}
		(V_{\alpha\beta})_{cd;ab} = \sqrt{\frac{d_A}{d_C}} \sum_{e} p_{ea;c}^{\alpha} q_{d;eb}^{\beta}.
	\end{align} 
	It is obvious that by
	Eqs.~\eqref{eq41} and~\eqref{eq53}, $V_{\alpha\beta}$ is real for any $\alpha,\beta$. Then we verify that $\mathcal G=\mathcal{E}_{\Theta}$.
	
	Direct calculations show that
	\begin{align}\label{eq54}
		\left[ \sum_{\alpha,\beta} V_{\alpha\beta} \rho V_{\alpha\beta}^{\dagger} \right]_{cd;c'd'} 
		&= \sum_{\alpha,\beta,a,b,a',b'} (V_{\alpha\beta})_{cd;ab} \, \rho_{ab;a'b'} \, \overline{(V_{\alpha\beta})_{c'd';a'b'}} \notag \\
		&= \frac{d_A}{d_C} \sum_{\alpha,\beta,a,b,a',b'} \left( \sum_{e} p_{ea;c}^{\alpha} q_{d;eb}^{\beta} \right) \rho_{ab;a'b'} \left( \sum_{f} \overline{p_{fa';c'}^{\alpha}} \, \overline{q_{d';fb'}^{\beta}} \right) \notag\\
		&= \frac{d_A}{d_C} \sum_{\alpha,\beta,e,f,a,a',b,b'} p_{ea;c}^{\alpha} \overline{p_{fa';c'}^{\alpha}} \, q_{d;eb}^{\beta} \overline{q_{d';fb'}^{\beta}} \rho_{ab;a'b'}.
	\end{align}
	Eqs.~\eqref{eq51} and~\eqref{eq54} coincide term by term. Therefore
	\begin{align}
		J(\Theta(\mathcal{N})) = \sum_{\alpha,\beta} V_{\alpha\beta} J(\mathcal{N}) V_{\alpha\beta}^{\dagger}=\mathcal G[J(\mathcal{N})].
	\end{align}
	The above calculation applies to every linear map
	$\mathcal N:A\to B$, with $\Theta$ acting through the
	pre- and post-processing realization in Eq.~~\eqref{eq39}.
	For any $X\in\mathcal{L}(\mathcal{H}_{A}\otimes \mathcal{H}_{B})$, let
	$\mathcal N_X=J_{A\to B}^{-1}(X)$. Then
	\begin{align}
		\mathcal G(X)
		&=\mathcal G\bigl(J_{A\to B}(\mathcal N_X)\bigr) \notag\\
		&=J_{C\to D}\bigl(\Theta(\mathcal N_X)\bigr) \notag\\
		&=\bigl(J_{C\to D}\circ\Theta\circ
		J_{A\to B}^{-1}\bigr)(X) \notag\\
		&=\mathcal E_\Theta(X).
	\end{align}
	Hence $\mathcal G=\mathcal E_\Theta$ on the entire
	operator space. Since
	$\Theta\in\mathcal{FSC}_{\mathrm{CPTP}}$,
	the induced map $\mathcal E_\Theta$ is CPTP.
	Consequently, the Kraus operators constructed above
	satisfy
	\begin{align}
		\sum_{\alpha,\beta}
		V_{\alpha\beta}^{\dagger}V_{\alpha\beta}=
		\mathbf I_A\otimes\mathbf I_B.
	\end{align}
	Moreover, each $V_{\alpha\beta}$ is real.
	Thus $\mathcal E_\Theta$ is a CPTP map admitting a real Kraus representation, and hence is a free channel.
\end{proof}
\end{appendices}

\begin{appendices}
	\renewcommand{\theequation}{\Alph{section}\arabic{equation}}
	\setcounter{equation}{0}
	\begin{center}
		\textbf{Appendix B: Proof of Section~\ref{secIII}} \label{app:B}
	\end{center}
	
	\renewcommand{\theequation}{B\arabic{equation}}
	\setcounter{equation}{0}
\begin{center}
	\textbf{B1. Proof of Proposition \ref{prop3}}\label{app:B1}
\end{center}
\begin{proof}
	Without loss of generality, we assume that $\mathcal{N}\in\mathcal{C}(A,B)$. First, since $M(\cdot)$ is an imaginarity measure for states, we know that for any state $\rho$ and channel $\mathcal{N}$, $M[\mathcal{N}(\rho)]\ge0$, which implies that $C^g_M(\mathcal{N})\ge0$. It follows from $\mathcal{F}_A \subseteq \mathcal{D}(\mathcal{H}_A)$ that $C^b_M(\mathcal{N}) \ge C^g_M(\mathcal{N})\ge 0$. When $\mathcal{N}$ is a free channel, we know that $\mathcal{N}(\rho)$ is still a free state for any free state $\rho$. Therefore, we have $C^g_M(\mathcal{N})=0$ for any free channel $\mathcal{N}$. Meanwhile, if $\mathcal{N}$ is a free channel, by monotonicity, it holds that $M[\mathcal{N}(\rho)]\le M(\rho)$, which means that $C^b_M(\mathcal{N})\le0$. Hence $C^b_M(\mathcal{N})=0$ for any free channel $\mathcal{N}$.
	
	Next, we prove that $C^g_M(\mathcal{N})$ and $C^b_M(\mathcal{N})$ are monotonic under $\mathcal{MLFSC}$. Let $\Theta\in\mathcal{MLFSC}(AB\to CD)$ be arbitrary. By definition,
	\begin{align}
		\Theta(\mathcal N_{A\to B})
		=
		\mathcal Q_{B\to D}\circ
		\mathcal N_{A\to B}\circ
		\mathcal P_{C\to A},
	\end{align}
	where $\mathcal P$ and $\mathcal Q$ are free channels.
	
	For any $\rho\in\mathcal F_C$, the state
	$\mathcal P(\rho)$ belongs to $\mathcal F_A$.
	Moreover, monotonicity of $M$ under the free channel
	$\mathcal Q$ gives
	\begin{align}
		M\bigl[\mathcal Q(\mathcal N(\mathcal P(\rho)))\bigr]
		\le
		M\bigl[\mathcal N(\mathcal P(\rho))\bigr].
	\end{align}
	Consequently,
	\begin{align}
		C_M^g(\Theta(\mathcal N))
		&=
		\max_{\rho\in\mathcal F_C}
		M\bigl[\mathcal Q(\mathcal N(\mathcal P(\rho)))\bigr]
		\nonumber\\
		&\le
		\max_{\rho\in\mathcal F_C}
		M\bigl[\mathcal N(\mathcal P(\rho))\bigr]
		\nonumber\\
		&\le
		\max_{\sigma\in\mathcal F_A}
		M[\mathcal N(\sigma)]
		\nonumber\\
		&=
		C_M^g(\mathcal N),
	\end{align}
	where the second inequality follows from
	$\mathcal P(\mathcal F_C)\subseteq\mathcal F_A$.
	
	We next consider $C_M^b$.
	For any $\rho\in\mathcal D(\mathcal H_C)$, monotonicity
	of $M$ under $\mathcal P$ and $\mathcal Q$ implies
	\begin{align}
		M\bigl[\mathcal Q(\mathcal N(\mathcal P(\rho)))\bigr]-M(\rho)&\le
		M\bigl[\mathcal N(\mathcal P(\rho))\bigr]-M(\rho)
		\nonumber\\
		&=\Bigl(
		M\bigl[\mathcal N(\mathcal P(\rho))\bigr]
		-M[\mathcal P(\rho)]\Bigr)+\Bigl(
		M[\mathcal P(\rho)]-M(\rho)\Bigr)
		\nonumber\\
		&\le M\bigl[\mathcal N(\mathcal P(\rho))\bigr]-M[\mathcal P(\rho)]
		\nonumber\\
		&\le C_M^b(\mathcal N),
	\end{align}
	where the second inequality uses
	$M[\mathcal P(\rho)]\le M(\rho)$, and the last one
	follows from
	$\mathcal P(\rho)\in\mathcal D(\mathcal H_A)$
	and the definition of $C_M^b(\mathcal N)$.
	Maximizing over $\rho\in\mathcal D(\mathcal H_C)$ yields
    $C_M^b(\Theta(\mathcal N))
    \le C_M^b(\mathcal N)$.
	Thus, both $C_M^g$ and $C_M^b$ are monotonic under
	$\mathcal{MLFSC}$.
\end{proof}

\begin{center}
	\textbf{B2. Proof of Proposition \ref{prop4}}\label{app:B2}
\end{center}

\begin{proof}
	The key to proving Proposition~\ref{prop4} lies in proving $2M_{G}(\rho)\le M_{r}(\rho)$, i.e.,
	\begin{align}\label{eq67}
		1-\sqrt{F(\rho,\rho^T)}\le S[\mathrm{Re}(\rho)]-S(\rho).
	\end{align}
	
	Since transposition preserves the spectrum of a matrix, we have $S(\rho^T)=S(\rho)$. 
	Using the definition of quantum relative entropy, a direct computation yields that
	\begin{align}\label{eq68}
		S[\mathrm{Re}(\rho)]-S(\rho)=\frac12 S(\rho\|\mathrm{Re}(\rho))+\frac12 S[\rho^T\|\mathrm{Re}(\rho)].
	\end{align}
	
	We begin by proving a stronger statement. Namely, for any $\rho$ and $\sigma$, we have
	\begin{align}\label{eq69}
		J(\rho,\sigma)\ge 1-\sqrt{F(\rho,\sigma)},
	\end{align}
	where $J(\rho,\sigma)=\frac12 S\left(\rho\|\frac{\rho+\sigma}{2}\right)+\frac12 S\left(\sigma\|\frac{\rho+\sigma}{2}\right)$ is the quantum Jensen-Shannon divergence between $\rho$ and $\sigma$. Eq.~\eqref{eq67} follows immediately from Eqs.~\eqref{eq68} and  \eqref{eq69} by setting $\sigma=\rho^T$.
	
	Now we prove Eq.~\eqref{eq69}. It follows from Ref.~\cite{CAFC} that for any two quantum states $\rho$ and $\sigma$,
	\begin{equation}\label{eq70}
		\sqrt{F(\rho,\sigma)}= \min_{\{E_i\}} \sum_i \sqrt{\operatorname{tr}(\rho E_i) \operatorname{tr}(\sigma E_i)},
	\end{equation}
	where the minimization is taken over all POVM measurements. Assume that $\{\tilde{E}_i\}$ is the POVM measurement that attains the minimum in Eq.~\eqref{eq70}, i.e., $\sqrt{F(\rho,\sigma)}=  \sum_i \sqrt{\operatorname{tr}(\rho \tilde{E}_i) \operatorname{tr}(\sigma \tilde{E}_i)}$. And further set $p_i=\operatorname{tr}(\rho \tilde{E}_i)$ and $q_i=\operatorname{tr}(\sigma \tilde{E}_i)$. Then we have $\sqrt{F(\rho,\sigma)}=\operatorname{tr}\sqrt{\sqrt{\rho}\,\sigma\sqrt{\rho}}=\sum_i \sqrt{p_i q_i}$.
	
	Define the quantum channel $\mathcal{W}$,
	\begin{align}
		\mathcal{W}(\tau) = \sum_i \operatorname{tr}(\tilde{E}_i \tau) |i\rangle\langle i|,
	\end{align}
	for any quantum state $\tau$. It can be found that
	\begin{align}
		\mathcal{W}(\rho)=\sum_i p_i|i\rangle\langle i|, \quad \mathcal{W}(\sigma)=\sum_i q_i|i\rangle\langle i|, \quad \text{and} \quad \mathcal{W}\left(\frac{\rho+\sigma}{2}\right)=\sum_i \frac{p_i+q_i}{2}|i\rangle\langle i|.
	\end{align}
	Since relative entropy obeys the data-processing inequality, one has
	\begin{align}\label{eq73}
		S\left(\rho\middle\|\frac{\rho+\sigma}{2}\right)\ge S\left[\mathcal{W}(\rho)\middle\|\mathcal{W}\left(\frac{\rho+\sigma}{2}\right)\right]=H\left(p\middle\|\frac{p+q}{2}\right),
	\end{align} and 
	\begin{align}\label{eq74}
		S\left(\sigma\middle\|\frac{\rho+\sigma}{2}\right)\ge S\left[\mathcal{W}(\sigma)\middle\|\mathcal{W}\left(\frac{\rho+\sigma}{2}\right)\right]=H\left(q\middle\|\frac{p+q}{2}\right),
	\end{align} 
	where $p$ and $q$ denote the probability distributions $\{p_i\}$ and $\{q_i\}$, respectively, and $H(m \| n) = \sum_i m_i \log(m_i / n_i)$ is  the classical relative entropy.
	
	Combining Eqs.~\eqref{eq73} and~\eqref{eq74}, one has
	\begin{align}
		J(\rho,\sigma)&=\frac{1}{2}S\left(\rho\middle\|\frac{\rho+\sigma}{2}\right)+	\frac{1}{2}S\left(\sigma\middle\|\frac{\rho+\sigma}{2}\right) \notag\\
		&\ge \frac{1}{2}H\left(p\middle\|\frac{p+q}{2}\right)+\frac{1}{2}H\left(q\middle\|\frac{p+q}{2}\right)=D(p,q), 
	\end{align}
	where $D(p,q)$ is the Jensen-Shannon divergence between the two classical probability distributions.
	
	Next, we show that 
	\begin{align}\label{eq76}
		D(p,q)\ge 1-\sum_i \sqrt{p_i q_i}=1-\sqrt{F(\rho,\sigma)}.
	\end{align}
	For this purpose, define $ w_i = \frac{p_i + q_i}{2}$, and
	for indices with \(p_i + q_i > 0\), set $x_i = \frac{p_i}{p_i + q_i}$.
	Then we have $ p_i = 2w_i x_i, q_i = 2w_i (1 - x_i)$, and $\sum_i w_i = 1$.
	The classical Jensen-Shannon divergence $D(p, q)$ is given by
	\begin{align}
		D(p, q) 
		&= \frac{1}{2} \sum_i p_i \log \frac{2p_i}{p_i + q_i} 
		+ \frac{1}{2} \sum_i q_i \log \frac{2q_i}{p_i + q_i} \notag\\
		&= \sum_i w_i \left[ x_i \log (2x_i) + (1 - x_i) \log (2(1 - x_i)) \right],
	\end{align}
	which yields that $	D(p, q) = 1 - \sum_i w_i h(x_i)$.
	
	On the other hand, it holds that
	\begin{align}
		\sum_i \sqrt{p_i q_i} 
		&= \sum_i \sqrt{2w_i x_i \cdot 2w_i (1 - x_i)} = \sum_i 2w_i \sqrt{x_i (1 - x_i)}.
	\end{align}
	Therefore, proving Eq.~\eqref{eq76} is equivalent to proving $h(x) \leq 2\sqrt{x(1-x)}$ for any $x\in[0,1]$. By elementary methods, we can prove that it holds, and omit it for brevity. Summarizing the above arguments, we obtain that Eq.~\eqref{eq69} holds.
	
	With the aid of Eq.~\eqref{eq67}, we immediately obtain that for any quantum channel $\mathcal{N}$,
	\begin{align}
		\max_{\rho\in\mathcal{F}} 2M_{G}[\mathcal{N}(\rho)]\le \max_{\rho\in\mathcal{F}} M_{r}[\mathcal{N}(\rho)],
	\end{align}
	that is, $2C^g_{M_G}(\mathcal{N})\le C^g_{M_r}(\mathcal{N})$.
\end{proof}

\begin{center}
	\textbf{B3. Proof of Proposition \ref{prop5}}\label{app:B3}
\end{center}

\begin{proof} Without loss of generality, we assume that $\mathcal{N}\in \mathcal{C}(A,B)$ and $\Theta\in\mathcal{FSC}_{\mathrm{CPTP}}(AB \to CD)$.
	Since $D$ is a distance measure, we see that $D(J(\mathcal{N}),J(\mathcal{M}))\ge0$, and $D(J(\mathcal{N}),J(\mathcal{M}))=0$ if $J(\mathcal{N})=J(\mathcal{M})$, which yields that $\mathcal{N}=\mathcal{M}\in \mathcal{FC}(A,B)$. So (C1) holds for $C^J_D(\mathcal{N})\ge 0$.
	
	For any free superchannel $\Theta$, we have
	\begin{align}\label{eq85-}
		C^J_D(\Theta(\mathcal{N}))
		&=\min_{\mathcal{M}\in \mathcal{FC}(C,D)}D(J(\Theta(\mathcal{N})),J(\mathcal{M})) \notag \\
		&=\min_{\mathcal{M}\in \mathcal{FC}(C,D)}D(\mathcal{E}_{\Theta}(J(\mathcal{N})),J(\mathcal{M})) \notag \\
		&\le\min_{\mathcal{\tilde{M}}\in \mathcal{FC}(A,B)}D(\mathcal{E}_{\Theta}(J(\mathcal{N})),\mathcal{E}_{\Theta}(J(\mathcal{\tilde{M}}))) \notag \\
		&\le\min_{\mathcal{\tilde{M}}\in \mathcal{FC}(A,B)}D(J(\mathcal{N}),J(\mathcal{\tilde{M}})) \notag \\
		&=C^J_D(\mathcal{N}),
	\end{align}
	where the second line comes from Eq.~\eqref{eq34}, the third one follows from the fact that $\mathcal{E}_{\Theta}$ transfers real Choi states to real Choi states since $\Theta$ is a free superchannel, and the fourth one is true because $D$ satisfies data-processing inequality. Therefore, $C^J_D(\Theta(\mathcal{N}))$ satisfies (C1) and (C2), i.e., it is an imaginarity measure of quantum channels.  
\end{proof}

\begin{center}
	\textbf{B4. Proof of Proposition \ref{prop6}}\label{app:B4}
\end{center}
\begin{proof}
	First of all, note that
	\begin{align}\label{eq91}
		C^J_S(\mathcal{N})&=\min_{\mathcal{M}\in \mathcal{FC}(A,B)}S(J(\mathcal{N})\|J(\mathcal{M}))\notag  \\
		&\ge \min_{\sigma\in \mathcal{F}_{AB}}S(J(\mathcal{N})\|\sigma)\notag  \\
		&=S[\mathrm{Re}(J(\mathcal{N}))]-S[J(\mathcal{N})].
	\end{align}
	
	On the other hand, let $\{N_k\}$ be a Kraus representation of $\mathcal{N}$, and write
	\begin{equation}
		N_k = A_k + \mathrm{i}B_k, \quad A_k = \frac{N_k + \overline{N_k}}{2}, \quad B_k = \frac{N_k - \overline{N_k}}{2\mathrm{i}},
	\end{equation}
	where $A_k$ and $B_k$ are real matrices. Consider the channel
	\begin{equation}\label{B23}
		\mathcal{M}_{\mathcal{N}}= \frac{\mathcal{N} + \overline{\mathcal{N}}}{2}.
	\end{equation}
	
	For any input state $\rho$, we have
	\begin{align}
		\mathcal{M}_{\mathcal{N}}(\rho) &= \frac{1}{2} \sum_k \left( N_k \rho N_k^\dagger + \overline{N_k} \rho \overline{N_k}^{\dagger} \right) = \sum_k \left( A_k \rho A_k^T + B_k \rho B_k^T \right).
	\end{align}
	Moreover, Since $\sum_k N_k^\dagger N_k = \mathbf{I}_A$, we have $	\sum_k \left( A_k^T A_k + B_k^T B_k \right) =\mathbf{I}_A$.
	Thus, $\mathcal{M}_{\mathcal{N}}$ admits the real Kraus representation $\{A_k, B_k\}$, and hence $\mathcal{M}_{\mathcal{N}} \in \mathcal{FC}$. At the same time, note that
	\begin{align}
		J(\overline{\mathcal N})
		&=\sum_k(\mathbb{I}_A\otimes\overline{N_k})\phi_{AA'}(\mathbb{I}_A\otimes \overline{N_k}^{\dagger})\notag\\
		&=\overline{\sum_k(\mathbb{I}_A\otimes N_k)\phi_{AA'}(\mathbb{I}_A\otimes N_k^\dagger)}\notag\\
		&=\overline{J(\mathcal N)},
	\end{align}
	which yields that
	\begin{align}
		J(\mathcal{M}_{\mathcal{N}}) &= J\left(\frac{\mathcal{N} + \overline{\mathcal{N}}}{2}\right) = \frac{J(\mathcal{N}) + J(\overline{\mathcal{N}})}{2} \notag\\
		&= \frac{J(\mathcal{N}) + \overline{J(\mathcal{N})}}{2} = \frac{J(\mathcal{N}) + J(\mathcal{N})^T}{2} = \mathrm{Re}J(\mathcal{N}).
	\end{align}
	Therefore, we get
	\begin{align}\label{eq99}
		C^J_S(\mathcal{N})&=\min_{\mathcal{M}\in \mathcal{FC}(A,B)}S(J(\mathcal{N})\|J(\mathcal{M}))\notag  \\
		&\le S(J(\mathcal{N})\|J(\mathcal{M}_{\mathcal{N}}))\notag  \\
		&=S(J(\mathcal{N})\|\mathrm{Re}J(\mathcal{N}))\notag  \\
		&=S[\mathrm{Re}(J(\mathcal{N}))]-S[J(\mathcal{N})].
	\end{align}
	Combining Eqs.~\eqref{eq91} and~\eqref{eq99} yields Eq.~\eqref{eq90}.
\end{proof}

\begin{center}
	\textbf{B5. Proof of Proposition \ref{prop10}}\label{app:B8}
\end{center}

\begin{proof}
	Since $\mathcal{FC^J}_{AB}\subseteq \mathcal{F}_{AB}$, we obtain that 
	\begin{align}\label{C0}
		C_G^J(\mathcal{N})\ge M_G[J(\mathcal{N})]=\frac{1-\sqrt{F[J(\mathcal{N}),J(\mathcal{N})^T]}}{2},
	\end{align}
	which means that the left-hand side of Eq.~\eqref{eq120} holds.	
	
	On the other hand, in Proposition~\ref{prop6} we show that $\mathcal{M}_{\mathcal{N}}= \frac{\mathcal{N} + \overline{\mathcal{N}}}{2}$ is a free channel, thus we have
	\begin{align}\label{eq122}
		C^J_G(\mathcal{N})&=1-\max_{\mathcal{M}\in \mathcal{FC}}F(J(\mathcal{N}),J(\mathcal{M}))\notag\\
		&\le 1-F[J(\mathcal{N}),J(\mathcal{M}_{\mathcal{N}})] \notag\\
		&= 1-F[J(\mathcal{N}),\mathrm{Re}(J(\mathcal{N}))].
	\end{align}
\end{proof}

\begin{center}
	\textbf{B6. Proof of Proposition \ref{prop11-}}\label{app:B9}
\end{center}
\begin{proof}
	By the semidefinite programming characterization of the fidelity~\cite{JW2012}, for any quantum states $\rho,\sigma$,
	\begin{equation}
		\sqrt{F(\rho,\sigma)}
		=
		\max_X\left\{
		\frac{1}{2}\operatorname{tr}X
		+\frac{1}{2}\operatorname{tr}X^\dagger:
		\begin{pmatrix}
			\rho & X\\
			X^\dagger & \sigma
		\end{pmatrix}\succeq0
		\right\},
	\end{equation}
	where $X$ ranges over linear operators on the underlying
	Hilbert space.
	
	By the characterization of free channels in terms of real Choi matrices, their Choi states form the set
	\begin{equation}
		\mathcal F^J=
		\left\{
		\begin{aligned}
			\sigma\in\mathcal{D}
			(\mathcal H_A\otimes\mathcal H_B):\quad&
			\sigma=\sigma^T,
			&\operatorname{tr}_B\sigma=\frac{\mathbf{I}_A}{d_A}
		\end{aligned}
		\right\}.
	\end{equation}
	
	Setting $\rho=J(\mathcal N)$ and optimizing over
	$\sigma\in\mathcal F^J$ yields
	\begin{equation}
		u=
		\max_{\mathcal M\in\mathcal{FC}(A,B)}
		\sqrt{F\bigl(J(\mathcal N),J(\mathcal M)\bigr)}.
	\end{equation}
	Since squaring is monotonic on the nonnegative real numbers,
	\begin{equation}
		u^2=
		\max_{\mathcal M\in\mathcal{FC}(A,B)}
		F\bigl(J(\mathcal N),J(\mathcal M)\bigr).
	\end{equation}
	The result thus follows from the definition of $C_G^J(\mathcal{N})$.
\end{proof}
\end{appendices}

\begin{appendices}
	\renewcommand{\theequation}{\Alph{section}\arabic{equation}}
	\setcounter{equation}{0}
	\begin{center}
		\textbf{Appendix C: Proof of Section~\ref{secIV}} \label{app:C}
	\end{center}
	
	\renewcommand{\theequation}{C\arabic{equation}}
	\setcounter{equation}{0}

\begin{center}
	\textbf{C1. Proof of Theorem \ref{thm1}}\label{app:C1}
\end{center}

\begin{proof}
	From the triangle inequality and the homogeneity property of the trace norm, it follows that
	\begin{align}
		p_{\mathrm{succ}}(\mathcal{N},\mathcal{M},\mathcal{F})
		&=\max_{\rho\in\mathcal{F}}p_{\mathrm{succ}}(\mathcal{N},\mathcal{M},\rho) \notag \\
		&=\frac{1}{2}+\frac{1}{4}\max_{\rho\in\mathcal{F}}\lVert\mathcal{N}(\rho)-\mathcal{M}(\rho)\rVert_1 \notag \\
		&=\frac{1}{2}+\frac{1}{8}\max_{\rho\in\mathcal{F}}\left\{\lVert\mathcal{N}(\rho)-\mathcal{M}(\rho)\rVert_1+\left\lVert[\mathcal{N}(\rho)]^T-[\mathcal{M}(\rho)]^T\right\rVert_1\right\} \notag \\
		&=\frac{1}{2}+\frac{1}{8}\max_{\rho\in\mathcal{F}}\left\{\lVert\mathcal{N}(\rho)-\mathcal{M}(\rho)\rVert_1+\left\lVert[\mathcal{M}(\rho)]^T-[\mathcal{N}(\rho)]^T\right\rVert_1\right\} \notag \\
		&\ge\frac{1}{2}+\frac{1}{8}\max_{\rho\in\mathcal{F}}\left\{\left\lVert\mathcal{N}(\rho)-\mathcal{M}(\rho)+[\mathcal{M}(\rho)]^T-[\mathcal{N}(\rho)]^T\right\rVert_1\right\} \notag \\
		&\ge\frac{1}{2}+\frac{1}{8}\max_{\rho\in\mathcal{F}}\left\{\left|\left\lVert\mathcal{N}(\rho)-[\mathcal{N}(\rho)]^T\right\rVert_1-\left\lVert[\mathcal{M}(\rho)]^T-\mathcal{M}(\rho)\right\rVert_1\right|\right\} \notag \\
		&\ge\frac{1}{2}+\frac{1}{8}\left\{\left|\max_{\rho\in\mathcal{F}}\left\lVert\mathcal{N}(\rho)-[\mathcal{N}(\rho)]^T\right\rVert_1-\max_{\rho\in\mathcal{F}}\left\lVert\mathcal{M}(\rho)-[\mathcal{M}(\rho)]^T\right\rVert_1\right|\right\} \notag \\
		&=\frac{1}{2}+\frac{1}{4}\left|C^g_{M_R}(\mathcal{N})-C^g_{M_R}(\mathcal{M})\right|, 
	\end{align}
	which means that Theorem~\ref{thm1} holds. 
\end{proof}

\begin{center}
	\textbf{C2. Proof of Theorem \ref{thm2}}\label{app:C2}
\end{center}

\begin{proof}
	Let $\mathcal{N} \in \mathrm{RNG} \setminus \mathrm{RKO}$, and $\{T_\mu\}$ be a Kraus representation of $\mathcal N$, i.e.,
	\begin{equation}\label{F1}
		\mathcal{N}(\cdot)=
		\sum_\mu T_\mu (\cdot) T_\mu^\dagger, \quad
		\sum_\mu T_\mu^\dagger T_\mu=\mathbf{I}.
	\end{equation}
	By the definition, the channel $\overline{\mathcal{N}}$ admits a Kraus representation $\{\overline{T_\mu}\}$. 
	
	It follows from Proposition~\ref{prop6} that $\mathcal{M}_{\mathcal{N}}\in\mathrm{RKO}$, where $\mathcal{M}_{\mathcal{N}}$ is defined in Eq.~\eqref{B23}.
	Write each Kraus operator as
	\begin{equation}
		T_\mu=A_\mu+\mathrm{i}B_\mu,
		\quad
		A_\mu=\operatorname{Re}T_\mu,
		\quad
		B_\mu=\operatorname{Im}T_\mu,
	\end{equation}
	where $A_\mu$ and $B_\mu$ are real matrices. Then $\{A_\mu,B_\mu\}$ is a set of real Kraus operators for
	$\mathcal{M}_{\mathcal{N}}$. 
	
	We next show that $\mathcal{N}$ and $\mathcal{M}_{\mathcal{N}}$ coincide
	on every real input state. Let $\sigma\in \mathcal{F}$. Since $\mathcal{N}\in \mathrm{RNG}$, the output
	$\mathcal{N}(\sigma)$ is real. Note that
	\begin{align}
		\mathcal{N}(\sigma)= \sum_{\mu} \left( A_{\mu} \sigma A_{\mu}^T + B_{\mu} \sigma B_{\mu}^T \right) 
		+ \mathrm{i} \sum_{\mu} \left( B_{\mu} \sigma A_{\mu}^T - A_{\mu} \sigma B_{\mu}^T \right),
	\end{align}
	which means that
	\begin{align}\label{eq146}
		\sum_{\mu} \left( B_{\mu} \sigma A_{\mu}^T - A_{\mu} \sigma B_{\mu}^T \right)=0.
	\end{align}
	Noting that
	\begin{align}
		\overline{\mathcal{N}}(\sigma)
		=\sum_\mu \overline{T_\mu} \sigma \overline{T_\mu}^{\dagger} =\sum_\mu \left( A_\mu \sigma A_\mu^T + B_\mu \sigma B_\mu^T \right)- \mathrm{i}\sum_\mu \left( B_\mu \sigma A_\mu^T - A_\mu \sigma B_\mu^T \right),
	\end{align}
	it follows from Eq.~\eqref{eq146} that $	\mathcal{N}(\sigma)=\overline{\mathcal{N}}(\sigma)$.
	Consequently, for any $\sigma\in \mathcal{F}$, we have
	\begin{equation}\label{eq149}
		\mathcal M_{\mathcal N}(\sigma)=\frac{1}{2}\left[
		\mathcal{N}(\sigma)+\mathcal{\overline{N}}(\sigma)\right]=
		\mathcal N(\sigma).
	\end{equation}
	Eq.~\eqref{eq149} then implies
	\begin{equation}\label{eq150}
		p_{\mathrm{succ}}(\mathcal{N},\mathcal{M}_{\mathcal{N}},\mathcal{F})
		=\frac{1}{2}+\max_{\rho\in\mathcal{F}}\frac{1}{4}\lVert\mathcal{N}(\rho)-\mathcal{M}_{\mathcal{N}}(\rho)\rVert_1=\frac{1}{2}.
	\end{equation}
	
	On the other hand, since $\mathcal{N}\in \mathrm{RNG}\setminus\mathrm{RKO}$ whereas
	$\mathcal{M}_{\mathcal{N}}\in \mathrm{RKO}$, we get $\mathcal N\neq\mathcal M_{\mathcal N}$,
	which means that there exists a state
	$\tilde{\rho}\in \mathcal{D}(\mathcal{H})$ such that
    $\mathcal N(\tilde{\rho})\neq\mathcal M_{\mathcal N}(\tilde{\rho}$.
	Therefore, $\left\|\mathcal N(\tilde{\rho})-\mathcal M_{\mathcal N}(\tilde{\rho})\right\|_1>0$.
	Using $\tilde{\rho}$ as an admissible input in
	Eq.~\eqref{eq137}, we obtain that
	\begin{align}\label{eq154}
		p_{\mathrm{succ}}\left(\mathcal{N},
		\mathcal{M}_{\mathcal{N}},\mathcal{D}(\mathcal{H})\right)&=\frac{1}{2}+\max_{\rho\in\mathcal{D}(\mathcal{H})}\frac{1}{4}\lVert\mathcal{N}(\rho)-\mathcal{M}_{\mathcal{N}}(\rho)\rVert_1 \notag \\
		&\geq \frac{1}{2}+\frac{1}{4}\left\|\mathcal N(\tilde{\rho})-\mathcal M_{\mathcal N}(\tilde{\rho})\right\|_1>\frac{1}{2}.
	\end{align}
	Combining Eqs.~\eqref{eq150} and \eqref{eq154} proves Eq.~\eqref{eq141}.
\end{proof}
\end{appendices}

\begin{appendices}
	\renewcommand{\theequation}{\Alph{section}\arabic{equation}}
	\setcounter{equation}{0}
	\begin{center}
		\textbf{Appendix D: Proof of Section~\ref{secV}} \label{app:D}
	\end{center}
	
	\renewcommand{\theequation}{D\arabic{equation}}
	\setcounter{equation}{0}
\begin{center}
	\textbf{D1. Proof of Proposition~\ref{prop11}}\label{app:D1}
\end{center}

\begin{proof}
	Suppose that $\Phi(\mathcal N_1)=\mathcal N_2$, where
	$\Phi\in\mathcal{FSC}$. Let
	$m/n\in R(\mathcal M,\mathcal N_1)$ be an arbitrary achievable
	conversion rate, with $m,n\in\mathbb N$. By the definition, there
	exists $\widetilde{\Theta}\in\mathcal{FSC}$ such that
    $\widetilde{\Theta}(\mathcal M^{\otimes n})
    =\mathcal N_1^{\otimes m}$.
	Applying $\Phi$ independently to the $m$ output copies, we obtain
	\begin{align}
		\left(\Phi^{\otimes m}\circ\widetilde{\Theta}\right)
		\left(\mathcal M^{\otimes n}\right)
		=
		\Phi^{\otimes m}\left(\mathcal N_1^{\otimes m}\right)
		=
		\mathcal N_2^{\otimes m}.
	\end{align}
	The class $\mathcal{FSC}$ is closed under tensor products and
	composition, since these channels preserve realizations with
	real pre-processing and post-processing. Hence,
	$\Phi^{\otimes m}\circ\widetilde{\Theta}\in\mathcal{FSC}$, which
	implies that $m/n\in R(\mathcal M,\mathcal N_2)$. Therefore,
	\begin{align}
		R(\mathcal M,\mathcal N_1)
		\subseteq
		R(\mathcal M,\mathcal N_2).
	\end{align}
	Taking supremum over the achievable rates gives
	\begin{align}
		r(\mathcal M\to\mathcal N_1)
		\le
		r(\mathcal M\to\mathcal N_2),
	\end{align}
	which proves Eq.~\eqref{eq159}.
	
	Next, let $m/n\in R(\mathcal N_2,\mathcal M)$ be an arbitrary
	achievable conversion rate, with $m,n\in\mathbb N$.
	Then there exists $\Theta^*\in\mathcal{FSC}$ such that
    $\Theta^*(\mathcal N_2^{\otimes n})
    =\mathcal M^{\otimes m}$.
	Precomposing this protocol with $\Phi^{\otimes n}$ yields
	\begin{align}
		\left(\Theta^*\circ\Phi^{\otimes n}\right)
		\left(\mathcal N_1^{\otimes n}\right)
		=
		\Theta^*\left(\mathcal N_2^{\otimes n}\right)
		=
		\mathcal M^{\otimes m}.
	\end{align}
	Since $\Theta^*\circ\Phi^{\otimes n}\in\mathcal{FSC}$, it follows
	that $m/n\in R(\mathcal N_1,\mathcal M)$. Consequently,
	\begin{align}
		R(\mathcal N_2,\mathcal M)
		\subseteq
		R(\mathcal N_1,\mathcal M).
	\end{align}
	Taking supremum gives
	\begin{align}
		r(\mathcal N_2\to\mathcal M)
		\le
		r(\mathcal N_1\to\mathcal M),
	\end{align}
	which establishes Eq.~\eqref{eq160}.
	
	For any $\Theta,\Phi \in\mathcal{FSC}_{\mathrm{CPTP}}$, by Eq.~\eqref{eq33}, we have
	\begin{align}
		&\mathcal{E}_{\Theta\circ \Phi}=J\circ (\Theta\circ \Phi)\circ J^{-1}\notag \\
		&=J\circ \Theta\circ J^{-1}\circ J\circ \Phi\circ J^{-1}\notag \\
		&=\mathcal{E}_{\Theta}\circ\mathcal{E}_{\Phi}.
	\end{align}
	Since $\mathcal{E}_{\Theta}$ and $\mathcal{E}_{\Phi}$ are both CPTP, we know that $\mathcal{E}_{\Theta\circ \Phi}$ is a CPTP, which means that $\Theta\circ \Phi\in\mathcal{FSC}_{\mathrm{CPTP}}$.
	
	Using the method employed in the proof of Eqs.~\eqref{eq159} and~\eqref{eq160}, together with the above fact, we readily obtain Eqs.~\eqref{eq161} and~\eqref{eq162}. Hence, for brevity, their proofs are omitted.
\end{proof}

\begin{center}
	\textbf{D2. Proof of Proposition~\ref{prop12}}\label{app:D2}
\end{center}

\begin{proof}
	If either $r(\mathcal N_1\to\mathcal N_2)$ or
	$r(\mathcal N_2\to\mathcal N_3)$ is zero, the right-hand side
	is zero under the above convention, and Eq.~\eqref{eq168} follows
	from the nonnegativity of $r(\mathcal N_1\to\mathcal N_3)$. Similar argument holds for Eq.~\eqref{eq169}.
	We therefore assume that both rates are strictly positive,
	possibly infinite.
	
	Take arbitrary
	\begin{equation*}
		a = \frac{m_1}{n_1} \in R(\mathcal{N}_1, \mathcal{N}_2), \quad b = \frac{m_2}{n_2} \in R(\mathcal{N}_2, \mathcal{N}_3),
	\end{equation*}
	where $n_1, n_2, m_1, m_2 \in \mathbb{N}$. According to the definition of the exact conversion rate, there exist $\Theta_1, \Theta_2 \in \mathcal{FSC}$, satisfying
	\begin{equation}
		\Theta_1(\mathcal{N}_1^{\otimes n_1}) = \mathcal{N}_2^{\otimes m_1} \text{  and  } \Theta_2(\mathcal{N}_2^{\otimes n_2}) = \mathcal{N}_3^{\otimes m_2}.
	\end{equation}
	
	Implementing the first protocol in parallel $n_2$ times yields
	\begin{equation}
		\begin{aligned}
			\Theta_1^{\otimes n_2}(\mathcal{N}_1^{\otimes n_1 n_2}) = [\Theta_1(\mathcal{N}_1^{\otimes n_1})]^{\otimes n_2} 
			= \mathcal{N}_2^{\otimes m_1 n_2}.
		\end{aligned}
	\end{equation}
	
	Subsequently, the obtained $m_1 n_2$ intermediate channels are divided into $m_1$ blocks, each containing $n_2$ copies of $\mathcal{N}_2$, and the second protocol is implemented on each block, yielding
	\begin{equation}
		\begin{aligned}
			\Theta_2^{\otimes m_1}(\mathcal{N}_2^{\otimes m_1 n_2})= [\Theta_2(\mathcal{N}_2^{\otimes n_2})]^{\otimes m_1}
			= \mathcal{N}_3^{\otimes m_1 m_2}.
		\end{aligned}
	\end{equation}
	
	Since $\mathcal{FSC}$ is closed under tensor products and composition, the entire process constitutes a free superchannel $\Xi=\Theta_2^{\otimes m_1}\circ\Theta_1^{\otimes n_2}$, satisfying
	\begin{equation}
		\Xi(\mathcal{N}_1^{\otimes n_1 n_2}) =\Theta_2^{\otimes m_1}\circ\Theta_1^{\otimes n_2}(\mathcal{N}_1^{\otimes n_1 n_2})= \mathcal{N}_3^{\otimes m_1 m_2}.
	\end{equation}
	
	Thus,
	\begin{equation}
		ab = \frac{m_1 m_2}{n_1 n_2} \in R(\mathcal{N}_1, \mathcal{N}_3).
	\end{equation}
	
	Consequently,
	\begin{equation}
		r(\mathcal{N}_1 \to \mathcal{N}_3) \ge ab.
	\end{equation}
	
	Directly taking the supremum, we have
	\begin{align}
		r(\mathcal{N}_1 \to \mathcal{N}_3) &\ge \sup_{\substack{a \in R({\mathcal{N}_1\to\mathcal{N}_2}) \\ b \in R({\mathcal{N}_2\to\mathcal{N}_3})}} ab \notag\\
		&= \left( \sup_{a \in R({\mathcal{N}_1\to\mathcal{N}_2})} a \right) \left( \sup_{b \in R({\mathcal{N}_2\to\mathcal{N}_3})} b \right) \notag\\
		&= r(\mathcal{N}_1 \to \mathcal{N}_2) r(\mathcal{N}_2 \to \mathcal{N}_3).
	\end{align}
\end{proof}

\begin{center}
	\textbf{D3. Proof of Theorem~\ref{thm3}}\label{app:D3}
\end{center}

\begin{proof}
	If $\mathcal{L}^J(\mathcal{N}_1) = +\infty$, then the upper bound holds trivially. In the following, we assume that $\mathcal{L}^J(\mathcal{N}_1) < +\infty$.
	
	Consider an arbitrary feasible exact conversion protocol, i.e., there exist $n, m \in \mathbb{N}$ and $\Theta \in \mathcal{FSC}_{\mathrm{CPTP}}$ such that $	\Theta(\mathcal{N}_1^{\otimes n}) = \mathcal{N}_2^{\otimes m}$.
	
	Let $\mathcal{E}_{\Theta}$ denote the map induced by superchannel $\Theta$. We thus have 
	\begin{equation}\label{eq181}
		\mathcal{E}_{\Theta}\!\left( J(\mathcal{N}_1)^{\otimes n} \right) =\mathcal{E}_{\Theta}\!\left( J(\mathcal{N}^{\otimes n}_1) \right)=J(\mathcal{N}^{\otimes m}_2)= J(\mathcal{N}_2)^{\otimes m}.
	\end{equation}
	In addition, from Proposition~\ref{prop2}, we know that $\mathcal{E}_{\Theta}$ is a free channel. It follows from Ref.~\cite[Corollary~1]{HAGG} that 
	\begin{align}\label{eq182}
		\mathcal{E}_{\Theta}(\rho^T)=[\mathcal{E}_{\Theta}(\rho)]^T.
	\end{align}
	
	Combining Eqs.~\eqref{eq181} and~\eqref{eq182}, we obtain that
	\begin{align}\label{eq183}
		\mathcal{E}_{\Theta}\!\left( \left[ J(\mathcal{N}_1)^T \right]^{\otimes n} \right) 
		= \mathcal{E}_{\Theta}\!\left( \left[ J(\mathcal{N}_1)^{\otimes n} \right]^T \right) = \left[ \mathcal{E}_{\Theta}\!\left( J(\mathcal{N}_1)^{\otimes n} \right) \right]^T = \left[ J(\mathcal{N}_2)^T\right]^{\otimes m}.
	\end{align}
	
	Therefore, one has
	\begin{align}\label{eq184}
		\left[ F\!\left( J(\mathcal{N}_2), J(\mathcal{N}_2)^T \right) \right]^m 
		&= F\!\left( J(\mathcal{N}_2)^{\otimes m}, \left[ J(\mathcal{N}_2)^T \right]^{\otimes m} \right) \notag \\
		&= F\!\left( \mathcal{E}_{\Theta}\!\left( J(\mathcal{N}_1)^{\otimes n} \right), \mathcal{E}_{\Theta}\!\left( \left[ J(\mathcal{N}_1)^T \right]^{\otimes n} \right) \right) \notag\\
		&\geq F\!\left( J(\mathcal{N}_1)^{\otimes n}, \left[ J(\mathcal{N}_1)^T \right]^{\otimes n} \right) \notag \\
		&= \left[ F\!\left( J(\mathcal{N}_1), J(\mathcal{N}_1)^T \right) \right]^n. 
	\end{align}
	Here, the first and last equalities follow from the tensor product property of the fidelity, i.e., $F(\rho_1 \otimes \rho_2, \sigma_1 \otimes \sigma_2) = F(\rho_1, \sigma_1) F(\rho_2, \sigma_2)$, and the second one follows from Eqs.~\eqref{eq181} and~\eqref{eq183}, while the inequality follows from the monotonicity of the fidelity under CPTP maps.
	
	Since $\mathcal{L}^J(\mathcal{N}_1) < +\infty$ and $0 < \mathcal{L}^J(\mathcal{N}_2) < +\infty$, the two fidelities involved in Eq.~\eqref{eq184} are both strictly greater than zero. Taking the logarithm on both sides of Eq.~\eqref{eq184}, we obtain
	\begin{equation}
		-m \log F\!\left( J(\mathcal{N}_2), J(\mathcal{N}_2)^T \right) \leq -n \log F\!\left( J(\mathcal{N}_1), J(\mathcal{N}_1)^T \right).
	\end{equation}
	
	According to the definition of $\mathcal{L}^J(\mathcal{N})$, the above inequality is equivalent to
	\begin{equation}
		m \mathcal{L}^J(\mathcal{N}_2) \leq n \mathcal{L}^J(\mathcal{N}_1).
	\end{equation}
	Since $\mathcal{L}^J(\mathcal{N}_2) > 0$, we obtain that
	\begin{equation}
		\frac{m}{n} \leq \frac{\mathcal{L}^J(\mathcal{N}_1)}{\mathcal{L}^J(\mathcal{N}_2)}.
	\end{equation}
	
	When $m=0$, Eq.~\eqref{eq179} also holds trivially. Therefore, taking the supremum over all feasible exact conversion ratios $m/n$, we obtain
	\begin{equation}
		r^*(\mathcal{N}_1 \to \mathcal{N}_2) \leq \frac{\mathcal{L}^J(\mathcal{N}_1)}{\mathcal{L}^J(\mathcal{N}_2)}. 
	\end{equation}
\end{proof}

\begin{center}
	\textbf{D4. Proof of Proposition~\ref{prop14}}\label{app:D4}
\end{center}

\begin{proof}
	The Choi state of the replacement channel $\mathcal{P}_s$ is
	\begin{equation}
		J(\mathcal{P}_s) = \frac{\mathbf{I}_d}{d} \otimes |\psi_s\rangle\langle\psi_s|.
	\end{equation}
	A direct calculation shows that
	\begin{equation}
		F\!\left( J(\mathcal{P}_s), J(\mathcal{P}_s)^T \right) = \left| \langle\psi_s^*|\psi_s\rangle \right|^2 = s^2.
	\end{equation}
	Therefore, $\mathcal{L}^J(\mathcal{P}_s) = -2 \log s$, which yields that
	\begin{align}
		\frac{\mathcal{L}^J(\mathcal{P}_{s_1})}{\mathcal{L}^J(\mathcal{P}_{s_2})} = \frac{\log s_1}{\log s_2}.
	\end{align}
	Next we claim that for any \(n, m \in \mathbb{N}\), under $\mathcal{FSC}$ or $\mathcal{FSC}_{\mathrm{CPTP}}$, we have
	\begin{equation}\label{eq193}
		\mathcal{P}_{s_1}^{\otimes n} \longrightarrow \mathcal{P}_{s_2}^{\otimes m}
		\iff s_1^n \leq s_2^m.
	\end{equation}
	
	First, we prove the necessity. For each fixed pair $n,m$, let $A$ and $B$ denote
	the input and output systems of
	$\mathcal P_{s_1}^{\otimes n}$, and $C$ and $D$
	denote those of $\mathcal P_{s_2}^{\otimes m}$. Suppose there exists $\Theta \in \mathcal{FSC}$ such that
    $\Theta(\mathcal{P}_{s_1}^{\otimes n}) = \mathcal{P}_{s_2}^{\otimes m}$,
	which implies that there exists free channels $\mathcal{Q}$ and $\mathcal{P}$ such that
	\begin{align}\label{eq195}
		\mathcal{Q} \circ (\mathbb{I}_E \otimes \mathcal{P}_{s_1}^{\otimes n}) \circ \mathcal{P} = \mathcal{P}_{s_2}^{\otimes m}.
	\end{align}
	Since Eq.~\eqref{eq195} holds for any input state,  we can choose it as $\tau_C=|0\rangle\langle0|_{C}$. Then set $\mathcal{P}(\tau_C)=\sigma_{EA}$, where $\sigma_{EA}$ is a real state on $\mathcal{H}_E \otimes \mathcal{H}_A$. Note that
	\begin{align}\label{eq196}
		\sigma_{EA} &= (\mathbf{I}_E \otimes \mathbf{I}_A)\sigma_{EA}(\mathbf{I}_E \otimes \mathbf{I}_A) \notag\\
		&= \sum_{a,a'} (\mathbf{I}_E \otimes |a\rangle\langle a|)\sigma_{EA}(\mathbf{I}_E \otimes |a'\rangle\langle a'|) \notag \\
		&= \sum_{a,a'}(\mathbf{I}_E \otimes |a\rangle) \left[ (\mathbf{I}_E \otimes \langle a|)\sigma_{EA}(\mathbf{I}_E \otimes |a'\rangle) \right] (\mathbf{I}_E \otimes \langle a'|) \notag\\
		&= \sum_{a,a'}X_{aa'}^E \otimes |a\rangle\langle a'|,
	\end{align}
	where $X_{aa'}^E=(\mathbf{I}_E \otimes \langle a|)\sigma_{EA}(\mathbf{I}_E \otimes |a'\rangle)$. And from Eq.~\eqref{eq188}, we have 
	\begin{equation}\label{eq197}
		\mathcal{P}_{s_1}^{\otimes n}(|a\rangle\langle a'|) = \delta_{aa'} \left[ |\psi_{s_1}\rangle\langle \psi_{s_1}| \right]^{\otimes n}.
	\end{equation}
	Combining Eqs.~\eqref{eq196} and~\eqref{eq197} yields that
	\begin{align}
		(\mathbb{I}_E \otimes \mathcal{P}_{s_1}^{\otimes n})(\sigma_{EA}) 
		&= \sum_{a,a'} X_{aa'}^E \otimes \delta_{aa'} \left[ |\psi_{s_1}\rangle\langle \psi_{s_1}| \right]^{\otimes n} \notag\\
		&= \left( \sum_{a} X_{aa}^E \right) \otimes \left[ |\psi_{s_1}\rangle\langle \psi_{s_1}| \right]^{\otimes n} \notag\\
		&= \mathrm{tr}_A(\sigma_{EA}) \otimes \left[ |\psi_{s_1}\rangle\langle \psi_{s_1}| \right]^{\otimes n}.
	\end{align}
	Therefore, one gets 
	\begin{align}
		(\mathbb{I}_E \otimes \mathcal{P}_{s_1}^{\otimes n}) \circ \mathcal{P}(\tau_C)=\mathrm{tr}_A(\sigma_{EA}) \otimes \left[ |\psi_{s_1}\rangle\langle \psi_{s_1}| \right]^{\otimes n}.
	\end{align}
	Then from Eq.~\eqref{eq195}, we know that
	\begin{align}\label{eq200}
		\mathcal{Q}(\mathrm{tr}_A(\sigma_{EA}) \otimes \left[ |\psi_{s_1}\rangle\langle \psi_{s_1}| \right]^{\otimes n})= \left[ |\psi_{s_2}\rangle\langle \psi_{s_2}| \right]^{\otimes m}.
	\end{align} 
	We define a channel $\mathcal{T}$ of appending real state $\mathrm{tr}_A(\sigma_{EA})$, i.e., $\mathcal{T}(\rho)=\mathrm{tr}_A(\sigma_{EA})\otimes \rho$ for any quantum state. It can be verified that $\mathcal{T}$ is a free channel. Then Eq.~\eqref{eq200} is equivalent to
	\begin{align}\label{eq201}
		\mathcal{Q}\circ\mathcal{T}(\left[ |\psi_{s_1}\rangle\langle \psi_{s_1}| \right]^{\otimes n})=\left[ |\psi_{s_2}\rangle\langle \psi_{s_2}| \right]^{\otimes m}.
	\end{align} That is, free channel $\mathcal{Q}\circ\mathcal{T}$ transforms $\left[ |\psi_{s_1}\rangle\langle \psi_{s_1}| \right]^{\otimes n}$ into $\left[ |\psi_{s_2}\rangle\langle \psi_{s_2}| \right]^{\otimes m}$. At the same time, it follows from Ref.~\cite[Theorem~1]{WKKT} that in stochastic transformations, the maximum probability for a pure state transformation $|\psi\rangle \rightarrow |\phi\rangle$ via stochastic free operations is given by
	\begin{equation}
		P(|\psi\rangle \rightarrow |\phi\rangle) = \min \left\{ \frac{1 - |\langle \psi^* | \psi \rangle|}{1 - |\langle \phi^* | \phi \rangle|}, 1 \right\}. 
	\end{equation}
	When the maximum probability is equal to 1, it becomes deterministic transformations, which means that there exists a free channel $\hat{\mathcal{N}}$ such that $\hat{\mathcal{N}}(|\psi\rangle)=|\phi\rangle$ iff $P(|\psi\rangle \rightarrow |\phi\rangle)=1$, i.e., 
	\begin{align}\label{eq203}
		|\langle \psi^* | \psi \rangle| \le |\langle \varphi^* | \varphi \rangle|.
	\end{align}
	Combining the above fact and Eq.~\eqref{eq201}, we have
	\begin{align}
		&|\langle \psi_{s_1}^* |^{\otimes n} |\psi_{s_1}\rangle^{\otimes n}| = | \langle \psi_{s_1}^* | \psi_{s_1} \rangle |^n = {s_1}^n\notag \\
		&\le|\langle \psi_{s_2}^* |^{\otimes m} |\psi_{s_2}\rangle^{\otimes m}| = | \langle \psi_{s_2}^* | \psi_{s_2} \rangle |^m = {s_2}^m.
	\end{align}
	In addition, this condition is necessary for the entire $\mathcal{FSC}$ and naturally applies to its subclass $\mathcal{FSC}_{\mathrm{CPTP}}$.
	
	To prove the sufficiency, suppose that $s_1^n \le s_2^m$. From Eq.~\eqref{eq203}, there exists a CPTP map $\mathcal{X}_{n,m}$ with a real Kraus representation, such that
	\begin{align}
		\mathcal{X}_{n,m} \left( \left[ |\psi_{s_1}\rangle\langle \psi_{s_1}| \right]^{\otimes n} \right) 
		=\left[ |\psi_{s_2}\rangle\langle \psi_{s_2}| \right]^{\otimes m}.
	\end{align}
	We construct the following superchannel $\Theta_{n,m} : \{\mathcal{N} : A \rightarrow B\} \longrightarrow \{\Theta_{n,m}(\mathcal{N}) : C \rightarrow D\}$. For any quantum channel $\mathcal{N}$, it holds that
	\begin{align}
		\Theta_{n,m}(\mathcal{N}) = \mathcal{X}_{n,m} \circ \mathcal{N} \circ \mathcal{A}_{n,m},
	\end{align}
	where $\mathcal{A}_{n,m}$ is defined by $\mathcal{A}_{n,m}(X) = \mathrm{tr}(X) \frac{\mathbf{I}_A}{d_A}$ for any $X\in\mathcal{L}(\mathcal{H}_C)$.
	It is realized via the following connection:
	\begin{align}
		C \xrightarrow{\mathcal{A}_{n,m}} A \xrightarrow{\mathcal{N}} B \xrightarrow{\mathcal{X}_{n,m}} D.
	\end{align} In addition, it is easy to see that $\mathcal{A}_{n,m}$ is a free channel which follows that $\Theta_{n,m}$ is a free superchannel. Therefore, we have 
	\begin{align}
		\Theta_{n,m}(\mathcal{P}_{s_1}^{\otimes n})(Z) &= \mathrm{tr}(Z) \, \mathcal{X}_{n,m} \left( \left[ |\psi_{s_1}\rangle\langle \psi_{s_1}| \right]^{\otimes n} \right) \notag\\
		&= \mathrm{tr}(Z) \left[ |\psi_{s_2}\rangle\langle \psi_{s_2}| \right]^{\otimes m} \notag\\
		&= \mathcal{P}_{s_2}^{\otimes m}(Z).
	\end{align}
	Since $Z$ is arbitrary, we have $\Theta_{n,m}(\mathcal{P}_{s_1}^{\otimes n})=\mathcal{P}_{s_2}^{\otimes m}$. Therefore, for $\mathcal{FSC}$, we complete the proof of sufficiency. 
	
	Next, we only need to prove that $\Theta_{n,m}\in\mathcal{FSC}_{\mathrm{CPTP}}$. In fact, by the definition of the Choi-Jamio\l kowski isomorphism, we have
	\begin{align}\label{eq209}
		J(\Theta_{n,m}(\mathcal{N})) = \frac{\mathbf{I}_{C}}{d_C} \otimes \mathcal{X}_{n,m} \left( \mathcal{N} \left( \frac{\mathbf{I}_A}{d_A} \right) \right).
	\end{align}
	Meanwhile, note that
	\begin{align}\label{eq210}
		\mathrm{tr}_{A} J(\mathcal{N}) = \mathcal{N}\left( \frac{\mathbf{I}_A}{d_A} \right).
	\end{align}
	Substituting Eq.~\eqref{eq210} into Eq.~\eqref{eq209}, we obtain
	\begin{align}
		J(\Theta_{n,m}(\mathcal{N})) = \frac{\mathbf{I}_{C}}{d_C} \otimes \mathcal{X}_{n,m}(\mathrm{tr}_{A} J(\mathcal{N})).
	\end{align}
	
	Therefore, the Choi map induced by $\Theta_{n,m}$ is
	\begin{align}\label{eq212}
		\mathcal{E}_{\Theta_{n,m}}(Y) = \frac{\mathbf{I}_{C}}{d_C} \otimes \mathcal{X}_{n,m}(\mathrm{tr}_{A} Y).
	\end{align}
	It is obvious that Eq.~\eqref{eq212} is a free CPTP map. Hence $\Theta_{n,m}\in\mathcal{FSC}_{\mathrm{CPTP}}$.
	
	Since $0<s_1,s_2<1$, the condition in
	Eq.~\eqref{eq193} is equivalent to
	\begin{align}
		\frac{m}{n}\le\frac{\log s_1}{\log s_2}.
	\end{align}
	Taking the supremum over all feasible rate $m/n$, we obtain Eq.~\eqref{eq189}.
\end{proof} 
\end{appendices}

\begin{appendices}
	\renewcommand{\theequation}{\Alph{section}\arabic{equation}}
	\setcounter{equation}{0}
	\begin{center}
		\textbf{Appendix E: Proof of Section~\ref{secVI}} \label{app:E}
	\end{center}
	
	\renewcommand{\theequation}{E\arabic{equation}}
	\setcounter{equation}{0}
	
\begin{center}
	\textbf{E1. Proof of Theorem~\ref{thm4}}\label{app:E1}
\end{center}

\begin{proof}
	We first prove $d(\mathcal{N})\le C_{\mathrm{tr}}^J(\mathcal{N})$, and then show that $C_{\mathrm{tr}}^J(\mathcal{N})$ can be achieved. For convenience, set $C_{\mathrm{tr}}^J(\mathcal{N})=c$.
	
	First of all, note that
	\begin{align}\label{eq230}
		J(\mathcal{G}_t) = \frac{1 + t}{2} |s_+\rangle\langle s_+| + \frac{1 - t}{2} |s_-\rangle\langle s_-|,
	\end{align}
	where $|s_\pm\rangle = \frac{|00\rangle \pm \mathrm{i}|11\rangle}{\sqrt{2}}$. Direct calculations show that
	\begin{align}
		\frac{1}{2} \| J(\mathcal{G}_t) - J(\mathcal{G}_t)^T \|_1 = t.
	\end{align}
	Assume that there exists $\Theta \in \mathcal{FSC}_{\mathrm{CPTP}}$ such that $\Theta(\mathcal{N}) = \mathcal{G}_t$. Therefore,
	\begin{align}
		2t &= \| J(\mathcal{G}_t) - J(\mathcal{G}_t)^T \|_1 \notag\\
		&= \| \mathcal{E}_\Theta(J(\mathcal{N})) - \mathcal{E}_\Theta(J(\mathcal{N}))^T \|_1 \notag\\
		&= \| \mathcal{E}_\Theta(J(\mathcal{N}) - J(\mathcal{N})^T) \|_1 \notag\\
		&\le \| J(\mathcal{N}) - J(\mathcal{N})^T \|_1 \notag \\
		&=2c,
	\end{align}
	where the second equality follows from Eq.~\eqref{eq34}, and the inequality holds because trace distance satisfies data-processing inequality. We thus have $t \le c$, which means $d(\mathcal{N})\le c$.
	
	Now we construct a $\Theta_c\in\mathcal{FSC}_{\mathrm{CPTP}}$ such that $\Theta_c(\mathcal{N})=\mathcal{G}_{c}$. 
	
	\textbf{Step 1}. Construct a real CPTP map $\mathcal{E}$ such that
	\begin{align}\label{eq233}
		\mathcal{E}\left[J(\mathcal{N})\right]=\omega_c
	\end{align}
	for any $J(\mathcal{N})$, where $\omega_c=\frac{1+c}{2}I_+ + \frac{1-c}{2}I_-$, and $I_\pm = |\pm \mathrm{i}\rangle\langle \pm \mathrm{i}|$ are the maximally imaginary states.
	
	Firstly, let 
	\begin{align}
		\Delta=J(\mathcal{N}) - J(\mathcal{N})^T,\quad H = \mathrm{sgn}(\Delta) \quad\text{and}\quad E_\pm = \frac{\mathbf{I}\pm H}{2}.
	\end{align}
	It is easy to verify that $H^T = -H$, $ E_+^T = E_-$ and $E_\pm \ge0$. For any $X\in\mathcal{L}(\mathcal{H}_A\otimes \mathcal{H}_{B})$, define
	\begin{align}\label{eq235}
		\mathcal{E}(X) = \mathrm{tr}(E_+ X)I_+ + \mathrm{tr}(E_- X)I_-.
	\end{align}
	It is clear that $\mathcal{E}$ is a CPTP map. Meanwhile, we find that
	\begin{align}
		J(\mathcal{E}) = \frac{1}{d_Ad_B} \left( E_+^T \otimes I_+ + E_-^T \otimes I_- \right),
	\end{align}
	which yields that $J(\mathcal{E})=J(\mathcal{E})^T$. Hence, $J(\mathcal{E})$ is a free state, which means that $\mathcal{E}$ is a free channel. 
	
	Note that
	\begin{align}\label{eq237}
		\mathrm{tr}[E_+J(\mathcal{N})] &=\frac{1}{2}\mathrm{tr}\left[J(\mathcal{N})\right]+\frac{1}{2}\mathrm{tr}\left[HJ(\mathcal{N})\right].
	\end{align}
	
	It follows from $H^T = -H$ that
	\begin{align}
		\mathrm{tr}\left[HJ(\mathcal{N})^T\right] =\mathrm{tr}\left[H^TJ(\mathcal{N})\right] =-\mathrm{tr}\left[HJ(\mathcal{N})\right],
	\end{align}
	which yields that $	2\,\mathrm{tr}\left[HJ(\mathcal{N})\right] = \mathrm{tr}(H\Delta)$.
	Eq.~\eqref{eq237} thus becomes 
	\begin{align}
		\mathrm{tr}[E_+J(\mathcal{N})]=\frac{1}{2} + \frac{1}{4}\mathrm{tr}(H\Delta).
	\end{align}
	Next, we prove that $\mathrm{tr}(H\Delta) = \|\Delta\|_1$. Since \(\Delta\) is a Hermitian matrix, let its spectral decomposition be
	\begin{align}
		\Delta = \sum_j \lambda_j |v_j\rangle\langle v_j|.
	\end{align}
	According to the definition of the matrix sign function,
	\begin{align}
		H = \sum_j \mathrm{sgn}(\lambda_j) |v_j\rangle\langle v_j|.
	\end{align}
	Therefore, one has
	\begin{align}
		H\Delta = \sum_j \mathrm{sgn}(\lambda_j)\lambda_j |v_j\rangle\langle v_j| 
		= \sum_j |\lambda_j| |v_j\rangle\langle v_j| 
		= |\Delta|,
	\end{align}
	which implies that
	\begin{align}
		\mathrm{tr}(H\Delta) = \sum_j |\lambda_j| = \|\Delta\|_1.
	\end{align}
	We thus have
	\begin{align}
		\mathrm{tr}\left[E_+J(\mathcal{N})\right]&=\frac{1}{2} + \frac{1}{4}\|\Delta\|_1= \frac{1 + c}{2}.
	\end{align}
	Similarly, one has $	\mathrm{tr}\left[E_-J(\mathcal{N})\right]=\frac{1-c}{2}$. Substituting them into Eq.~\eqref{eq235}, we obtain Eq.~\eqref{eq233}.
	
	\textbf{Step 2}. With the assistance of quantum state $\omega_c$, construct a free channel $\mathcal{R}$ that realizes the optimal unit resource channel, i.e., $\mathcal{R}(\omega_c \otimes \rho)=\mathcal{G}_c(\rho)$ for any $\rho\in\mathcal{D}(\mathcal{H}_2)$.
	
	Let
	\begin{equation}\label{eq246}
		L = \begin{pmatrix} 0 & 1 \\ -1 & 0 \end{pmatrix}.
	\end{equation}
	Direct calculations show that
	\begin{equation}
		L^T L =\mathbf{I}_2, \quad\text{and}\quad L|\pm \mathrm{i}\rangle = \pm\mathrm{i}|\pm \mathrm{i}\rangle.
	\end{equation}
	We further define a real matrix
	\begin{equation}
		V =\mathbf{I}_2\otimes |0\rangle\langle 0| + L\otimes |1\rangle\langle 1|.
	\end{equation}
	Then we have
	\begin{equation}
		V^T V =\mathbf{I}_2 \otimes |0\rangle\langle 0| + L^T L \otimes |1\rangle\langle 1| = \mathbf{I}_{4},
	\end{equation}
	which means that $V$ is a real orthogonal matrix. For any state $	|\psi\rangle = \alpha|0\rangle + \beta|1\rangle$ with $|\alpha|^2+|\beta|^2=1$, one gets
	\begin{align}
		V(|+\mathrm{i}\rangle \otimes |\psi\rangle) &= (\mathbf{I}_2 \otimes |0\rangle\langle 0|)(|+\mathrm{i}\rangle \otimes |\psi\rangle)  + (L \otimes |1\rangle\langle 1|)(|+\mathrm{i}\rangle \otimes |\psi\rangle) \notag\\
		&= |+\mathrm{i}\rangle \otimes |0\rangle\langle 0|\psi\rangle + L|+\mathrm{i}\rangle \otimes |1\rangle\langle 1|\psi\rangle \notag\\
		&= \alpha|+\mathrm{i}\rangle \otimes |0\rangle + \mathrm{i}\beta|+\mathrm{i}\rangle \otimes |1\rangle \notag\\
		&= |+\mathrm{i}\rangle \otimes (\alpha|0\rangle + \mathrm{i}\beta|1\rangle).
	\end{align}
	Meanwhile, note that
	\begin{equation}
		S|\psi\rangle = \alpha|0\rangle + \mathrm{i}\beta|1\rangle,
	\end{equation}
	which yields that
	\begin{equation}
		V(|+\mathrm{i}\rangle \otimes |\psi\rangle) = |+\mathrm{i}\rangle \otimes S|\psi\rangle.
	\end{equation}
	Similarly, we obtain that
	\begin{equation}
		V(|-\mathrm{i}\rangle \otimes |\psi\rangle) = |-\mathrm{i}\rangle \otimes \overline{S}|\psi\rangle.
	\end{equation}
	
	Therefore, for any $\rho\in\mathcal{D}(\mathcal{H}_2)$, we know that
	\begin{align}
		V(\omega_c \otimes \rho)V^\dagger &= \frac{1 + c}{2} I_+ \otimes S\rho S^\dagger + \frac{1 - c}{2} I_- \otimes \overline{S} \rho \overline{S}^\dagger.
	\end{align}
	
	Based on the above arguments, we define the following CPTP map 
	\begin{equation}\label{eq255}
		\mathcal{R}(Y) = \mathrm{tr}_1(V Y V^{\dagger}),
	\end{equation}
	where the partial trace is over the first system. Since $\mathrm{tr}\,I_+ = \mathrm{tr}\,I_- = 1$,
	we obtain that
	\begin{align}
		\mathcal{R}(\omega_c \otimes \rho) &= \mathrm{tr}_1\left[ V(\omega_c \otimes \rho)V^\dagger \right] \notag\\
		&= \frac{1 + c}{2} S \rho S^\dagger + \frac{1 - c}{2} \overline{S}\rho \overline{S}^\dagger \notag\\
		&=\mathcal{G}_c(\rho).
	\end{align}
	In addition, since $V$ is a real matrix and partial trace is a free channel, we know that $\mathcal{R}$ is a free channel.
	
	\textbf{Step 3}. Based on Step 1 and Step 2, for an arbitrary channel $\mathcal{N}\in\mathcal{C}(A,B)$, define
	\begin{equation}\label{eq257}
		\Theta_c(\mathcal{N})(\rho) = \mathcal{R}\left( \mathcal{E}(J(\mathcal{N})) \otimes \rho \right)=\mathcal{G}_c(\rho)
	\end{equation}
	for any $\rho\in\mathcal{D}(\mathcal{H}_2)$. 
	
	The superchannel $\Theta_c$ admits the following realization. Firstly apply channel 
	\begin{equation}
		\mathcal{P}_0(\rho) = \phi_{AA'} \otimes \rho,
	\end{equation}
	and secondly apply channel 
	\begin{align}
		(\mathbb{I}_{A} \otimes \mathcal{N}\otimes \mathbb{I}_2)(\phi_{AA'} \otimes\rho) 
		=(\mathbb{I}_{A} \otimes \mathcal{N})(\phi_{AA'}) \otimes\rho =J(\mathcal{N}) \otimes \rho.
	\end{align}
	Next apply channel 
	\begin{align}
		(\mathcal{E} \otimes \mathbb{I}_2)(J(\mathcal{N}) \otimes \rho) 
		= \mathcal{E}(J(\mathcal{N})) \otimes \rho,
	\end{align}
	where $\mathcal{E}$ is defined by Eq.~\eqref{eq235}. Finally, apply channel
	\begin{align}
		\mathcal{R}\left[\mathcal{E}(J(\mathcal{N})) \otimes \rho\right]=\mathcal{G}_c(\rho).
	\end{align}
	where $\mathcal{R}$ is defined by Eq.~\eqref{eq255}.
	Therefore, combining the above steps,
	\begin{equation}
		\Theta_c(\mathcal{N}) = \mathcal{Q}_0 \circ (\mathbb{I}_{A} \otimes \mathcal{N} \otimes \mathbb{I}_2) \circ \mathcal{P}_0,
	\end{equation}
	where $\mathcal{Q}_0=\mathcal{R}\circ(\mathcal{E}\otimes \mathbb{I}_2)$. It is evident that $\mathcal{P}_0$ and $\mathcal{Q}_0$ are free channels.
	
	Next, we verify that $\Theta_c\in\mathcal{FSC}_{\mathrm{CPTP}}$. By the Choi definition, we know that
	\begin{equation}\label{eq263}
		J(\Theta_c(\mathcal{N})) = \frac{1}{2} \sum_{i,j=0}^{1} |i\rangle\langle j| \otimes \Theta_c(\mathcal{N})(|i\rangle\langle j|).
	\end{equation}
	
	Substituting Eq.~\eqref{eq257} into Eq.~\eqref{eq263}, we obtain that
	\begin{equation}
		J(\Theta_c(\mathcal{N})) = \frac{1}{2} \sum_{i,j=0}^{1} |i\rangle\langle j| \otimes \mathcal{R}(\mathcal{E}(J(\mathcal{N})) \otimes |i\rangle\langle j|).
	\end{equation}
	
	Therefore, for an arbitrary operator $Z \in \mathcal{L}(\mathcal{H}_A\otimes \mathcal{H}_B)$, the map induced by superchannel $\Theta_c$ is
	\begin{equation}\label{eq265}
		\mathcal{E}_{\Theta_c}(Z) = \frac{1}{2} \sum_{i,j=0}^{1} |i\rangle\langle j| \otimes \mathcal{R}(\mathcal{E}(Z) \otimes |i\rangle\langle j|).
	\end{equation}
	
	Since in Section~~\ref{secII} we pointed out that the induced maps $\mathcal{E}_{\Theta_c}$ of superchannel $\Theta_c$ is CP, we only need to verify that it is TP. Note that, from Eq.~\eqref{eq265}, 
	\begin{align}
		\mathrm{tr}\mathcal{E}_{\Theta_c}(Z) &= \frac{1}{2} \sum_{i=0}^{1} \mathrm{tr}\mathcal{R}\left(\mathcal{E}(Z) \otimes |i\rangle\langle i| \right) \notag\\
		&= \frac{1}{2} \sum_{i=0}^{1} \mathrm{tr}\left( \mathcal{E}(Z) \otimes |i\rangle\langle i| \right) \notag\\
		&= \mathrm{tr}\mathcal{E}(Z) \notag\\
		&= \mathrm{tr}Z,
	\end{align}
	where we use the trace-preserving property of $\mathcal{R}$ and $\mathcal{E}$ in the second and fourth equality, respectively. Hence $\Theta_c\in\mathcal{FSC}_{\mathrm{CPTP}}$. In summary, we have constructed a superchannel $\Theta_c\in\mathcal{FSC}_{\mathrm{CPTP}}$ such that $\Theta_c(\mathcal{N})=\mathcal{G}_{c}$, which implies that $d(\mathcal{N})=c$.
\end{proof}

\begin{center}
	\textbf{E2. Proof of Theorem~\ref{thm5}}\label{app:E2}
\end{center}

	\begin{proof}
	We first prove the necessity. Suppose that there exists $\Theta \in \mathcal{FSC}_{\mathrm{CPTP}}$ such that
	$\Theta(\mathcal{G}_t) = \mathcal{N}$.
	
	Let $\mathcal{M} = \Theta(\mathcal{S})$. Since $\Theta \in \mathcal{FSC}_{\mathrm{CPTP}}$, by Proposition~\ref{prop1}, one has
	\begin{equation}\label{eq272}
		\Theta(\overline{\mathcal{S}}) =\overline{\Theta(\mathcal{S})} =\overline{\mathcal{M}}.
	\end{equation}
	Therefore, combining Eqs.~\eqref{eq222} and~\eqref{eq272} yields 
	\begin{equation}
		\mathcal{N} = \frac{1 + t}{2} \mathcal{M} + \frac{1 - t}{2} \overline{\mathcal{M}}.
	\end{equation}
	
	Let $\tau = J(\mathcal{M})$. Since $J(\overline{\mathcal{M}})=\tau^T$, we obtain that
	\begin{equation}
		J(\mathcal{N})= \frac{\tau + \tau^T}{2} + t \frac{\tau - \tau^T}{2},
	\end{equation}
	which yields that
	\begin{equation}
		W=\frac{\tau + \tau^T}{2}\quad\text{and}\quad \frac{V}{t}=\frac{\tau - \tau^T}{2} .
	\end{equation}
	
	Thus, one has
	\begin{equation}
		\tau = W+ \frac{V}{t}.
	\end{equation}
	
	Since $\tau$ is a Choi state, it holds that
	\begin{equation}
		W+ \frac{V}{t} \succeq \textbf{0}.
	\end{equation}
	
	On the other hand, for sufficiency, fix an arbitrary $t\in(0,1]$ and suppose that the right-hand side of Eq.~\eqref{eq268} holds.
	It can be verified that 
	\begin{align}
		\mathrm{tr}_{B} W = \frac{\mathbb{I}_{A}}{d_{A}}, \quad \text{and}\quad \mathrm{tr}_{B} V =\mathbf{0},
	\end{align}
	which implies that $\mathrm{tr}_{B}\left(W+ \frac{V}{t}\right)=\frac{\mathbb{I}_{A}}{d_{A}}$. Therefore, there exists a channel $\mathcal{V}$ such that $W+ \frac{V}{t}=J(\mathcal{V})$. We can construct $\Theta \in \mathcal{FSC}_{\mathrm{CPTP}}$ via the following process such that $\Theta(\mathcal{G}_t) = \mathcal{N}$. Let system $C$ be associated with $\mathcal{H}_2$. 
	
	\textbf{Step 1}. For any $\rho_A\in\mathcal{D}(\mathcal{H}_A)$, we define the following channel
	\begin{align}
		\mathcal{P}(\rho_A) =\rho_A \otimes \phi_{CC'}.
	\end{align}
	For any qubit channel $\mathcal{K} : \mathcal{H}_{C} \to \mathcal{H}_{C'}$, one has
	\begin{equation}\label{eq280}
		(\mathbb{I}_{AC} \otimes \mathcal{K})(\mathcal{P}(\rho_A)) = \rho_A \otimes J(\mathcal{K})_{CC'}.
	\end{equation}
For any $Y\in\mathcal{L}(\mathcal{H}_{C}\otimes \mathcal{H}_{C'})$, define the channel
	\begin{equation}
		\mathcal{D}(Y) = \mathrm{tr}_{C'} \left[ U_{\mathrm{CNOT}} Y U_{\mathrm{CNOT}}^\dagger \right],
	\end{equation}
	where $U_{\mathrm{CNOT}}$ is the controlled-NOT gate, $C$ is the control qubit and $C'$ is the target qubit. Since
	\begin{align}
		U_{\mathrm{CNOT}}|s_\pm\rangle = |\pm \mathrm{i}\rangle_C \otimes |0\rangle_{C'}, 
	\end{align}
	from Eq.~\eqref{eq230}, it follows that
	\begin{equation}\label{eq283}
		\mathcal{D}(J(\mathcal{G}_t)) = \omega_t = \frac{1 + t}{2} I_+ + \frac{1 - t}{2} I_-.
	\end{equation}
	
	\textbf{Step 2}. We show that for channel $\mathcal{V} : \mathcal{H}_{A} \to \mathcal{H}_{B}$, there exists a free channel $\mathcal{R}_{\mathcal{V}} : \mathcal{H}_C \otimes \mathcal{H}_A \to \mathcal{H}_B$ such that, for every input state $\rho_A$,
	\begin{equation}\label{eq284}
		\mathcal{R}_{\mathcal{V}}(\omega_t \otimes \rho_{A}) = \frac{1 + t}{2} \mathcal{V}(\rho_{A}) + \frac{1 - t}{2} \overline{\mathcal{V}}(\rho_{A}).
	\end{equation}
	This construction is analogous to Step 2 in the proof of Theorem~\ref{thm4}. Suppose that the quantum channel $	\mathcal{V}$ has Kraus operators $\{K_a\}$, i.e.,
	\begin{equation}
		\mathcal{V}(\cdot) = \sum_a K_a (\cdot) K_a^\dagger \quad\text{and}\quad \sum_a K_a^\dagger K_a = \mathbf{I}_{A}.
	\end{equation}
	Write each Kraus operator as
	\begin{equation}
		K_a = C_a + \mathrm{i}D_a,
	\end{equation}
	where $C_a$ and $D_a$ are real matrices.
	We further define the real operator from $\mathcal{H}_C \otimes \mathcal{H}_A$ to $\mathcal{H}_C \otimes \mathcal{H}_B$ as
	\begin{equation}\label{eq287}
		R_a = \mathbf{I}_2 \otimes C_a + L \otimes D_a,
	\end{equation}
	where $L$ is defined by Eq.~\eqref{eq246}. It is easy to verify that $\sum_a R_a^\dagger R_a=\mathbf{I}_{CA}$. For an arbitrary state $|\psi\rangle \langle\psi|\in\mathcal{D}(\mathcal{H}_{A})$, using Eq.~\eqref{eq287}, we have
	\begin{align}
		R_a(|+\mathrm{i}\rangle \otimes |\psi\rangle) = |+\mathrm{i}\rangle \otimes C_a|\psi\rangle + \mathrm{i}|+\mathrm{i}\rangle \otimes D_a|\psi\rangle = |+\mathrm{i}\rangle \otimes K_a|\psi\rangle,
	\end{align}
	and
	\begin{align}
		R_a(|-\mathrm{i}\rangle \otimes |\psi\rangle) = |-\mathrm{i}\rangle \otimes C_a|\psi\rangle - \mathrm{i}|-\mathrm{i}\rangle \otimes D_a|\psi\rangle = |-\mathrm{i}\rangle \otimes \overline{K_a}|\psi\rangle.
	\end{align} 
	Then, discarding the output reference system $C$, for any $\rho_{CA}\in\mathcal{D}(\mathcal{H}_{CA})$,
	\begin{align}
		\mathcal{R}_{\mathcal{V}}(\rho_{CA}) = \mathrm{tr}_C \left( \sum_a R_a \rho_{CA} R_a^\dagger \right)
	\end{align}
	realizes Eq.~\eqref{eq284}.
	
	\textbf{Step 3}. Let $U_{\mathrm{SW}}$ denote the real swap channel that rearranges the system order from $AC$ to $CA$, and let
	\begin{equation}\label{eq291}
		\mathcal{Q}= \mathcal{R}_{\mathcal{V}} \circ \mathcal{U}_{\mathrm{SW}} \circ (\mathbb{I}_A \otimes \mathcal{D}).
	\end{equation}
	Since $\mathcal{D}$, $U_{\mathrm{SW}}$, and $\mathcal{R}_{\mathcal{V}}$ are all free channels, $\mathcal{Q}$ is also a free channel.
	
	Now define the superchannel
	\begin{equation}\label{eq292}
		\Theta(\mathcal{K})= \mathcal{Q} \circ (\mathbb{I}_{AC} \otimes \mathcal{K}) \circ \mathcal{P}.
	\end{equation}
	Eq.~\eqref{eq292} has a physical realization with real pre-processing and real post-processing, which means that $\Theta \in \mathcal{FSC}$. Combining Eqs.~\eqref{eq280},~\eqref{eq291} and~\eqref{eq292}, for an arbitrary input state $\rho_A$, we have
	\begin{equation}\label{eq293}
		\Theta(\mathcal{K})(\rho_A) = \mathcal{R}_{\mathcal{V}}\big( \mathcal{D}(J(\mathcal{K})) \otimes \rho_A \big).
	\end{equation}
	Next, we further verify that its induced Choi map $\mathcal{E}_{\Theta}$ is CPTP. By calculation, we obtain that
	\begin{align}\label{eq294}
		J(\Theta(\mathcal{K})) = (\mathbb{I}_{A'} \otimes \mathcal{R}_{\mathcal{V}}) \left[ \Pi \left( \phi_{A'A} \otimes \mathcal{D}(J(\mathcal{K})) \right) \Pi^\dagger \right],
	\end{align}
	where $\Pi$ is the real permutation unitary that rearranges the system order from $A'AC$ to $A'CA$. From Eq.~\eqref{eq294}, the induced Choi map $\mathcal{E}_{\Theta}$ can be written as, for any $Y\in\mathcal{L}(\mathcal{H}_C\otimes \mathcal{H}_{C'})$,
	\begin{equation}
		\mathcal{E}_{\Theta}(Y) = (\mathbb{I}_{A'} \otimes \mathcal{R}_{\mathcal{V}}) \left[ \Pi (\phi_{A'A} \otimes \mathcal{D}(Y)) \Pi^\dagger \right].
	\end{equation}
	
	It remains to verify that $\Theta\in\mathcal{FSC}_{\mathrm{CPTP}}$. As is pointed out in Section~\ref{secII}, the induced map $\mathcal{E}_{\Theta}$ of a superchannel $\Theta$ is CP, so we only need to verify that it is TP.
	Because $\mathcal{D}$ and $\mathcal{R}_{\mathcal{V}}$ are TP and $\operatorname{tr}\phi_{A'A} = 1$, for any operator $Y$, one has
	\begin{align}
		\operatorname{tr}\mathcal{E}_\Theta(Y)
		&= \operatorname{tr} \left[ \Pi \bigl( \phi_{A'A} \otimes \mathcal{D}(Y) \bigr) \Pi^\dagger \right] \nonumber \\
		&= \operatorname{tr} \left[ \phi_{A'A} \otimes \mathcal{D}(Y) \right] \nonumber \\
		&= \operatorname{tr}\phi_{A'A} \, \operatorname{tr}\mathcal{D}(Y) \nonumber \\
		&= \operatorname{tr}Y. 
	\end{align}
	Thus, $\mathcal{E}_\Theta$ is CPTP, which implies that $\Theta\in\mathcal{FSC}_{\mathrm{CPTP}}$.
	
	Now, taking $\mathcal{K} = \mathcal{G}_t$ in Eq.~\eqref{eq293}. From Eqs.~\eqref{eq283} and~\eqref{eq284}, for an arbitrary input state $\rho_A$,
	\begin{align}
		\Theta(\mathcal{G}_t)(\rho_A) = \mathcal{R}_{\mathcal{V}}(\mathcal{D}(J(\mathcal{G}_t)) \otimes \rho_A) = \mathcal{R}_{\mathcal{V}}(\omega_t \otimes \rho_A) = \frac{1 + t}{2} \mathcal{V}(\rho_A) + \frac{1 - t}{2} \overline{\mathcal{V}}(\rho_A),
	\end{align}
	which means that
	\begin{equation}
		\Theta(\mathcal{G}_t) = \frac{1 + t}{2} \mathcal{V} + \frac{1 - t}{2} \overline{\mathcal{V}}.
	\end{equation} 
	
	From $W^T = W$ and $V^T = -V$, we have
	\begin{equation}
		J(\overline{\mathcal{V}}) = J(\mathcal{V})^T = W - \frac{V}{t},
	\end{equation}
	which yields that
	\begin{align*}
		J(\Theta(\mathcal{G}_t)) &= \frac{1 + t}{2} J(\mathcal{V}) + \frac{1 - t}{2} J(\overline{\mathcal{V}}) \\
		&= \frac{1 + t}{2} \left( W + \frac{V}{t} \right) + \frac{1 - t}{2} \left( W - \frac{V}{t} \right) \\
		&= W + V \\
		&= J(\mathcal{N}).
	\end{align*}
	
	Since the Choi correspondence is a one-to-one map, one gets $	\Theta(\mathcal{G}_t) = \mathcal{N}$.
	This completes the proof of sufficiency for Eq.~\eqref{eq268}.
	
	For $t \in (0,1]$, it holds that
	\begin{align}
		W + \frac{V}{t} \succeq\textbf{0} \quad \Longleftrightarrow \quad tW + V \succeq \textbf{0}.
	\end{align}
	At $t = 0$, the latter condition becomes $V \succeq 0$. Since $V$ is Hermitian and traceless, this holds if and only if $V = \textbf{0}$, equivalently, if and only if $J(\mathcal{N})$ is real, that is, $\mathcal{N}$ is a free channel.
	
	Therefore, when $t = 0$, we need to prove that
	\begin{align}
		\mathcal{G}_0 \xrightarrow{\mathcal{FSC}_{\mathrm{CPTP}}} \mathcal{N} \iff \mathcal{N} \text{ is a free channel.}
	\end{align}
	First, the channel $\mathcal G_0$ is free and cannot be transformed into an imaginary channel by a free superchannel. Hence $\mathcal{N}$ must be a free channel. On the other hand, suppose $\mathcal{N}$ is a free channel. Let
	\begin{align}\label{eq303}
		\Theta^*(\mathcal{K})= \mathcal{Q}^* \circ (\mathbb{I}_{AC} \otimes \mathcal{K}) \circ \mathcal{P},
	\end{align}
	where $\mathcal{Q}^*(X) = \mathcal{N}(\mathrm{tr}_{CC'} X)$ for any $X\in\mathcal{L}(\mathcal{H}_{A}\otimes \mathcal{H}_{C}\otimes \mathcal{H}_{C'})$. That is, we replace $\mathcal{Q}$ in Eq.~\eqref{eq292} with $\mathcal{Q}^*$. Since $\mathcal{N}$ and the partial trace are both free, $\mathcal{Q}^*$ is free, which implies that $\Theta^* \in \mathcal{FSC}$.
	
	Letting $\mathcal{K}=\mathcal{G}_0$, for an arbitrary input state $\rho_A$, one has
	\begin{align}
		\Theta^*(\mathcal{G}_0)(\rho_A) = \mathcal{N}\left( \mathrm{tr}_{CC'}[\rho_A \otimes J(\mathcal{G}_0)] \right)= \mathcal{N}\left( \rho_A \mathrm{tr} J(\mathcal{G}_0) \right) = \mathcal{N}(\rho_A),
	\end{align}
	which implies that $\Theta^*(\mathcal{G}_0) = \mathcal{N}$.

	Next we show that $\Theta^*\in\mathcal{FSC}_{\mathrm{CPTP}}$, and only need to check that $\mathcal{E}_{\Theta^*}$ is TP. For an arbitrary input channel $\mathcal K$, Eq.~\eqref{eq303} gives
	\begin{align}
		J(\Theta^*(\mathcal{K})) = (\mathbb{I}_A \otimes \mathcal{Q}^*)(\phi_{AA'} \otimes J(\mathcal{K})) =(\mathbb{I}_A \otimes \mathcal{N}) \left[ \mathrm{tr}_{CC'} (\phi_{AA'} \otimes J(\mathcal{K})) \right].
	\end{align}
	Therefore, for any $Y\in \mathcal{L}(\mathcal{H}_{C}\otimes \mathcal{H}_{C'})$, one gets
	\begin{align}
		\mathcal{E}_{\Theta^*}(Y)&=(\mathbb{I}_A \otimes \mathcal{N}) \left[ \mathrm{tr}_{CC'} (\phi_{AA'} \otimes Y) \right] \notag\\
		&= (\mathbb{I}_A \otimes \mathcal{N}) \left[ \mathrm{tr}(Y) \phi_{AA'} \right] \notag\\
		&= \mathrm{tr}(Y) (\mathbb{I}_A \otimes \mathcal{N})(\phi_{AA'}) \notag\\
		&= \mathrm{tr}(Y) J(\mathcal{N}).
	\end{align}
	Hence, 
	\begin{align}
		\mathrm{tr}\,\mathcal{E}_{\Theta^*}(Y) = \mathrm{tr}(Y)\mathrm{tr}J(\mathcal{N}) = \mathrm{tr}(Y),
	\end{align}
	which means that $\mathcal{E}_{\Theta^*}$ is TP. We thus have $\Theta^*\in\mathcal{FSC}_{\mathrm{CPTP}}$.
	
	In summary, it follows that
	\begin{equation}
		c(\mathcal{N}) = \inf\{t \in [0,1] : tW + V \succeq \textbf{0}\}.
	\end{equation}
	The feasible set is a closed subset of $[0,1]$ and is nonempty, since
	\begin{equation}
		W + V = J(\mathcal{N}) \succeq \textbf{0},
	\end{equation}
	so $t = 1$ is always feasible. Thus, the feasible set is compact and the infimum is attained, which follows that
	\begin{equation}
		c(\mathcal{N}) = \min\{t \in [0,1] : tW + V \succeq \textbf{0}\}.
	\end{equation}
	This proves Eq.~\eqref{eq269}. Since the objective is linear in $t$, and the constraints consist of a linear matrix inequality and linear scalar inequalities, Eq.~\eqref{eq270} gives the corresponding semidefinite programming.
\end{proof}

\begin{center}
	\textbf{E3. Proof of Theorem \ref{thm6}}\label{app:E3}
\end{center}

\begin{proof}
	We first provide the analytical expressions for $C_S^J(\mathcal{G}_t)$ and $C_G^J(\mathcal{G}_t)$. It follows from Eq.~\eqref{eq230} that in the computational basis $	\{|00\rangle, |01\rangle, |10\rangle, |11\rangle\}$,
	\begin{equation}
		J(\mathcal{G}_t) = \frac{1}{2} \begin{pmatrix} 1 & 0 & 0 & -\mathrm{i}t \\ 0 & 0 & 0 & 0 \\ 0 & 0 & 0 & 0 \\ \mathrm{i}t & 0 & 0 & 1 \end{pmatrix}.
	\end{equation}
	Therefore, $\mathrm{Re}[J(\mathcal{G}_t)]=\mathrm{diag}(1,0,0,1)/2$.
	In addition, it can be verified that Eq.~\eqref{eq230} coincides with the spectral decomposition of $J(\mathcal{G}_t)$. By a direct calculation, we have
	\begin{align}\label{eq319}
		C^J_S(\mathcal{G}_t)=S[\mathrm{Re}(J(\mathcal{G}_t))]-S[J(\mathcal{G}_t)]=1 - h\left(\frac{1+t}{2}\right).
	\end{align}
	Meanwhile, since $\left( |s_+\rangle\langle s_+| \right)^T = |s_-\rangle\langle s_-|$,
	we have
	\begin{equation}
		J(\mathcal{G}_t)^T = \frac{1-t}{2} |s_+\rangle\langle s_+| + \frac{1+t}{2} |s_-\rangle\langle s_-|.
	\end{equation}
	Therefore, one gets
	\begin{equation}
		\sqrt{J(\mathcal{G}_t)} = \sqrt{\frac{1+t}{2}} |s_+\rangle\langle s_+| + \sqrt{\frac{1-t}{2}} |s_-\rangle\langle s_-|,
	\end{equation}
	and
	\begin{equation}
		\sqrt{J(\mathcal{G}_t)} J(\mathcal{G}_t)^T \sqrt{J(\mathcal{G}_t)} = \frac{(1+t)(1-t)}{4} \left( |s_+\rangle\langle s_+| + |s_-\rangle\langle s_-| \right),
	\end{equation}
	which yields that
	\begin{align}
		F[J(\mathcal{G}_t), J(\mathcal{G}_t)^T] &= \left( \mathrm{tr} \sqrt{ \sqrt{J(\mathcal{G}_t)} J(\mathcal{G}_t)^T \sqrt{J(\mathcal{G}_t)} } \right)^2 \notag\\
		&= \left(2\sqrt{\frac{(1+t)(1-t)}{4}}\right)^2 \notag\\
		&= 1 - t^2.
	\end{align}
	We thus obtain that
	\begin{align}\label{eq324}
		\frac{1-\sqrt{F[J(\mathcal{G}_t),J(\mathcal{G}_t)^T]}}{2}=\frac{1-\sqrt{1 - t^2}}{2}.
	\end{align}
	Noting that $\mathrm{Re}(J(\mathcal{G}_t))=\frac{1}{2}(|s_+\rangle\langle s_+| + |s_-\rangle\langle s_-|)$, one has
	\begin{align}
		&\sqrt{J(\mathcal{G}_t)}\mathrm{Re}[J(\mathcal{G}_t)]\sqrt{J(\mathcal{G}_t)}\notag\\
		&\quad=\frac{1}{2}
		\left(
		\sqrt{\frac{1+t}{2}}\,|s_+\rangle\langle s_+|
		+\sqrt{\frac{1-t}{2}}\,|s_-\rangle\langle s_-|
		\right)
		\left(
		|s_+\rangle\langle s_+|+|s_-\rangle\langle s_-| 
		\right)\notag \\
		&\qquad\quad\left(
		\sqrt{\frac{1+t}{2}}\,|s_+\rangle\langle s_+|
		+\sqrt{\frac{1-t}{2}}\,|s_-\rangle\langle s_-|
		\right)\notag\\
		&\quad=\frac{1}{2}
		\left(
		\sqrt{\frac{1+t}{2}}\,|s_+\rangle\langle s_+|
		+\sqrt{\frac{1-t}{2}}\,|s_-\rangle\langle s_-|
		\right)^2\notag\\
		&\quad=\frac{1}{2}
		\left(
		\frac{1+t}{2}|s_+\rangle\langle s_+|
		+\frac{1-t}{2}|s_-\rangle\langle s_-|
		\right)\notag\\
		&\quad=\frac{J(\mathcal{G}_t)}{2}.
	\end{align}
	Therefore, we get
	\begin{align}\label{eq326}
		F[J(\mathcal{G}_t), \mathrm{Re}(J(\mathcal{G}_t))] &= \left( \mathrm{tr} \sqrt{ \sqrt{J(\mathcal{G}_t)} (\mathrm{Re}(J(\mathcal{G}_t))) \sqrt{J(\mathcal{G}_t)} } \right)^2 \notag\\
		&= \left( \mathrm{tr} \sqrt{ \frac{J(\mathcal{G}_t)}{2} } \right)^2 \notag\\
		&= \left( \sqrt{ \frac{1+t}{4} } + \sqrt{ \frac{1-t}{4} } \right)^2 \notag\\
		&= \frac{1 + \sqrt{1-t^2}}{2}.
	\end{align}
	
	Combining Eqs.~\eqref{eq324} and~\eqref{eq326}, and with the help of Proposition~\ref{prop10}, we obtain that 
	\begin{equation}\label{eq327}
		C_G^J(\mathcal{G}_t) = \frac{1 - \sqrt{1-t^2}}{2} = g(t).
	\end{equation}
	
	Let~$t$~be an achievable one-shot exact dynamic imaginarity distillation rate, i.e., there exists $\Theta_t \in \mathcal{FSC}_{\mathrm{CPTP}}$ such that
	\begin{equation}\label{eq328}
		\Theta_t(\mathcal{N}) = \mathcal{G}_t.
	\end{equation}
	From Eqs.~\eqref{eq319} and~\eqref{eq327}, we have
	\begin{equation}\label{eq329}
		f(t) = C_S^J(\mathcal{G}_t) = C_S^J(\Theta_t(\mathcal{N})) \le C_S^J(\mathcal{N}),
	\end{equation}
	and
	\begin{equation}\label{eq330}
		g(t) = C_G^J(\mathcal{G}_t) = C_G^J(\Theta_t(\mathcal{N})) \le C_G^J(\mathcal{N}),
	\end{equation}
	where the inequalities in Eqs.~\eqref{eq329} and~\eqref{eq330} hold because $C_S^J(\mathcal{N})$ and $C_G^J(\mathcal{N})$ are both monotonic under $\mathcal{FSC}_{\mathrm{CPTP}}$. 
	
	According to the definition of the supremum, one can choose a sequence of achievable parameters $t_n$ such that
	$	t_n \to d(\mathcal{N})$ when $n\to\infty$.
	Since $f(t)$ and $g(t)$ are continuous, taking the limit of Eqs.~\eqref{eq329} and~\eqref{eq330}, we obtain that
	\begin{equation}\label{eq332}
		f(d(\mathcal{N})) \le C_S^J(\mathcal{N}),
	\end{equation}
	and
	\begin{equation}\label{eq333}
		g(d(\mathcal{N})) \le C_G^J(\mathcal{N}).
	\end{equation}
	
	On the other hand, Let $s$ be an achievable one-shot exact dynamic imaginarity dilution cost, i.e., there exists $\widetilde{\Theta}_s \in \mathcal{FSC}_{\mathrm{CPTP}}$ such that $	\widetilde{\Theta}_s(\mathcal{G}_s) = \mathcal{N}$.
	From Eqs.~\eqref{eq319} and~\eqref{eq327}, and with the help of the monotonicity under $\mathcal{FSC}_{\mathrm{CPTP}}$, one has
	\begin{equation}\label{eq335}
		C_S^J(\mathcal{N}) = C_S^J(\widetilde{\Theta}_s(\mathcal{G}_s)) \le C_S^J(\mathcal{G}_s) = f(s),
	\end{equation}
	and
	\begin{equation}\label{eq336}
		C_G^J(\mathcal{N}) = C_G^J(\widetilde{\Theta}_s(\mathcal{G}_s)) \le C_G^J(\mathcal{G}_s) = g(s).
	\end{equation}
	According to the definition of the infimum, one can choose a sequence of achievable parameters $s_n$ such that
    $s_n \to c(\mathcal{N})$ when $n\to\infty$.
	Taking the limits of Eqs.~\eqref{eq335} and~\eqref{eq336}, we obtain that
	\begin{equation}\label{eq338}
		C_S^J(\mathcal{N}) \le f(c(\mathcal{N})),
	\end{equation}
	and
	\begin{equation}\label{eq339}
		C_G^J(\mathcal{N}) \le g(c(\mathcal{N})).
	\end{equation}
	Therefore, combining Eqs.~\eqref{eq332} and\eqref{eq338} yields Eq.\eqref{eq313}, and combining Eqs.~\eqref{eq333} and~\eqref{eq339} yields Eq.\eqref{eq314}. 
	
	Finally, it is easy to see that $f(t)$ and $g(t)$ are strictly increasing on $[0,1]$, and
	\begin{equation}
		g^{-1}(x) = 2\sqrt{x(1-x)}.
	\end{equation}
	Taking the inverse function yields Eqs.~\eqref{eq315} and~\eqref{eq316}. 
\end{proof}

\end{appendices}

\end{widetext}

\end{document}